\documentclass[10pt]{article}
\usepackage[BOE]{express}

\usepackage{amsmath,graphicx}
\usepackage{lipsum}
\usepackage{amssymb}
\usepackage{siunitx}
\usepackage{algorithmic}
\usepackage{commath}
\usepackage{color}
\usepackage{threeparttable}
\usepackage{url}
\usepackage{booktabs}
\usepackage{tabularx}
\usepackage{media9}
\usepackage{mathtools}
\usepackage{subcaption}
\usepackage{multicol}
\usepackage{scalerel}

\graphicspath{ {fig_original_images/} {fig_cornea_prior_algo_comparison/} {fig_modelFlow/} {fig_compare_unet_cornet/} {fig_arch/} {fig_binaryMask/} {fig_results_cornea/} {fig_results_limbus/} {fig_results_limbus_bad/} {validationGood/} {miscFig/} {validationAll/} {fig_results_limbusToCornea/}}

\begin{document}

\title{Accurate Tissue Interface Segmentation via Adversarial Pre-Segmentation of Anterior Segment OCT Images}
%\title{Adversarial Pre-Segmentation of Anterior Segment OCT Images Enables Accurate Surface Segmentation}
%\title{OCT Image Denoising for Surface Segmentation by Generative Adversarial Network}

\author{Jiahong Ouyang $^{1*}$, Tejas Sudharshan Mathai $^{1*}$, Kira Lathrop $^{2,3}$, John Galeotti $^{1,2}$}

\address{$^{1}$ The Robotics Institute, Carnegie Mellon University, USA	\\
$^{2}$ Department of Bioengineering, University of Pittsburgh, USA	\\
$^{3}$ Department of Ophthalmology, University of Pittsburgh, USA \\
$^{*}$ Equal contribution}

\email{\{jiahongo, tmathai\}@andrew.cmu.edu, lathropkl@upmc.edu, jgaleotti@cmu.edu} %% email address is required



%-------------------------------------------------------------------
%-------------------------------------------------------------------
\begin{abstract}
Optical Coherence Tomography (OCT) is an imaging modality that has been widely adopted for visualizing corneal, retinal and limbal tissue structure with micron resolution. It can be used to diagnose pathological conditions of the eye, and for developing pre-operative surgical plans. In contrast to the posterior retina, imaging the anterior tissue structures, such as the limbus and cornea, results in B-scans that exhibit increased speckle noise patterns and imaging artifacts. These artifacts, such as shadowing and specularity, pose a challenge during the analysis of the acquired volumes as they substantially obfuscate the location of tissue interfaces. To deal with the artifacts and speckle noise patterns and accurately segment the shallowest tissue interface, we propose a cascaded neural network framework, which comprises of a conditional Generative Adversarial Network (cGAN) and a Tissue Interface Segmentation Network (TISN). The cGAN pre-segments OCT B-scans by removing undesired specular artifacts and speckle noise patterns just above the shallowest tissue interface, and the TISN combines the original OCT image with the pre-segmentation to segment the shallowest interface. We show the applicability of the cascaded framework to corneal datasets, demonstrate that it precisely segments the shallowest corneal interface, and also show its generalization capacity to limbal datasets. We also propose a hybrid framework, wherein the cGAN pre-segmentation is passed to a traditional image analysis-based segmentation algorithm, and describe the improved segmentation performance. To the best of our knowledge, this is the first approach to remove severe specular artifacts and speckle noise patterns (prior to the shallowest interface) that affects the interpretation of anterior segment OCT datasets, thereby resulting in the accurate segmentation of the shallowest tissue interface. 
\end{abstract}

%For a complete list of OCIS codes, visit: https://www.osapublishing.org/oe/submit/ocis/

%-------------------------------------------------------------------
%-------------------------------------------------------------------

%\clearpage

%%%%%%%%%%%%%%%%%%%%%%% References %%%%%%%%%%%%%%%%%%%%%%%%%

%-------------------------------------------------------------------
%-------------------------------------------------------------------
\section{Introduction}
\label{sec:intro}
%-------------------------------------------------------------------
%-------------------------------------------------------------------

Optical coherence tomography (OCT) is a non-invasive and non-contact imaging technique that has been widely adopted for imaging sub-surface tissue structures with micrometer depth resolution in clinical ophthalmology \cite{Huang1991, Fujimoto2009}. OCT is a popular method to visualize structures in the eye, especially those in the retina \cite{Huang1991}, cornea \cite{Izatt1994}, and limbus \cite{Lathrop2012}. Specific to the anterior segment of the eye, OCT has been clinically used to characterize the changes that occur during the progression of Keratoconus \cite{Kuo2012, Venkateswaran2018}, diagnose benign and malignant conjunctival and corneal pathologies, such as Ocular Surface Squamous Neoplasia \cite{Lathrop2012,Venkateswaran2018}, and monitor potential complications for many anterior segment surgical procedures, such as Deep Anterior Lamellar Keratoplasty (DALK) \cite{Keller2018} and Descemet Membrane Endothelial Keratoplasty (DMEK) \cite{Venkateswaran2018}. Furthermore, OCT has been used to image the limbus \cite{Lathrop2012,Bizheva2011,Bizheva2017}, and enabled the analysis of the Palisades of Vogt (POV) \cite{Haagdorens2017}.

\begin{figure}[!h]
\centering
\begin{subfigure}[b]{0.3\columnwidth}
\centering
\includegraphics[height=3.5cm,width=0.9\columnwidth]{dc9_i25}\\
\centerline{(a)}
\end{subfigure}
\begin{subfigure}[b]{0.125\columnwidth}
\centering
\includegraphics[height=3.5cm,width=0.9\columnwidth]{dc31_i0212}\\
\centerline{(b)}
\end{subfigure}
\begin{subfigure}[b]{0.125\columnwidth}
\centering
\includegraphics[height=3.5cm,width=0.9\columnwidth]{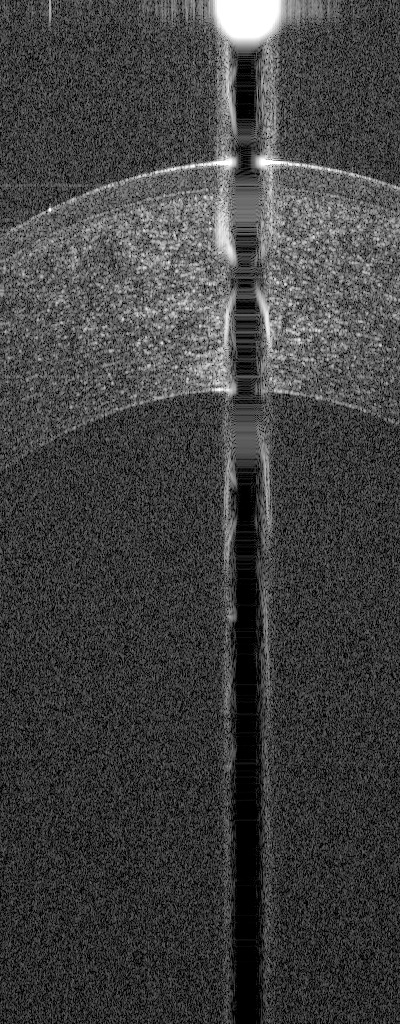}\\
\centerline{(c)}
\end{subfigure}
\begin{subfigure}[b]{0.145\columnwidth}
\centering
\includegraphics[height=3.5cm,width=0.9\columnwidth]{dp2_i0026}\\
\centerline{(d)}
\end{subfigure}
\begin{subfigure}[b]{0.125\columnwidth}
\centering
\includegraphics[height=3.5cm,width=0.9\columnwidth]{dp3_i00100}\\
\centerline{(e)}
\end{subfigure}
\begin{subfigure}[b]{0.125\columnwidth}
\centering
\includegraphics[height=3.5cm,width=0.9\columnwidth]{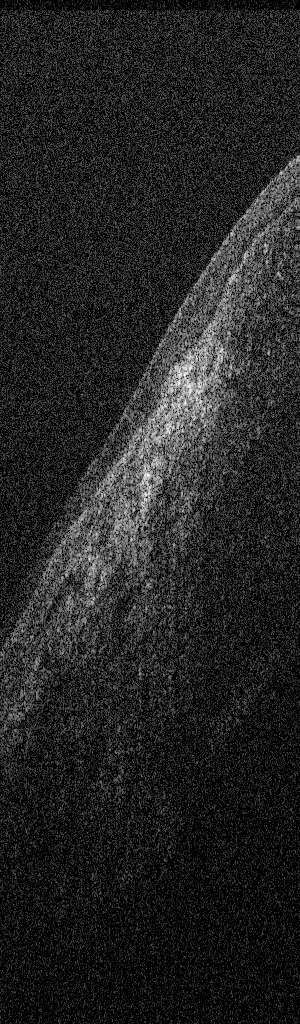}\\
\centerline{(f)}
\end{subfigure}
\caption{Original B-scans from (a) a 6$\times$6mm corneal volume acquired by a custom SD-OCT scanner, (b) a 6$\times$6mm corneal volume and (c) a 3$\times$3mm corneal volume acquired by a UHR-OCT scanner, (d) a 4$\times$4mm limbal volume acquired by a hand-held Leica SD-OCT scanner, and (e)-(f) 4$\times$4mm limbal volumes acquired by a UHR-OCT scanner. Specular artifacts in (a)-(d) and poor visibility in (e)-(f) affect the precise delineation of the tissue interfaces.}
\label{fig:fig_original_images}
\end{figure}

In all these applications, accurate estimation of the corneal or limbal tissue boundaries is required to determine a quantitative parameter for diagnosis or treatment. For example, in \cite{Kuo2012}, the corneal tissue interfaces were identified to estimate corneal biometric parameters. In \cite{Haagdorens2017}, the shallowest limbal interface was first identified, and then the tissue structure visualized in the image was ``flattened'' \cite{LaRocca2011,Haagdorens2017,Mathai2018} to enable the measurement of the palisade density. However, precise estimation of the corneal and limbal tissue interface location is challenging in anterior segment OCT imaging. The low signal-to-noise ratio (SNR), increased speckle noise patterns, and predominant specular artifacts pose barriers towards automatic delineation of the tissue interfaces (see Fig. \ref{fig:fig_original_images}). Furthermore, datasets are typically acquired in a clinical setting using different OCT scanners (including custom-built OCT scanners for clinical research) from different vendors as shown in Fig. \ref{fig:fig_original_images}. The scan settings of these OCT machines are usually different, thereby resulting in datasets with different image dimensions, SNR, speckle noise patterns, and specular artifacts. 

%--------------
%--------------

\begin{figure}[!h]
\centering
\begin{subfigure}[b]{0.125\columnwidth}
\centering
\includegraphics[height=3.5cm,width=0.9\columnwidth]{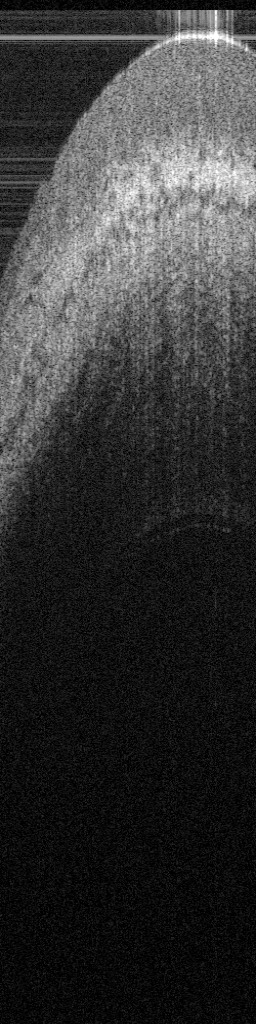}\\
\centerline{(a)}
\end{subfigure}
\begin{subfigure}[b]{0.125\columnwidth}
\centering
\includegraphics[height=3.5cm,width=0.9\columnwidth]{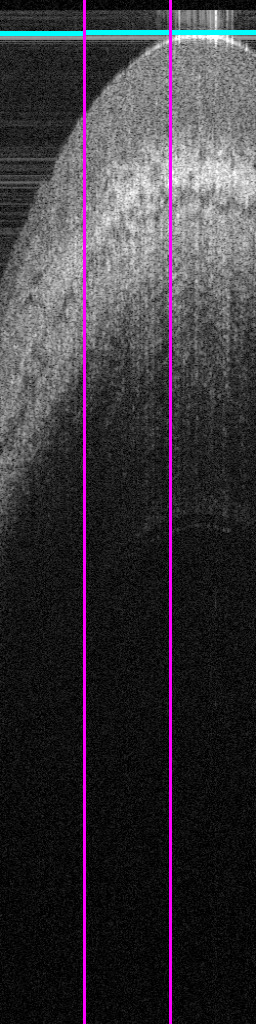}\\
\centerline{(b)}
\end{subfigure}
\begin{subfigure}[b]{0.125\columnwidth}
\centering
\includegraphics[height=3.5cm,width=0.9\columnwidth]{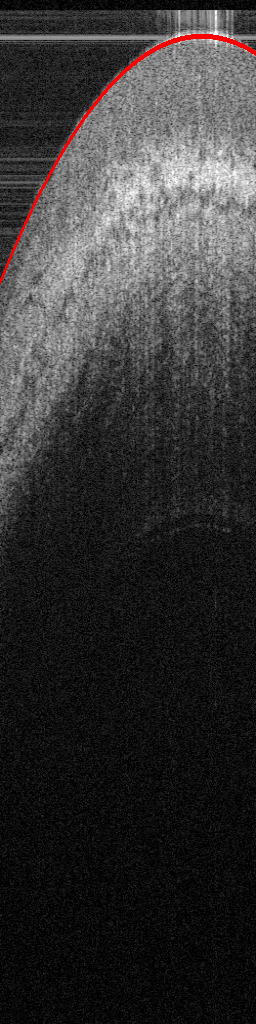}\\
\centerline{(c)}
\end{subfigure}
\begin{subfigure}[b]{0.125\columnwidth}
\centering
\includegraphics[height=3.5cm,width=0.9\columnwidth]{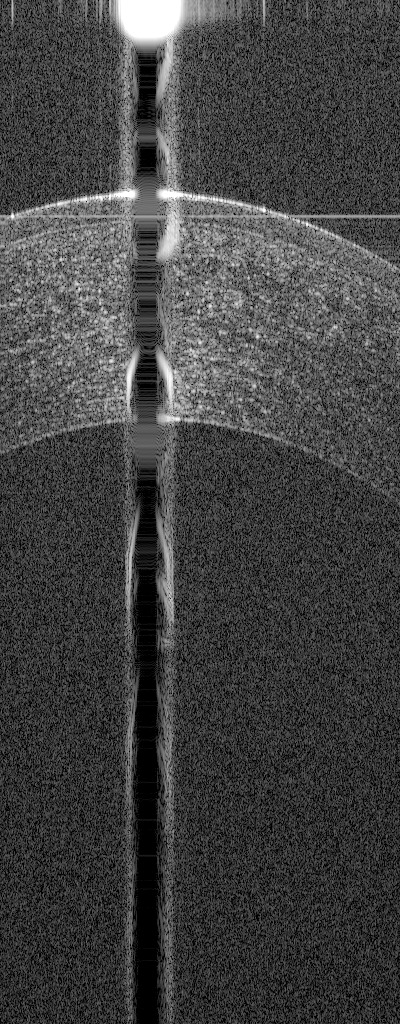}\\
\centerline{(d)}
\end{subfigure}
\begin{subfigure}[b]{0.125\columnwidth}
\centering
\includegraphics[height=3.5cm,width=0.9\columnwidth]{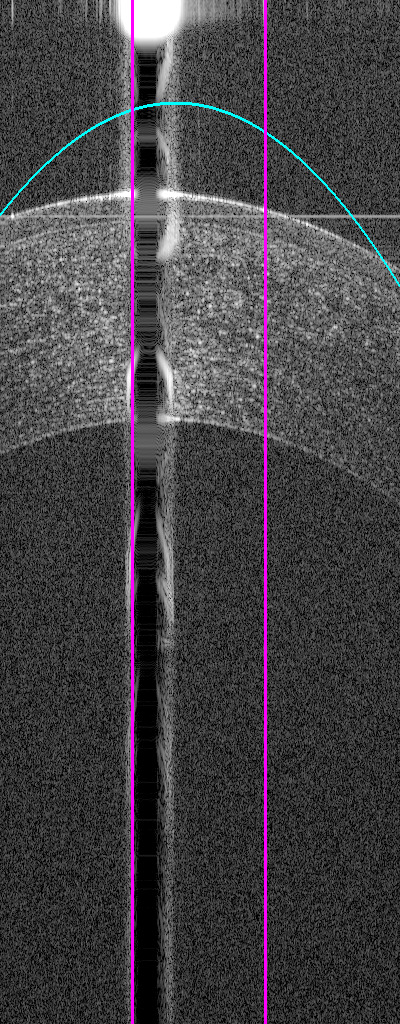}\\
\centerline{(e)}
\end{subfigure}
\begin{subfigure}[b]{0.125\columnwidth}
\centering
\includegraphics[height=3.5cm,width=0.9\columnwidth]{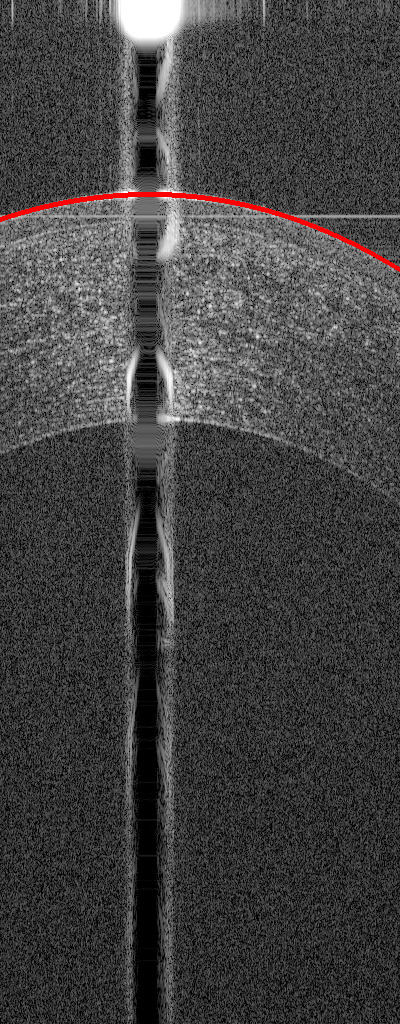}\\
\centerline{(f)}
\end{subfigure}
\caption{(a),(d) Original B-scans from a 4$\times$4mm limbal dataset acquired using a hand-held SD-OCT scanner and from a 3$\times$3mm corneal dataset acquired using a UHR-OCT scanner   respectively. As proposed in previous algorithms \cite{Kuo2012,Keller2018,LaRocca2011,Ge2012,Williams2013,Li_2_2006,Williams2016,Rabbani2016,Jahromi2014,Schmoll2012,Zhang2017}, vertical lines (magenta) denote the division of the image into three regions in order to deal with specular artifacts. (b),(e) Segmentation of the shallowest interface (cyan contour) by these algorithms failed due to presence of specular artifacts in different regions in the image. (c),(f) Segmentation result (red curve) from the proposed cascaded framework that accurately determined the location of shallowest tissue interface.}
\label{fig:fig_prior_algo_comparison}
\end{figure}

Speckle noise patterns and specular artifacts are major factors that influence the correct interpretation of anterior segment OCT images. To mitigate these degradations, there are many hardware- and software-based approaches that process each B-scan before they are analyzed in a segmentation pipeline. Hardware-based speckle noise reduction techniques \cite{Szkulmowski2012,Desjardins2007,Hughes2010} rely on the acquisition of multiple tomograms with decorrelated speckle patterns, such that they can be averaged to obtain images with lower speckle contrast. These techniques usually require modification of the OCT system's optical configuration and/or its scanning protocols. Software-based methods include wavelet transformations \cite{Adler2004,Ozcan2007,Puvanathasan2007,Gargesha2008,Chitchian2009,Hongwei2011}, local averaging and median filtering \cite{Pircher2003,Rogowska2002}, percentile and bilateral filtering \cite{Mathai2018}, regularization \cite{Marks2005}, local Bayesian estimation \cite{Wong2010}, and diffusion filtering \cite{Bernardes2010}. Efforts were also made to remove artifacts by using the reference spectrum \cite{Moon2010,LaRocca2011}, and piezoelectric fiber stretchers \cite{Vergnole2008} in the Fourier domain. However, these methods only work when a fixed type of artifact is encountered, such as the horizontal artifacts in \cite{Moon2010,LaRocca2011}, and they do not generalize to datasets where the assumption of the artifact presence is violated \cite{Mathai2018_2} as seen in Fig. \ref{fig:fig_prior_algo_comparison}. Furthermore, all the prior work are not robust when the SNR dropoff is substantial, which is typically the case while imaging the limbus; the anatomic curvature (and thus orientation toward the OCT scanner) changes when moving away from the cornea and towards the limbus, thereby causing a significant decrease in visibility of tissue boundaries as seen in Figs. \ref{fig:fig_original_images}(d) - \ref{fig:fig_original_images}(f). Particularly in our case, datasets were acquired by OCT scanners that imaged the limbal junction; the OCT scanner commenced scanning at the limbus and crossed over to the cornea, thereby incorporating the limbal junction during image acquisition. At the limbus, often only the shallowest interface is visible, and as the scanner crosses the limbal junction to image the cornea, different interfaces are gradually seen, such as the Bowman's Layer etc. In this work, we focus on delineating the shallowest tissue interface in all corneal and limbal datasets.  %On a related note, speckle patterns have been used to segment different tissue interfaces in posterior retinal OCT images \cite{Szkulmowski2012}. 

Towards the goal of mitigating these image degrading factors, a recent learning-based method featuring a conditional Generative Adversarial Network (GAN) \cite{Goodfellow2014,Isola2017} was proposed to remove the speckle noise patterns in retinal OCT images \cite{Radford2016, Ma2018}. It also generalized to datasets acquired from multiple OCT scanners. Although qualitatively good results were obtained, the central premise in their approach was based on limited (little to none) eye motion between frames during imaging. The ground truth data was generated using a compounding technique; the same tissue area was imaged multiple times, and individual volumes were registered yielding averaged B-scans for training, which corresponded to the gold standard despeckled images. However, in our case, this methodology to generate ground truth data for training is not feasible as corneal datasets exhibit large motion when acquired in-vivo, which makes registration and compounding challenging. In addition, existing research databases, from which corneal datasets can be extracted for use in algorithmic development, rarely contain multiple scans of the same tissue area for compounding. Moreover, the authors in \cite{Ma2018} opined that it was difficult to judge the efficacy of a despeckling algorithm using existing metrics, such as SNR or Contrast-to-Noise Ratio (CNR), as any one metric is not a good determining factor of the quality of the denoised image. They suggested that an alternate way to analyze the utility of a despeckling method was to estimate the improvement in segmentation accuracy following denoising.

To deal with these challenging scenarios, it is desirable for a tissue-interface segmentation algorithm to possess the following characteristics: 1) Robustness in the presence of speckle noise and artifacts, 2) Generalization capacity across datasets acquired from multiple OCT scanners with different scan settings, and 3) Applicability to different (anterior segment) anatomical regions. Currently, there are a myriad of prior approaches that directly segment corneal and retinal tissue interfaces. They can be broadly grouped into four categories: 1) Traditional image analysis-based segmentation algorithms, 2) Graph-based segmentation methods, 3) Contour modeling-based segmentation methods, and 4) Machine learning-based (including deep learning-based) segmentation algorithms. Traditional image analysis-based approaches filter the individual B-scans to enhance the contrast of tissue interfaces, and then threshold the image to segment the corneal \cite{Davidson2010,Shen2011,Mathai2018} and retinal \cite{Fernandez2005,Ishikawa2005,Fabritius2009} interface boundaries. These filters are typically hand-tuned and chosen for the explicit purpose of reducing speckle noise patterns and enhancing edges in the image for easier segmentation. Graph-based methods \cite{Li2006,Dufour2013,Shah2015,Boykov2006,Garvin2009,Shi2015,Lee2010,Song2013,Shah2014} pose the segmentation of the interfaces as an optimization problem, wherein tissue interfaces are detected subject to surface smoothness priors and the distance constraint between interfaces. Other graph-based methods \cite{Tian2015,Chiu2015} involve posing the boundary segmentation problem as a shortest-path finding approach, wherein the shortest path between a source node and sink node is deduced, given costs assigned to the nodes between them. Contour modeling approaches utilize active contours that dynamically change their shape based on shape metrics, such as deviation from a second order polynomial \cite{Yazdanpanah2009,Niu2016}, edge gradients \cite{Sisternes2017} underlying the contour etc.

Machine learning techniques express the segmentation problem as a classification task; features related to the tissue interfaces to be segmented are extracted, and then classified as belonging to the tissue boundary or background \cite{Lang2013,Ma2009,Kafieh2013}. In other cases, learning-based methods are an element of a hybrid system \cite{Antony2013,Fang2017}, wherein the generated output, or the intermediate learned features, improve/assist the performance of traditional/graph-based/contour modeling approaches. Currently, deep neural networks are the state-of-the-art algorithms \cite{Fang2017,Chen2017,Sui2017,Venhuizen2017,Roy2017,Shah2017,Lee2017} of choice for the segmentation task as they can learn highly discriminative multi-scale features from training data, thereby outperforming all other segmentation approaches. These neural network models are alluring because key algorithm parameters are learned from the training data, which are often manually tuned in other approaches - for example, hand-crafted parameters in traditional image analysis-based \cite{Davidson2010,Shen2011,Mathai2018,Fernandez2005,Ishikawa2005,Fabritius2009} and active contour-based approaches \cite{Yazdanpanah2009,Niu2016,Sisternes2017}. They can also be applied to pathological patients if appropriate datasets were introduced during the training procedure.

%--------------
%--------------

However, among all the aforementioned methods, the majority of traditional methods \cite{Dufour2013,Shah2015,Garvin2009,Shi2015,Lee2010,Song2013,Shah2014,Tian2015,Chiu2015} and learning-based methods \cite{Fang2017,Chen2017,Sui2017,Venhuizen2017,Roy2017,Shah2017,Lee2017,Apo2017} are focused on retinal interface segmentation. Corneal interface segmentation algorithms are predominately based on traditional approaches \cite{Kuo2012,Keller2018,LaRocca2011,Ge2012,Williams2013,Li_2_2006,Williams2016,Rabbani2016,Jahromi2014,Schmoll2012,Zhang2017}, with limited learning-based approaches \cite{Mathai2018_2,Santos2019} being proposed. Similarly, prior work on limbal interface segmentation is limited to a traditional image analysis-based approach \cite{Mathai2018}. Moreover, most of the prior work is suited towards the task of segmenting tissue interfaces of only one particular type of anatomy, such as retina or cornea, and these prior approaches are not easily generalizable across different types of anatomy. As shown in Fig. \ref{fig:fig_prior_algo_comparison}, most of the traditional approaches were not resilient when the methodology was transferred to our datasets obtained from different OCT scanners, which contained bulk tissue motion, severe specular artifacts and speckle noise patterns. 

As seen in Figs. \ref{fig:fig_prior_algo_comparison}(b) and \ref{fig:fig_prior_algo_comparison}(e), previous segmentation approaches would divide (A-scan-wise) the OCT image into three sections, and assume that the location of the central specular artifact was limited to the center of the OCT image (region between the vertical magenta lines) \cite{Kuo2012,Keller2018,LaRocca2011,Ge2012,Williams2013,Li_2_2006,Williams2016,Rabbani2016,Jahromi2014,Schmoll2012,Zhang2017}. But as seen in Fig. \ref{fig:fig_prior_algo_comparison}, this assumption can be violated when the central artifacts are located in different image regions \cite{Mathai2018_2}. In such cases, prior approaches failed to accurately segment the tissue interface as shown in Figs. \ref{fig:fig_prior_algo_comparison}(b) and \ref{fig:fig_prior_algo_comparison}(e). From our experiments, we postulated that most traditional algorithms are confounded by the presence of these strong specular artifacts and speckle noise patterns. Yet, once the shallowest interface is identified, these traditional approaches were able to delineate other interfaces, such as Bowman's Layer, Endothelium etc. 

Furthermore, there were two independent and concurrently published deep learning-based corneal interface segmentation approaches \cite{Mathai2018_2,Santos2019}. One of these approaches \cite{Santos2019} acquired data from a single OCT scanner, and focused only on the region centered around the corneal apex in these OCT sequences as the drop in SNR was greater when moving away from this region. The other approach is our recent publication \cite{Mathai2018_2}, where we utilized the entire OCT sequence from multiple scanners containing strong specular artifacts and low SNR regions, and successfully segmented three tissue interfaces. Yet, our previously proposed approach did not readily provide intermediate outputs, wherein the specular artifacts and speckle noise patterns were ameliorated, which could be used as input to the traditional approaches \cite{Kuo2012,Keller2018,LaRocca2011,Ge2012,Williams2013,Li_2_2006,Williams2016,Rabbani2016,Jahromi2014,Schmoll2012,Zhang2017, Mathai2018} for segmentation.  %It was also unclear if their methodology could precisely predict the interfaces in the presence of severe specular artifacts in datasets from different OCT devices manufactured by different vendors.

To this end, in this paper, we propose the first approach (to the best of our knowledge) to accurately identify the shallowest tissue interface in OCT images by mitigating speckle noise patterns and severe specular artifacts. We propose the creation of an intermediate OCT image representation that can influence the performance of a segmentation approach. Our major contributions in this paper are three-fold:

\begin{enumerate} %itemize
\item \textbf{Cascaded Framework}: We present a cascaded neural network framework, which comprises of a conditional Generative Adversarial Network (cGAN) and a Tissue Interface Segmentation Network (TISN). The cGAN pre-segments OCT images by removing undesired specular artifacts and speckle noise patterns just prior to the shallowest tissue interface. The pre-segmentation output of the cGAN is an intermediate output. Following pre-segmentation, the TISN predicts the final segmentation using both the original and pre-segmented images, and the shallowest interface is extracted and fitted with a curve.

\item \textbf{Hybrid Framework}: The intermediate pre-segmentation output yielded by the cGAN is used as the image input to another tissue-interface segmentation algorithm, e.g. \cite{Mathai2018}. In general, the pre-segmentation can be used by any segmentation algorithm, but in the Hybrid Framework the second-stage segmentation algorithm does not have access to the original OCT image.

\item \textbf{cGAN Weighted Loss}: We propose a task-specific weighted loss for the cGAN, which enforces the preservation of details related to the tissue structure, while removing specular artifacts and speckle noise patterns just prior to the shallowest interface in a context-aware manner.

\end{enumerate}

Our cascaded framework was first applied to corneal datasets, which were acquired using two different OCT systems and different scan protocols. Encouraged by our cascaded framework's performance on corneas, we diversified our training to also include limbal datasets (also acquired with different OCT systems).  It seemed reasonable to seek generalized learning since the characteristics of limbal datasets are similar to corneal datasets in terms of low SNR, speckle noise patterns, and specular artifacts. In all these datasets, we segmented the shallowest interface that could be extracted in each B-scan. 

A key motivation for the proposed hybrid framework was to directly integrate the output of the cGAN into the image acquisition pipeline of custom-built OCT scanners. As we postulated earlier, the varying degrees of specular artifacts and speckle noise patterns confound traditional segmentation algorithms. If the cGAN were integrated into the imaging pipeline and OCT B-scans were pre-segmented after acquisition, then we hypothesized that previously proposed segmentation algorithms should benefit from the removal of specular artifacts and speckle noise patterns just above the shallowest interface. Thus, our goal with the development of the hybrid framework was to show that the pre-segmented OCT image enabled one of these segmentation algorithms \cite{Mathai2018} to generate lower segmentation errors.

To quantify the performance of our proposed frameworks, we compared the results of the following baselines: 1) A traditional image analysis-based algorithm \cite{Mathai2018} that directly segmented the tissue interface, 2) The hybrid framework, 3) A deep learning-based approach \cite{Mathai2018_2} that directly segmented the tissue interface, and 4) The cascaded framework. We provide a summary of the major results below: 

\begin{enumerate} 

\item We show that our approach is generalizable to datasets acquired from multiple scanners displaying varying degrees of specular noise, artifacts, and bulk tissue motion.

\item Our proposed frameworks segment the shallowest interface in datasets where the scanner starts by imaging the limbus, crosses over the limbal junction, and images the cornea.

\item By executing a traditional image analysis-based algorithm on the pre-segmentation, the segmentation error was always reduced.

\item We always accurately segmented the shallowest interface in corneal datasets using our proposed frameworks. 

\item In a majority of limbal datasets (15/18), we were able to precisely delineate the shallowest interface with our proposed frameworks.

\end{enumerate} %itemize

%-------------------------------------------------------------------
%-------------------------------------------------------------------
\section{Methods}
\label{sec:method}
%-------------------------------------------------------------------
%-------------------------------------------------------------------

%-------------------------------------------------------------------
%-------------------------------------------------------------------
\subsection{Problem Statement}
%-------------------------------------------------------------------
%-------------------------------------------------------------------

Given an OCT image $\mathcal{I}$, the task of a conditional Generative Adversarial Network (cGAN) is to find a function $\mathcal{F}_{G}$ : $\{\mathcal{I}, z\} \rightarrow \mathcal{P}$ that maps a pixel in $\mathcal{I}$ using a random noise vector $z$ to a pre-segmented output image $\mathcal{P}$. The pixels in $\mathcal{P}$ just prior to the tissue interface are mapped to $0$ (black), while those at and below the interface are retained. $\mathcal{P}$ can then be used in a hybrid framework by any other segmentation algorithm.

\begin{figure}[!h]
\centering
\includegraphics[height=8.5cm,width=0.9\columnwidth]{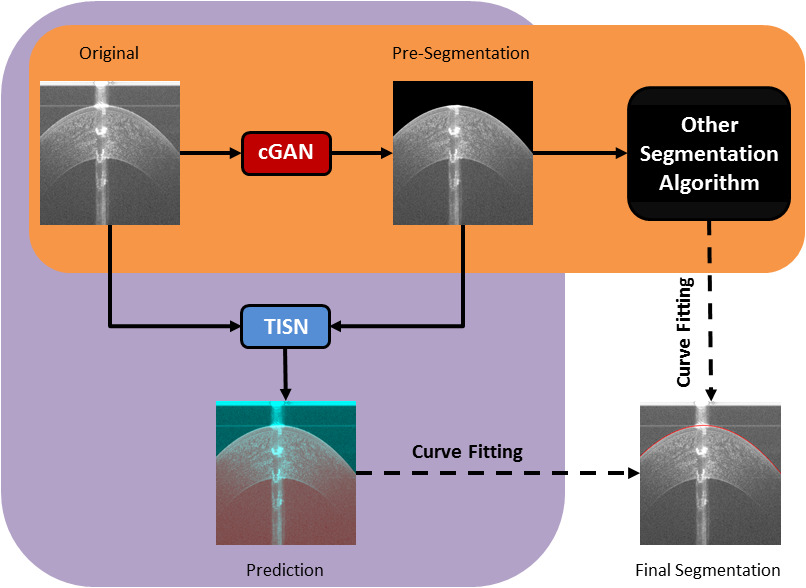}
\caption{Our proposed approach contains two frameworks: a cascaded framework (purple) and a hybrid framework (orange). First, a conditional Generative Adversarial Network (cGAN) takes an input OCT image, and produces an intermediate pre-segmentation image. In the pre-segmentation, pixels just prior to the shallowest tissue interface are set to 0 (black), while others are retained. In the cascaded framework, the pre-segmentation, along with the input image, are passed to a Tissue Interface Segmentation Network (TISN). The TISN predicts the location of shallowest interface by generating a binary segmentation mask (overlaid on the original image with a false color overlay; red - foreground, turquoise - background). In the hybrid framework, the pre-segmentation can be utilized by other segmentation algorithms. Ultimately, both frameworks fit a curve to the interface to produce the final segmentation.}
\label{fig:modelFlow}
\end{figure}

Next, the task of the Tissue Interface Segmentation Network (TISN) is to determine a mapping $\mathcal{F}_{O}$ : $ \{ \mathcal{I}, \mathcal{P} \} \rightarrow \mathcal{S}$, wherein every corresponding pixel in $\mathcal{I}$ and $\mathcal{P}$ is assigned a label $\mathcal{L} \in \{0,1\}$ in the final segmentation $\mathcal{S}$. In this paper, we only segment the shallowest tissue interface in the image, and thus assign pixels in $\mathcal{S}$ as: (0) pixels just above the tissue interface, (1) pixels at and below the tissue interface. Our frameworks are pictorally shown in Fig. \ref{fig:modelFlow}.

%--------------
%--------------

%-------------------------------------------------------------------
%-------------------------------------------------------------------
\subsection{Architecture}
\label{arch}
%-------------------------------------------------------------------
%-------------------------------------------------------------------

We first describe the neural network architecture that was used as the base for both the cGAN (generator), and the TISN. As mentioned in Sec. \ref{sec:intro}, images of the anterior segment of the eye acquired using OCT contain low SNR, strong specular artifacts, and faintly discernable interfaces that are corrupted by speckle noise patterns. In our previous work \cite{Mathai2018_2}, we have shown that the CorNet architecture captures faintly visible features across multiple scales. It produced state-of-the-art results on corneal datasets acquired using different OCT systems and using different scan protocols. The errors were 2$\times$ lower than non-proprietary state-of-the-art segmentation algorithms, including traditional image analysis-based \cite{LaRocca2011,Zhang2017} and deep learning-based approaches \cite{Roy2017,Ronneberger2015,Apo2017}. 

The CorNet architecture was built upon the BRUNET \cite{Apo2017} architecture, and enhanced the reuse of features generated in the network through residual connections \cite{He2016}, dense connections \cite{Huang2017}, and dilated convolutions \cite{Koltun2016,Devalla2018,Szegedy2015}. It alleviated the vanishing gradient problem, and prevented the holes in the segmentation generated by current deep learning-based approaches \cite{Ronneberger2015,Roy2017,Apo2017}. It could accurately extract poorly defined corneal interfaces, such as the Endothelium, which is very common in anterior segment OCT imaging \cite{Mathai2018_2}. 

%Yet, the original UNET \cite{Ronneberger2015} architecture, which is the base network incorporated in many prior approaches \cite{Venhuizen2017,Roy2017,Lee2017}, generated holes in the final segmentation as shown in Fig. \ref{fig:fig_compare_unet_cornet}, which lead to the development of new types of network architectures \cite{Apo2017,Shah2018}. We believe that the observed holes are due to gradients related to the desired region of interest (RoI) becoming smaller as they pass through the various decoding and encoding layers of the network during backpropagation. The gradients ``vanish'' before they reach the input layer, and do not allow the network to learn that the RoI is part of the final segmentation. These new networks incorporated current ideas in deep learning, such as dilated convolutions \cite{Koltun2016,Devalla2018,Szegedy2015}, batch normalization \cite{Ioffe2015}, residual connections \cite{He2016}, and bottleneck connections \cite{Szegedy2015}, culminating in segmentation outputs that were able accurately estimate tissue interfaces. 

\begin{figure}[!h]
\centering
\includegraphics[height=10cm,width=\columnwidth]{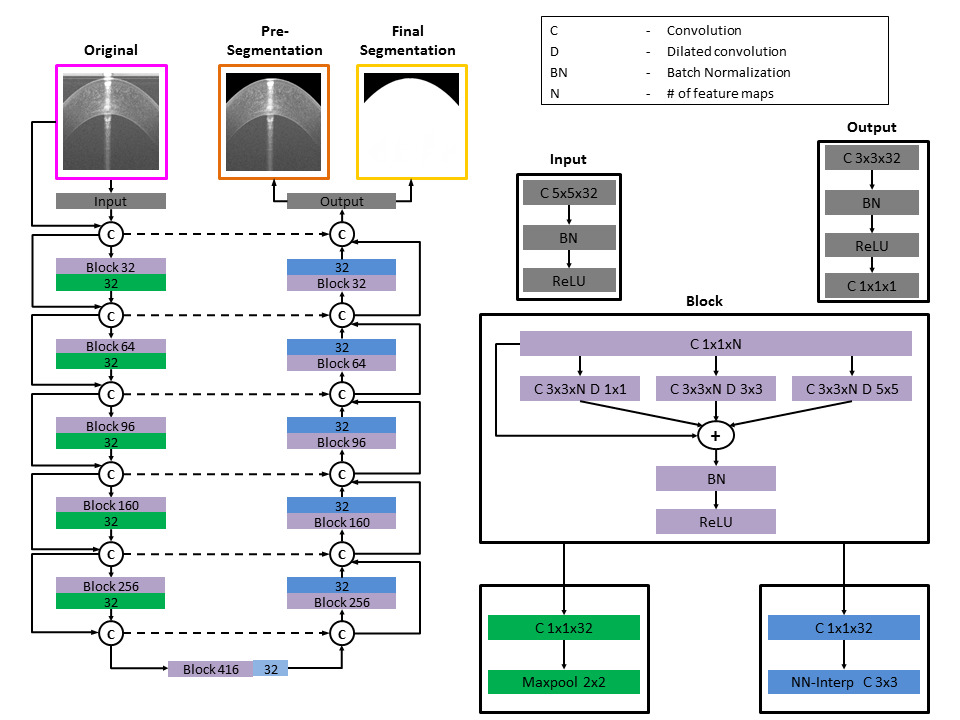}
\caption{The CorNet model is the base architecture used for training both the cGAN and TISN. The input to the cGAN is a two-channel image, the input OCT image and binary mask $w$ (see Sec. \ref{data_prep}), and the output is a pre-segmented OCT image (orange box). The TISN gets a two-channel input (magenta and orange boxes), and the output is a binary mask (yellow box). The dark green blocks in the contracting path represent downsampling operations, while the blue blocks constitute upsampling computations. This model uses residual and dense connections to efficiently pre-segment the OCT image, and  predict the location of the shallowest interface in the final output. The light blue module at the bottom of the model did not upsample feature maps, instead it functioned as a bottleneck to create outputs with the same size as those from the last layer.}
\label{fig:fig_arch}
\end{figure}

As shown in Fig. \ref{fig:fig_arch}, the CorNet architecture comprised of contracting and expanding branches; each branch consisted of a building block, which was inspired by the Inception block \cite{Szegedy2015}, followed by a bottleneck block. The building block extracted features related to edges and boundaries at different resolutions. The bottleneck block compactly represented the salient attributes, and these properties (even from earlier layers) were encouraged to be reused throughout the network. Thereby, faint tissue boundaries essential to our segmentation task were distinguished from speckle noise patterns, and pixels corresponding to the tissue interface and those below it were correctly predicted. In addition, extensive experiments were conducted in \cite{Mathai2018_2} to determine the right feature selection mechanisms \cite{Noh2015,Long2015,Odena2016,Roy2017,Khosravan2018} for segmentation, such as max-pooling \cite{Khosravan2018} for downsampling and nearest neighbor interpolation + 3$\times$3 convolution \cite{Odena2016} for upsampling.

%--------------
%--------------

%It was found that max-pooling picked the most discriminative trait from the feature maps that were learned at each layer. Similarly for feature map upsampling, nearest neighbor interpolation + 3$\times$3 convolution performed better than other upsampling methods, such as bilinear interpolation based upsampling, bilinear interpolation + 3$\times$3 convolution \cite{Odena2016}, unpooling \cite{Roy2017,Noh2015}, and fractionally-strided convolutions \cite{Long2015}. 

%As the PatchGAN is a Markovian discriminator \cite{Isola2017}, it could be applied to arbitrary sized images with only specific patches of the image of size p$\times$p pixels being considered, and pixels outside the patch were independent from those in the current local patch. Its performance can be understood in terms of penalizing the loss in texture; it punished deviations from the gold standard when the generator incorrectly removed high frequency structural information, such as edges and boundaries.

%-------------------------------------------------------------------
%-------------------------------------------------------------------
\subsection{Conditional Generative Adversarial Network (cGAN)}
%-------------------------------------------------------------------
%-------------------------------------------------------------------

%-------------------------------------------------------------------
%-------------------------------------------------------------------
\subsubsection{Original cGAN}
%-------------------------------------------------------------------
%-------------------------------------------------------------------

Conditional Generative Adversarial Networks \cite{Isola2017} are currently popular choices for image-to-image translation tasks, such as image super-resolution and painting style transfer. In these tasks, the cGAN learns to generate an output by being introduced to (conditioned on) an input image. The cGAN framework consists of two entities: a Generator (G) and a Discriminator (D). The generator G takes an input image $x$ and a random noise vector $z$, and generates a prediction $y_f$ that is similar to the desired gold standard output $y_t$. Next, the input $x$ is paired with $y_t$ and $y_f$, thereby creating two pairs of images respectively; the true gold standard pair ($x$, $y_t$) and the predicted pair ($x$, $y_f$). Then, the discriminator D recognizes the pair that most accurately represents the gold standard output desired. These two entities are trained in conjunction, such that they compete with each other; G tries to fool D by producing an output that closely resembles the gold standard, while D tries improve its ability to distinguish the two pairs of images. 

Initially, G generates a prediction $y_f$ that poorly resembles $y_t$. It learns to produce more realistic predictions by minimizing an objective function shown in Eq. \eqref{cGAN_final_Objective}. On the other hand, D tries to maximize this objective by accurately distinguishing the generated prediction $y_f$ from the true gold standard $y_t$. The objective function comprises of two losses: $L_{cGAN}$ in Eq. \eqref{cGAN_loss}, and $L_{1}$ in Eq. \eqref{l1_loss}, with $\lambda$ being a hyper-parameter. The ${L}_{1}$ loss penalizes regions in the generated output that differ from the ground truth image provided, thereby making the loss a ``structured'' loss \cite{Isola2017}. It forces the output of the generator to be close to the ground truth in the ${L}_{1}$ sense. This loss proved to result in less blurry outputs as opposed to the original GAN formulation \cite{Goodfellow2014}, which utilized an ${L}_{2}$ loss. The PatchGAN \cite{Isola2017} discriminator was employed to output the probability of a pair of images being real or fake. 

%The cGAN \cite{Isola2017} differs from the originally proposed GAN \cite{Goodfellow2014} as the latter learns to generate an output from the random noise distribution $z$ directly, \textit{without} the discrimintor being conditioned on (does not observe) the input $x$. In a cGAN, the input $x$ is observed by both the generator and discriminator \cite{Goodfellow2014,Isola2017}.

% ^^^^^^^^
\begin{equation}
\label{cGAN_final_Objective}
G^* = arg \ \underset{G}{\min} \ \underset{D}{\max} \ L_{cGAN}(G,D)+\lambda L_1(G)
\end{equation}
% ^^^^^^^^
\begin{equation}
\label{cGAN_loss}
L_{cGAN}(G, D) = E_{x,{y}_{t}}\bigg[log \: D(x,y_t)\bigg] + E_{x,z}\bigg[log(1-D(x, G(x,z))\bigg]
\end{equation}
% ^^^^^^^^
% ^^^^^^^^
\begin{equation}
\label{l1_loss}
L_1 = E_{x,{y}_{t},z} \bigg[ {\left\lVert {y}_{t} - G(x,z) \right\rVert}_{1} \bigg]
\end{equation}
% ^^^^^^^^

%The objective function in Eq. \eqref{cGAN_final_Objective} is the traditional objective  that has been used to generate impressive results for retinal denoising \cite{Isola2017,Ma2018}. 

\begin{figure}[h]
\centering
\begin{subfigure}[b]{0.12\columnwidth}
\centering
\includegraphics[height=3.5cm,width=0.9\columnwidth]{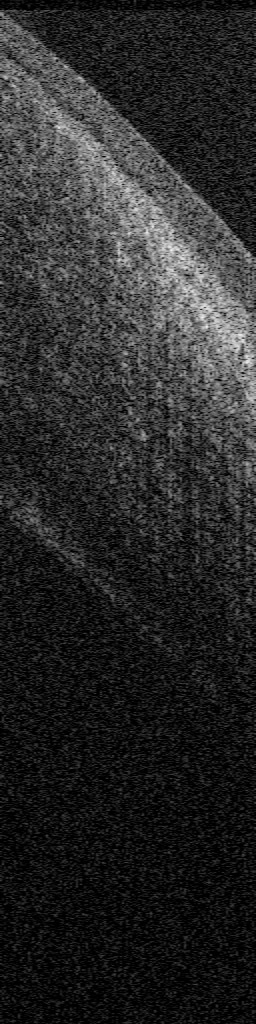}\\
\centerline{(a)}
\end{subfigure}
\begin{subfigure}[b]{0.12\columnwidth}
\centering
\includegraphics[height=3.5cm,width=0.9\columnwidth]{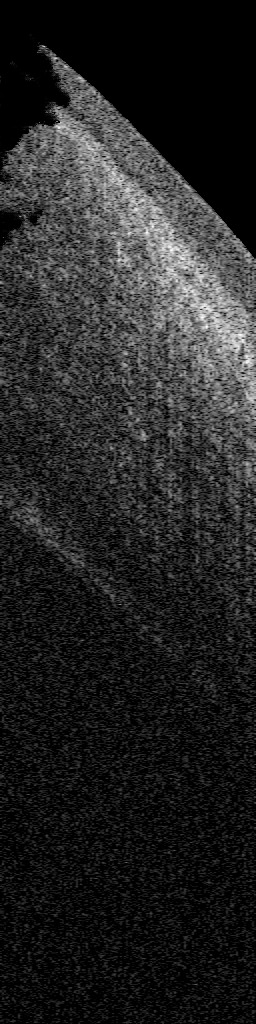}\\
\centerline{(b)}
\end{subfigure}
\begin{subfigure}[b]{0.12\columnwidth}
\centering
\includegraphics[height=3.5cm,width=0.9\columnwidth]{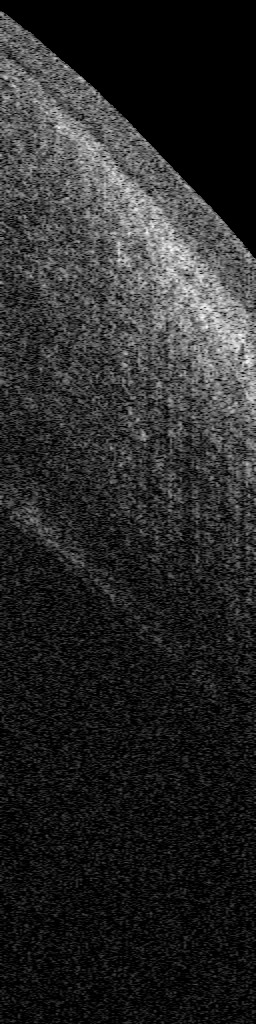}\\
\centerline{(c)}
\end{subfigure}
\begin{subfigure}[b]{0.12\columnwidth}
\centering
\includegraphics[height=3.5cm,width=0.9\columnwidth]{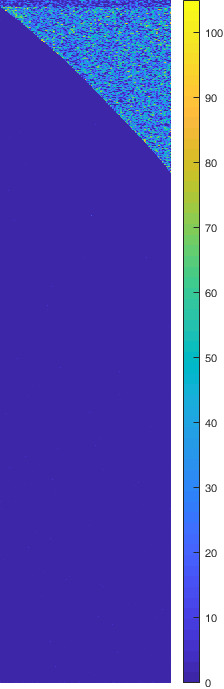}\\
\centerline{(d)}
\end{subfigure}
\begin{subfigure}[b]{0.12\columnwidth}
\centering
\includegraphics[height=3.5cm,width=0.9\columnwidth]{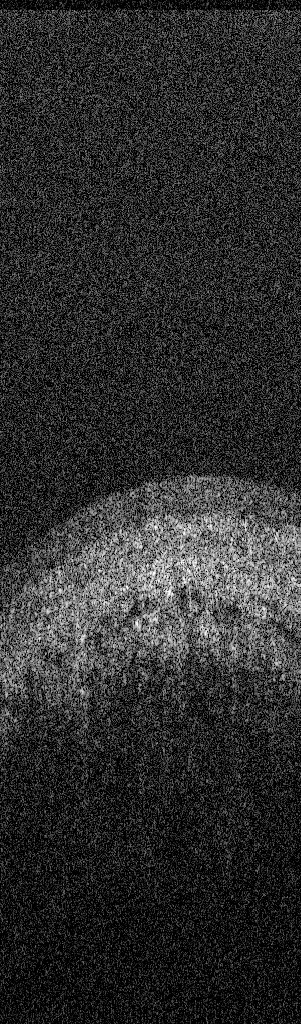}\\
\centerline{(e)}
\end{subfigure}
\begin{subfigure}[b]{0.12\columnwidth}
\centering
\includegraphics[height=3.5cm,width=0.9\columnwidth]{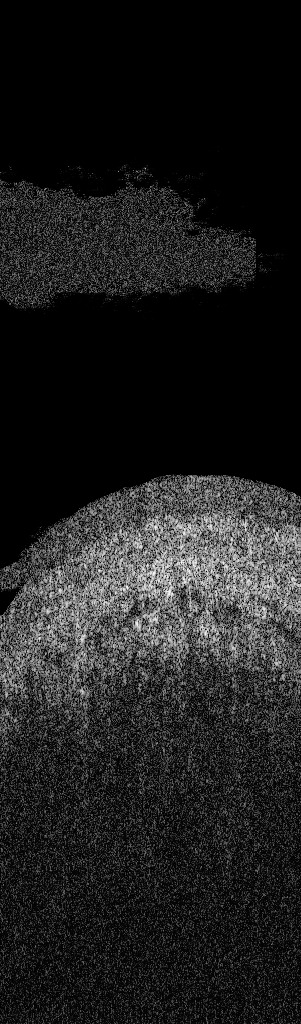}\\
\centerline{(f)}
\end{subfigure}
\begin{subfigure}[b]{0.12\columnwidth}
\centering
\includegraphics[height=3.5cm,width=0.9\columnwidth]{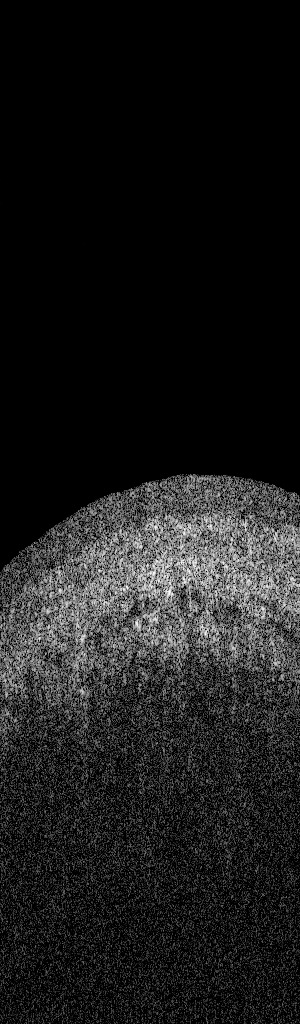}\\
\centerline{(g)}
\end{subfigure}
\begin{subfigure}[b]{0.12\columnwidth}
\centering
\includegraphics[height=3.5cm,width=0.9\columnwidth]{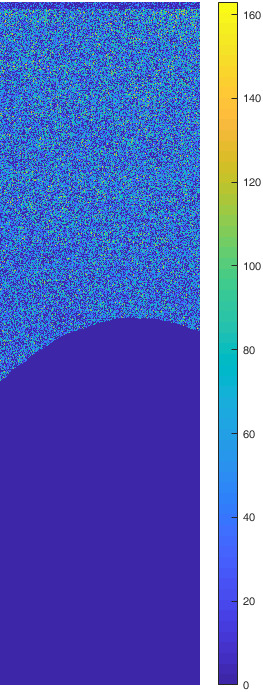}\\
\centerline{(h)}
\end{subfigure}
\caption{Comparing generated pre-segmentations between the U-Net architecture used in the original cGAN implementation \cite{Isola2017} against those generated by the CorNet architecture \cite{Mathai2018_2}. Original B-scans for two different limbal datasets are shown in (a) and (e) respectively, while the generated pre-segmentations for the cGAN U-Net is shown in (b) and (f), and the generated pre-segmentations for the CorNet are shown in (c) and (g). Note that in (b) and (f), the U-Net did not remove the speckle patterns above the shallow tissue interface, while also encroaching upon the tissue boundaries without preserving them accurately. (d) and (h) show heat maps of the difference between original and pre-segmented OCT B-scans by CorNet.}
\label{fig:fig_compare_unet_cornet}
\end{figure}

Directly transferring the full cGAN implementation with the cGAN loss in Eq. \eqref{cGAN_final_Objective} to our OCT datasets resulted in checkerboard artifacts \cite{Odena2016} in the generated predictions. Moreover, as shown in Fig. \ref{fig:fig_compare_unet_cornet}, parts of the tissue boundary that needed to be preserved were removed instead. From our experiments, we made two empirical observations: 1) The U-Net generator architecture \cite{Ronneberger2015} that was utilized in the cGAN paper \cite{Isola2017} created checkerboard artifacts in the generated pre-segmentation and did not preserve tissue boundaries correctly; it has been shown in prior work \cite{Odena2016,Apo2017,Mathai2018_2} that the original U-Net implementation is not the optimal choice; 2) The ${L}_{1}$ loss in Eq. \eqref{l1_loss} penalizes all pixels in the image equally. 

%--------------
%--------------

%-------------------------------------------------------------------
%-------------------------------------------------------------------
\subsubsection{Modified cGAN with Weighted Loss}
%-------------------------------------------------------------------
%-------------------------------------------------------------------

The required output of the cGAN is a pre-segmented OCT image, wherein the background pixels just prior to the shallowest tissue interface are to be eliminated, and the region at and below the interface is to be preserved. As mentioned before, the L1 loss in Eq. \eqref{l1_loss} equally penalized all pixels in the image without imparting a higher penalty to the background pixels, which contains specular artifacts and speckle noise patterns hindering segmentation, above the shallowest tissue interface. To mitigate this problem, a novel task-specific weighted ${L}_{1}$ loss, defined in Eq. \eqref{tswL1}, is proposed in this paper. In Eq. \eqref{tswL1}, $\circ$ denotes the pixel-wise product, and $\alpha$ is the hyper-parameter that imparts higher weight to the background pixels over the foreground pixels.

% ^^^^^^^^
\begin{equation}
\label{tswL1}
L_{w1} = E_{x,y,z} \bigg[ \alpha w \: \circ {\left\lVert {y}_{t} - G(x,z) \right\rVert}_{1} \: + \: (1 - w) \: \circ {\left\lVert {y}_{t} - G(x,z) \right\rVert}_{1} \bigg]
\end{equation}
% ^^^^^^^^

As the preservation of pixels at and below the interface is paramount, our loss function incorporated a binary mask $w$, which imparted different weights to the foreground and background pixels. This mask was generated from the gold standard annotation of an expert grader for each image in the training dataset, and its design is further described in Sec. \ref{data_prep}. We replaced the ${L}_{1}$ loss in Eq. \eqref{cGAN_final_Objective} with our weighted ${L}_{1}$ loss in Eq. \eqref{tswL1}, and it eliminated the speckle patterns and specular artifacts just prior to the shallowest interface. 

%-------------------------------------------------------------------
%-------------------------------------------------------------------
\subsection{Tissue Interface Segmentation Network (TISN)}
\label{tisn}
%-------------------------------------------------------------------
%-------------------------------------------------------------------

As mentioned in Sec. \ref{arch}, the CorNet architecture was used as the base model in order to segment the shallowest tissue interface. The intermediate pre-segmented OCT image from the cGAN, along with the original OCT image, is passed to the TISN to delineate the shallowest tissue interface. The output of the TISN is a binary mask, wherein pixels corresponding to the tissue interface and those below it were labeled as the foreground (1) and those above the interface were labeled as the background (0). As shown in Figs. \ref{fig:modelFlow} and \ref{fig:fig_arch}, the shallowest interface was extracted from this binary mask \cite{Mathai2014} and fitted with a curve \cite{Lowess1981}.

\section{Experiments and Results}
%-------------------------------------------------------------------
%-------------------------------------------------------------------

%-------------------------------------------------------------------
%-------------------------------------------------------------------
\subsection{Data}
%-------------------------------------------------------------------
%-------------------------------------------------------------------

%-------------------------------------------------------------------
%-------------------------------------------------------------------
\subsubsection{Acquisition}
%-------------------------------------------------------------------
%-------------------------------------------------------------------

25 corneal datasets and 25 limbal datasets, totaling 50 datasets, were randomly selected from an existing research database \cite{Mathai2018_2}. These datasets were acquired using different scan protocols from three different OCT scanners: a custom Bioptigen Spectral Domain OCT (SD-OCT) scanner (Device 1) that has been described before \cite{Wang2014}, a high-speed ultra-high resolution OCT (hsUHR-OCT) scanner (Device 2) \cite{Srinivasan2014}, and a Leica (formerly Bioptigen) Envisu C2300 SD-OCT system (Device 3) \cite{Leica}. Device 1 had a $3.4 \mu m$ axial and $6 \mu m$ lateral spacing, and it was used to scan an area of size $6 \times 6$mm on the cornea. Device 2 was used to scan two areas of sizes 6$\times$6mm and 3$\times$3mm respectively. This system had a 1.3\SI{}{\micro\meter} axial and a 15\SI{}{\micro\meter} lateral spacing while interrogating the 6$\times$6mm tissue area. It had the same axial spacing, but a different lateral spacing of 7.5\SI{}{\micro\meter} while imaging the 3$\times$3mm area. Device 3 had a $\sim$2.44\SI{}{\micro\meter} axial and 12\SI{}{\micro\meter} lateral spacing when fitted with the 18mm anterior imaging lens. Devices 1 and 2 were solely used to scan the cornea, with the former producing datasets of dimensions 1024$\times$1000$\times$50 pixels, and the latter generating datasets of dimensions 400$\times$1024$\times$50 pixels. Devices 2 and 3 were used to scan the limbus, resulting in volumes that had varying dimensions; the number of A-scans across all limbal datasets varied between 256 and 1024, with a constant 1024 pixels axial resolution, and the number of B-scans across all datasets varied between 25 and 375.

%-------------------------------------------------------------------
%-------------------------------------------------------------------
\subsubsection{Data Preparation}
\label{data_prep}
%-------------------------------------------------------------------
%-------------------------------------------------------------------

From the 50 datasets, we had a total of 1250 corneal images and 4437 limbal images respectively. Of the 50 corneal and limbal datasets, 14 datasets were randomly chosen for training the cGAN, and the remaining were used for testing. These datasets were chosen such that they came from both eyes; the number of patients that were imaged could not be ascertained as the database contained deidentified datasets. From the total set, we chose the training set to comprise of a balanced number of limbal and corneal datasets (7 each) that exhibited different magnitudes of specular artifacts, shadowing, and speckle. The training set contained 350 corneal and 1382 limbal images respectively, and the remaining were set aside in the testing set. Considering the varying dimensions of the OCT images acquired from three OCT systems that were used in this work, along with the limited GPU RAM available for training, it was challenging to train a framework using full-width images while preserving the pixel resolution. Similar to previous approaches \cite{Roy2017,Mathai2018_2}, we sliced the input images width-wise to produce a set of images of dimensions 256$\times$1024 pixels, and in this way, we preserved the OCT image resolution. We used the same datasets that were selected in the training set for training both the cGAN and the TISN.

\begin{figure}[h]
\centering
\begin{subfigure}[b]{0.245\columnwidth}
\centering
\includegraphics[height=3cm,width=0.9\columnwidth]{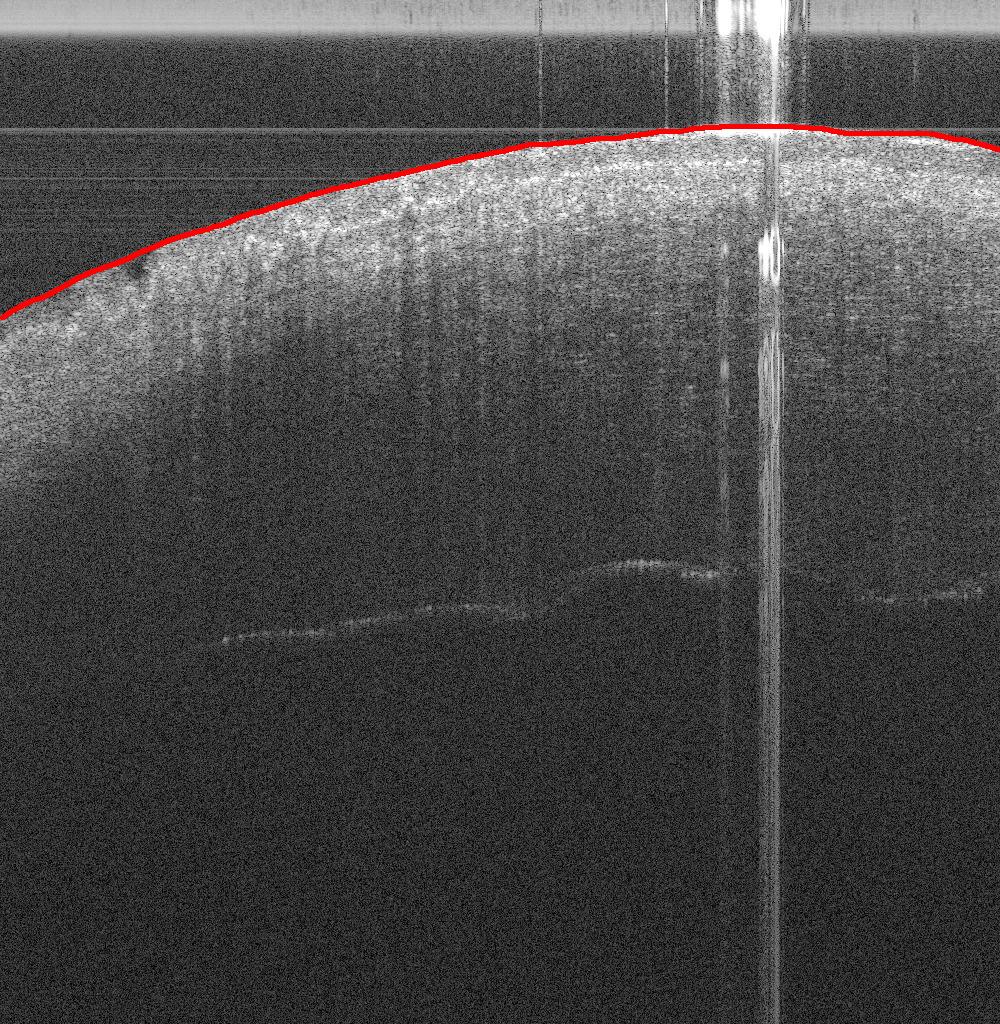}\\
\centerline{(a)}
\end{subfigure}
\begin{subfigure}[b]{0.245\columnwidth}
\centering
\includegraphics[height=3cm,width=0.9\columnwidth]{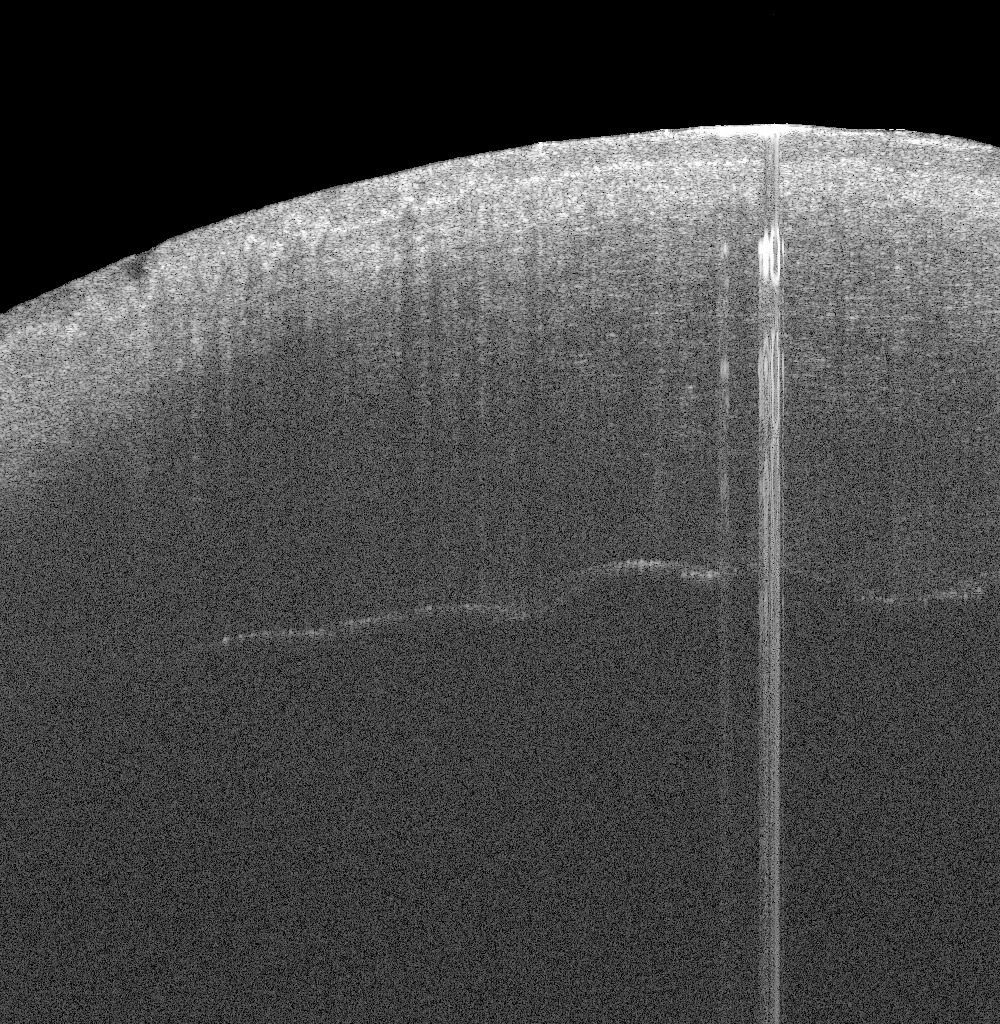}\\
\centerline{(b)}
\end{subfigure}
\begin{subfigure}[b]{0.245\columnwidth}
\centering
\includegraphics[height=3cm,width=0.9\columnwidth]{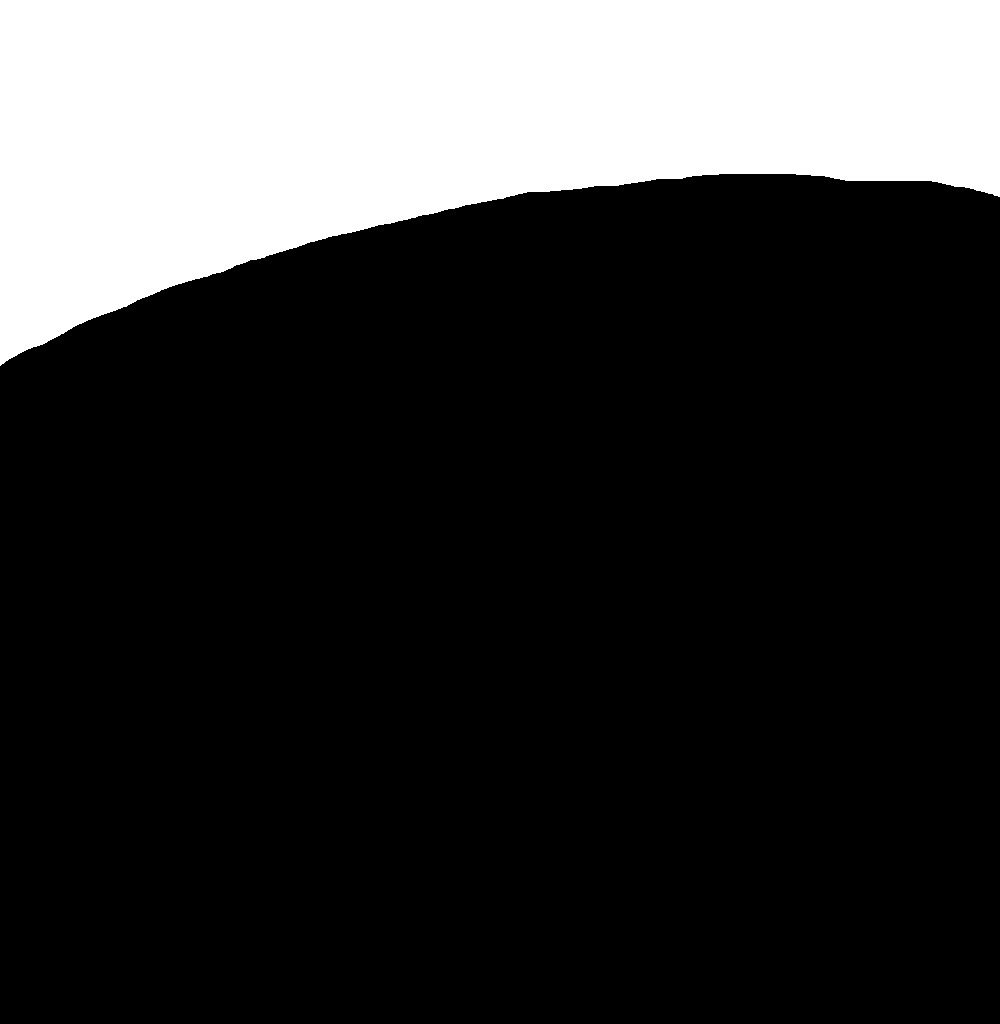}\\
\centerline{(c)}
\end{subfigure}
\begin{subfigure}[b]{0.245\columnwidth}
\centering
\includegraphics[height=3cm,width=0.9\columnwidth]{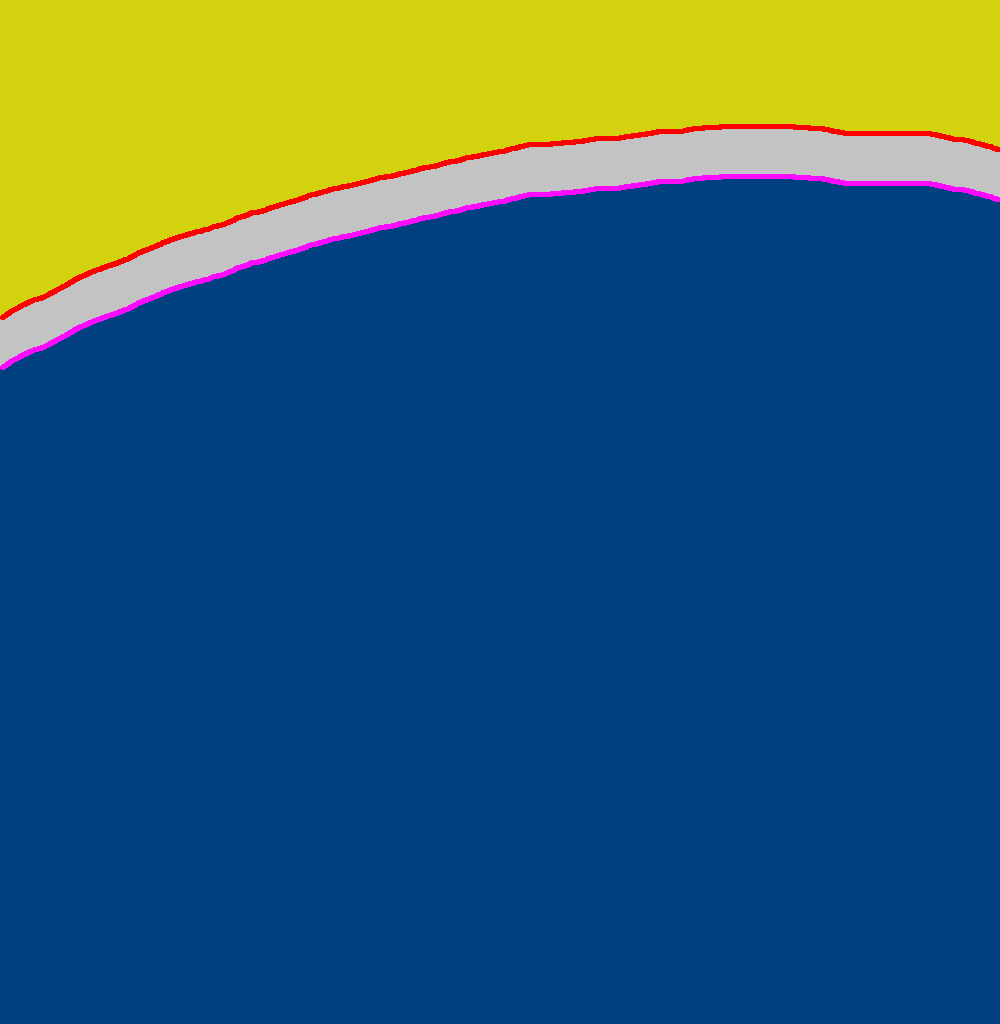}\\
\centerline{(d)}
\end{subfigure}
\caption{(a) Expert annotation of an original B-scan in a 6$\times$6mm limbal volume acquired by Device 3, (b) Gold standard pre-segmentation image for training, (c) Binary mask $w$ used in Eq. \eqref{tswL1} for training the cGAN, (d) Label map detailing the process of generating $w$ (see Sec. \ref{data_prep}).}
\label{fig:fig_anno_shiftUp}
\end{figure}

An example annotation by an expert grader is shown in Fig. \ref{fig:fig_anno_shiftUp}(a). To generate the gold standard pre-segmentation images for training, we eliminated the speckle noise and specular artifacts by setting the region just above the annotated surface to 0 (black), and kept the same pixel intensities corresponding to the tissue structure at the annotation contour and for all pixels below it - see Fig. \ref{fig:fig_anno_shiftUp}(b). The binary mask $w$ that was used in the Eq. \eqref{tswL1} is shown in Fig. \ref{fig:fig_anno_shiftUp}(c). Using the image in Fig. \ref{fig:fig_anno_shiftUp}(d) as reference, we detail the process of obtaining $w$. In Fig. \ref{fig:fig_anno_shiftUp}(d), the original annotation of the tissue interface boundary by the grader is shown in red, and this red annotated contour was shifted down by 50 pixels to the position of the magenta contour. The magenta contour, along with the blue region below the contour, was considered the foreground, while all pixels above the magenta contour belong to the background. The background in the binary mask was set to 1 and the foreground was set to 0, with the background being weighted $\alpha$ times higher than the foreground. 

In order to understand the effect of the proposed mask design, let us consider an alternate binary mask design ${w}^{*}$. Let ${w}^{*}$ represent the mask of the expert annotation in Fig. \ref{fig:fig_anno_shiftUp}(a), wherein the pixels above the annotation (without shifting it down/up) are the background and those at and below the annotation are the foreground, with the background weighted $\alpha$ times higher than the foreground. When the cGAN used this mask ${w}^{*}$, it mistakenly eroded the tissue interface and regions below it similar to the image in Fig. \ref{fig:fig_compare_unet_cornet}(b). In such a scenario, there is no large penalty applied to the erosion of pixels as detailed in Eq. \eqref{tswL1}. In order to correct this mistake, it would be necessary to impart a higher penalty to the region that was eroded. To do so, we measured the maximum extent of structural erosion (at the tissue interface and/or pixels below it) from the shallowest interface in the UNET pre-segmentation outputs. Using this value (rounded up to a nearest multiple of 10), we shifted expert annotation down (by 50 pixels) in our binary mask ${w}$, and conferred the same weight $\alpha$ to the regions (green + red + gray) to avoid the erosion of the tissue interface.

%--------------
%--------------

%-------------------------------------------------------------------
%-------------------------------------------------------------------
\subsubsection{Data Augmentation}
%-------------------------------------------------------------------
%-------------------------------------------------------------------

As our training datasets were smaller in number in contrast to those from datasets typically available in computer vision tasks, such as image recognition \cite{Russakovsky2015}, we augmented our datasets to increase the variety of the images that were seen during the training. These augmentations \cite{Patrice2003} included horizontal flips, gamma adjustment, elastic deformations, Gaussian blurring, median blurring, bilateral blurring, Gaussian noise addition, cropping, and affine transformations. The full set of augmented images was used to train the TISN as it required substantially larger amounts of data to generalize to new test inputs. On the other hand, the cGAN can be trained with smaller quantities of input training data as it has been shown to perform well on small training datasets \cite{Isola2017}. For the cGAN, augmentation was done by simply flipping each input slice horizontally along the X-axis.

%-------------------------------------------------------------------
%-------------------------------------------------------------------
\subsection{Experimental Setup}
%-------------------------------------------------------------------
%-------------------------------------------------------------------

%-------------------------------------------------------------------
%-------------------------------------------------------------------
\subsubsection{cGAN Training}
%-------------------------------------------------------------------
%-------------------------------------------------------------------

Training of the cGAN commenced from scratch using the architecture shown in Fig. \ref{fig:fig_arch}. The input to the generator was a two-channel image; the first channel corresponds to the input OCT image, and the second channel corresponds to the binary mask ${w}$. We used $\lambda$ = 100, and $\alpha$ = 10 in final objective function, and optimized the network parameters using the ADAM optimizer \cite{Kingma2015}. We used 90\% of the input data for training, and the remaining 10\% for validation. We trained the network for 100 epochs with the learning rate set to $2\times10^{-3}$. In order to prevent the network from over-fitting to the training data, early stopping was applied when the validation loss did not decrease for 10 epochs. At the last layer of the generator, a convolution operation, followed by a TanH activation, was used to convert the final feature maps into the desired output pre-segmentation with pixel values mapped to the range of $[-1,1]$. A NVIDIA Tesla V100 16GB GPU was used for training the cGAN with a batch size of 4. During test time, the input OCT image is replicated to produce a two-channel input to the cGAN. 

%The single output channel instead of four for the denoised image is modified with tanh activation to suit the pixel value's range of $[-1,1]$ in GAN.

%-------------------------------------------------------------------
%-------------------------------------------------------------------
\subsubsection{TISN Training}
%-------------------------------------------------------------------
%-------------------------------------------------------------------

The same datasets from cGAN training were used for training the TISN from scratch. The input to the TISN is a two-channel image; the first channel corresponds to the original input image, and the second channel corresponds to the predicted pre-segmentation obtained from the cGAN. The two-channel input allowed the TISN to focus on the high frequency regions, corresponding to the interface, in the image. The Mean Squared Error (MSE) loss, along with the ADAM optimizer \cite{Kingma2015}, was used for training. In this work, we used MSE loss to be consistent with the original CorNet implementation \cite{Mathai2018_2}, but the MSE loss can easily be substituted for the cross entropy loss \cite{Ronneberger2015} or the dice loss \cite{Milletari2016}. The batch size used for training was set to 2 slices as we fully wanted to utilize memory on a NVIDIA Titan Xp GPU. Validation data comprised of 10\% of the training data. We trained the network for a total of 150 epochs with the learning rate set to $10^{-3}$. When the validation loss did not improve for 5 epochs, the learning rate was decreased by a factor of 2. Finally, in order to prevent over-fitting, the training of the TISN was halted through early stopping when the validation loss did not improve for 10 consecutive epochs. 

\begin{figure}[!h]
\centering
\begin{subfigure}[b]{0.245\columnwidth}
\centering
\includegraphics[height=3cm,width=0.9\columnwidth]{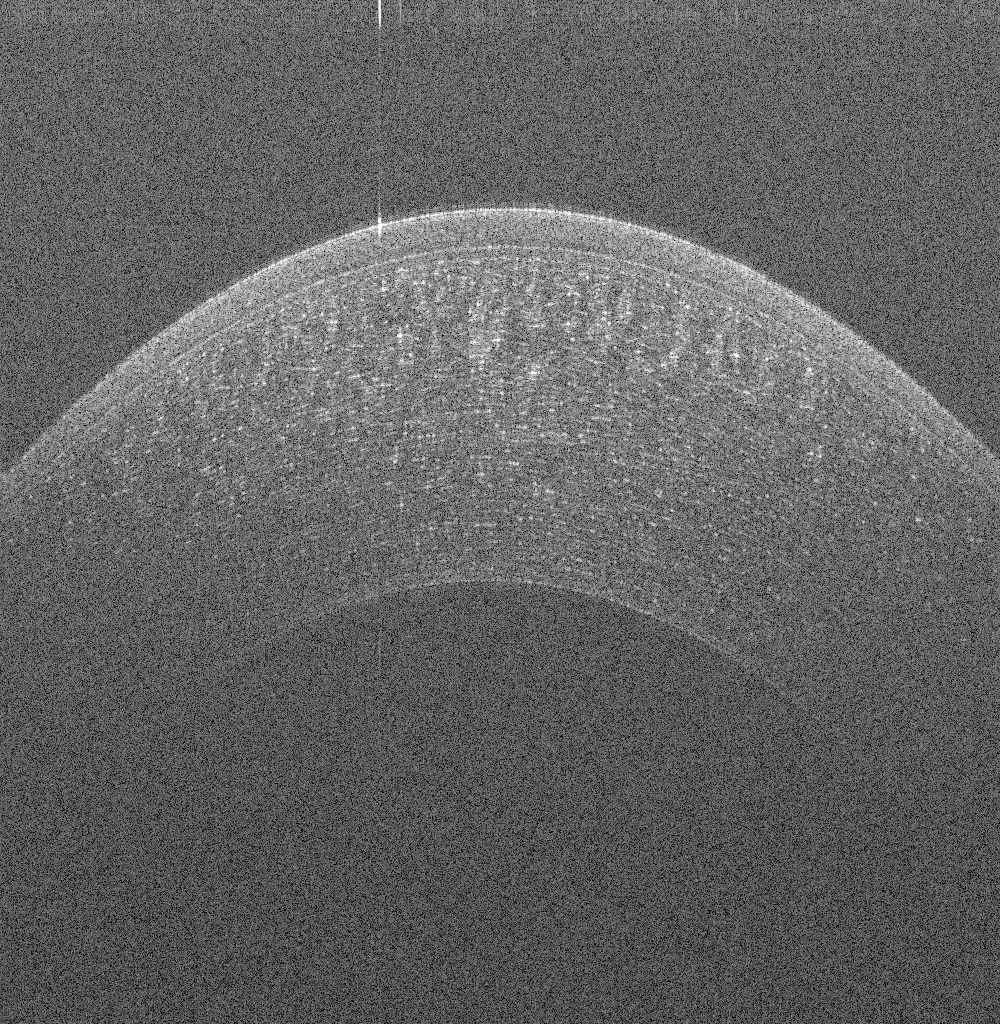}\\\medskip
\includegraphics[height=3cm,width=0.9\columnwidth]{dc9_i25}\\\medskip
\includegraphics[height=3cm,width=0.4\columnwidth]{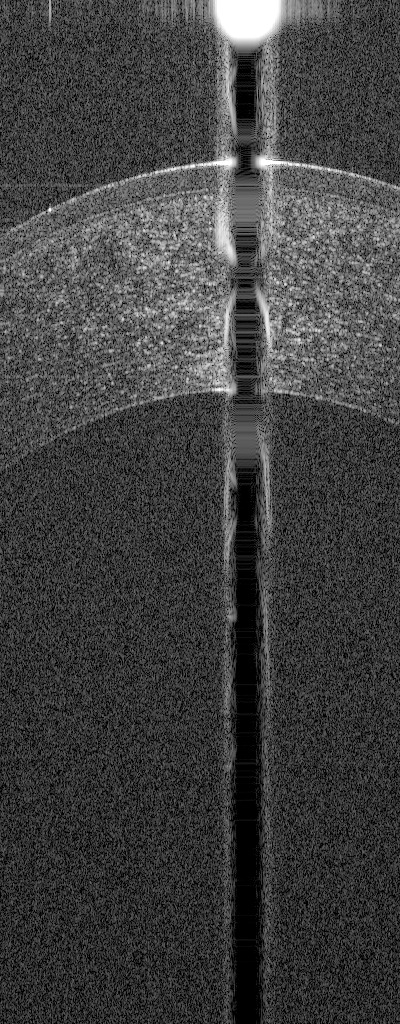}\\\medskip
\includegraphics[height=3cm,width=0.4\columnwidth]{dc31_i0212}
\centerline{(a)}
\end{subfigure}
\begin{subfigure}[b]{0.245\columnwidth}
\centering
\includegraphics[height=3cm,width=0.9\columnwidth]{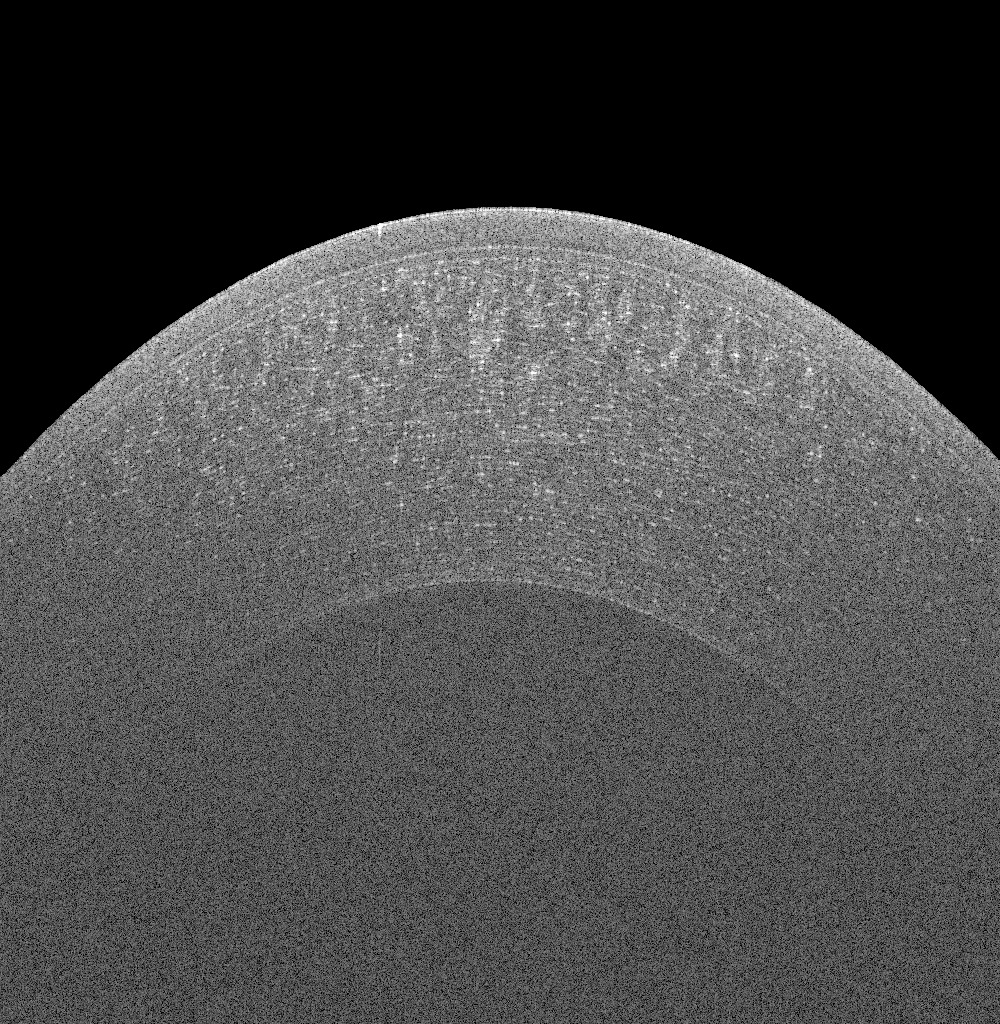}\\\medskip
\includegraphics[height=3cm,width=0.9\columnwidth]{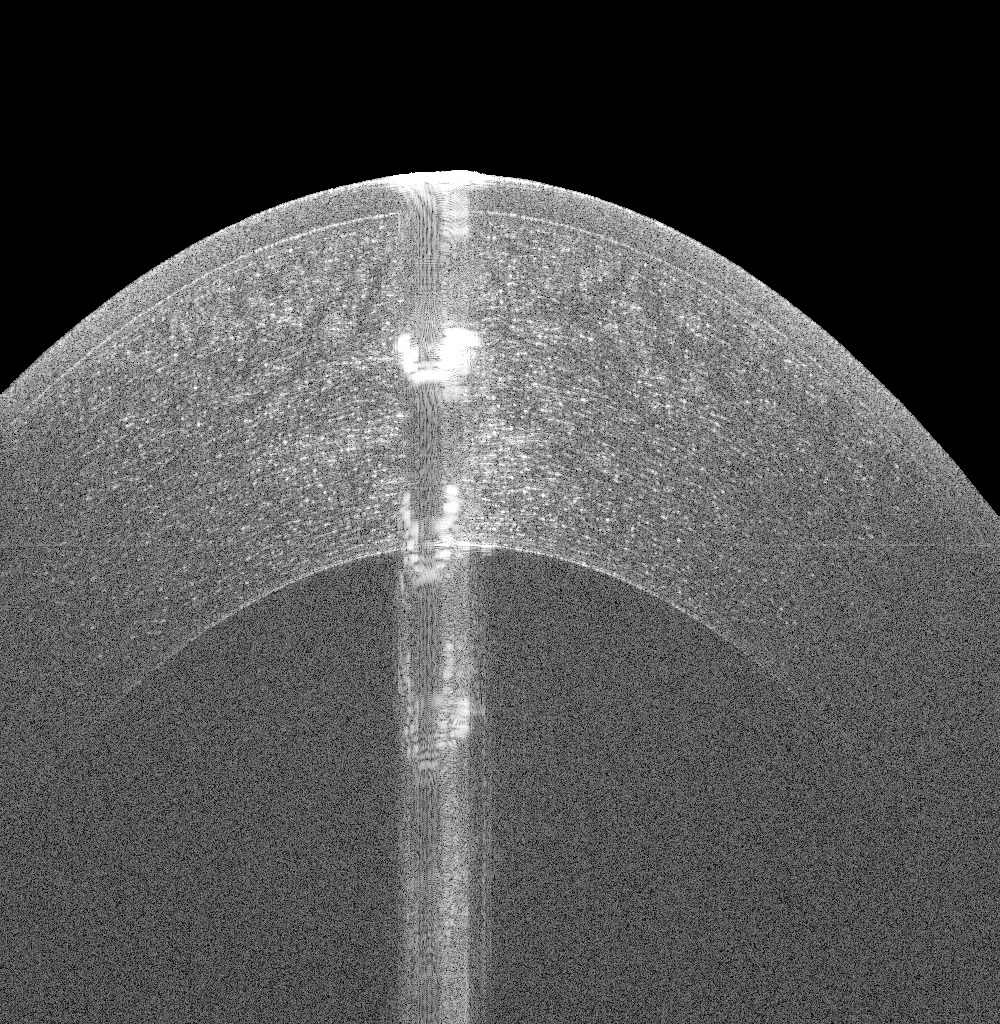}\\\medskip
\includegraphics[height=3cm,width=0.4\columnwidth]{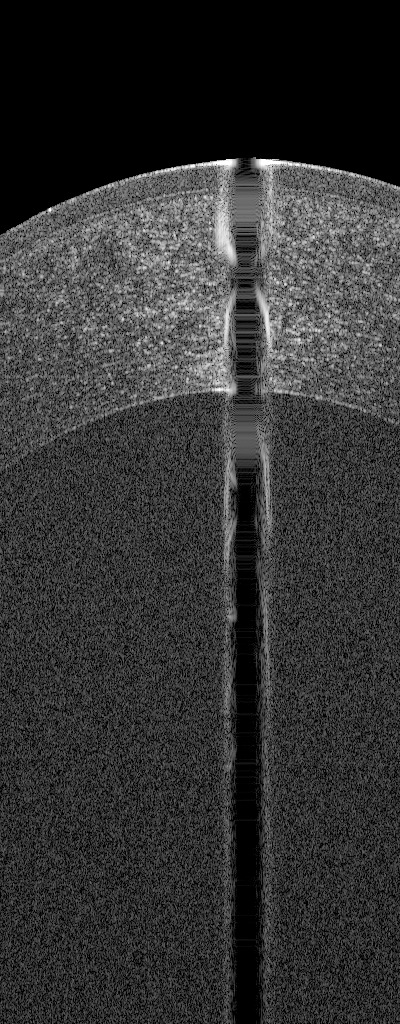}\\\medskip
\includegraphics[height=3cm,width=0.4\columnwidth]{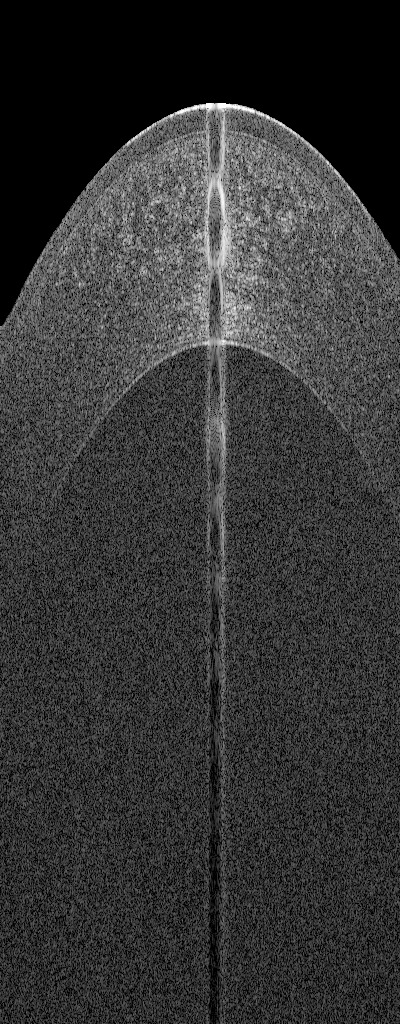}
\centerline{(b)}
\end{subfigure}
\begin{subfigure}[b]{0.245\columnwidth}
\centering
\includegraphics[height=3cm,width=0.9\columnwidth]{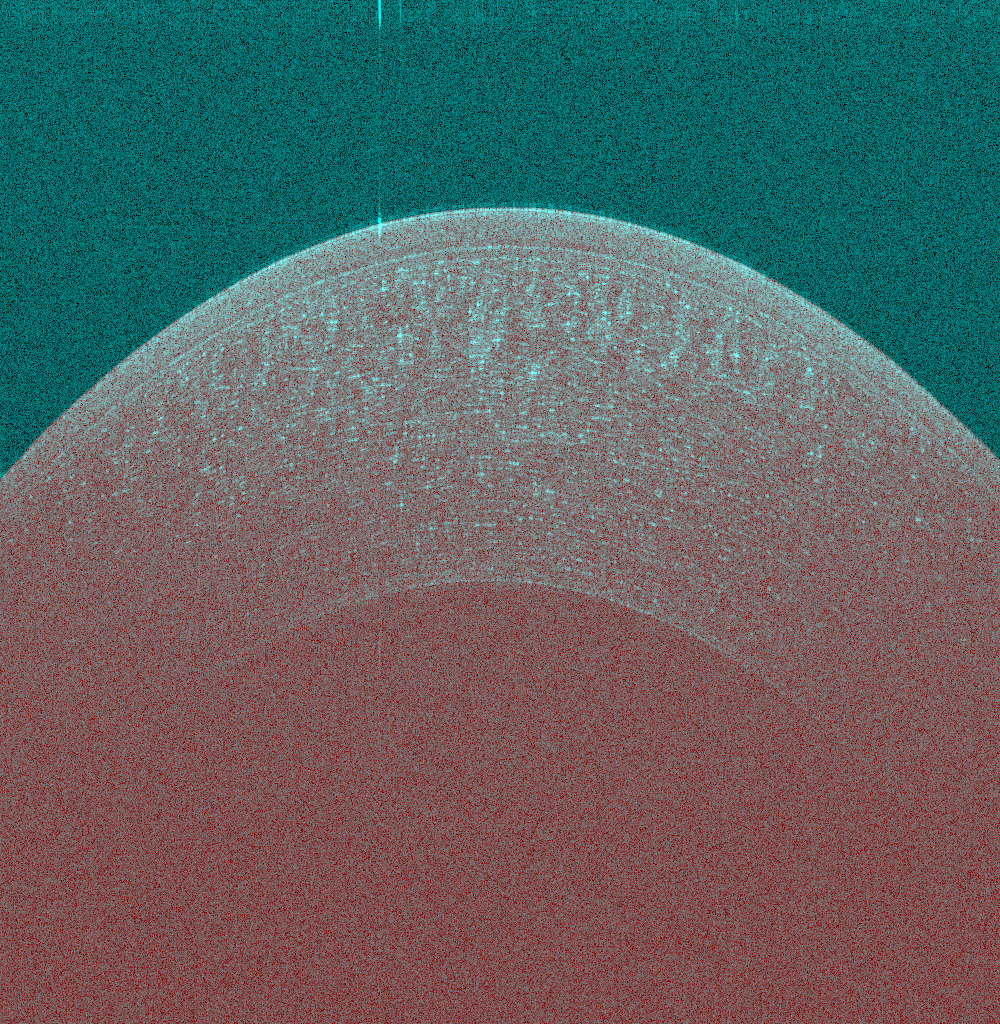}\\\medskip
\includegraphics[height=3cm,width=0.9\columnwidth]{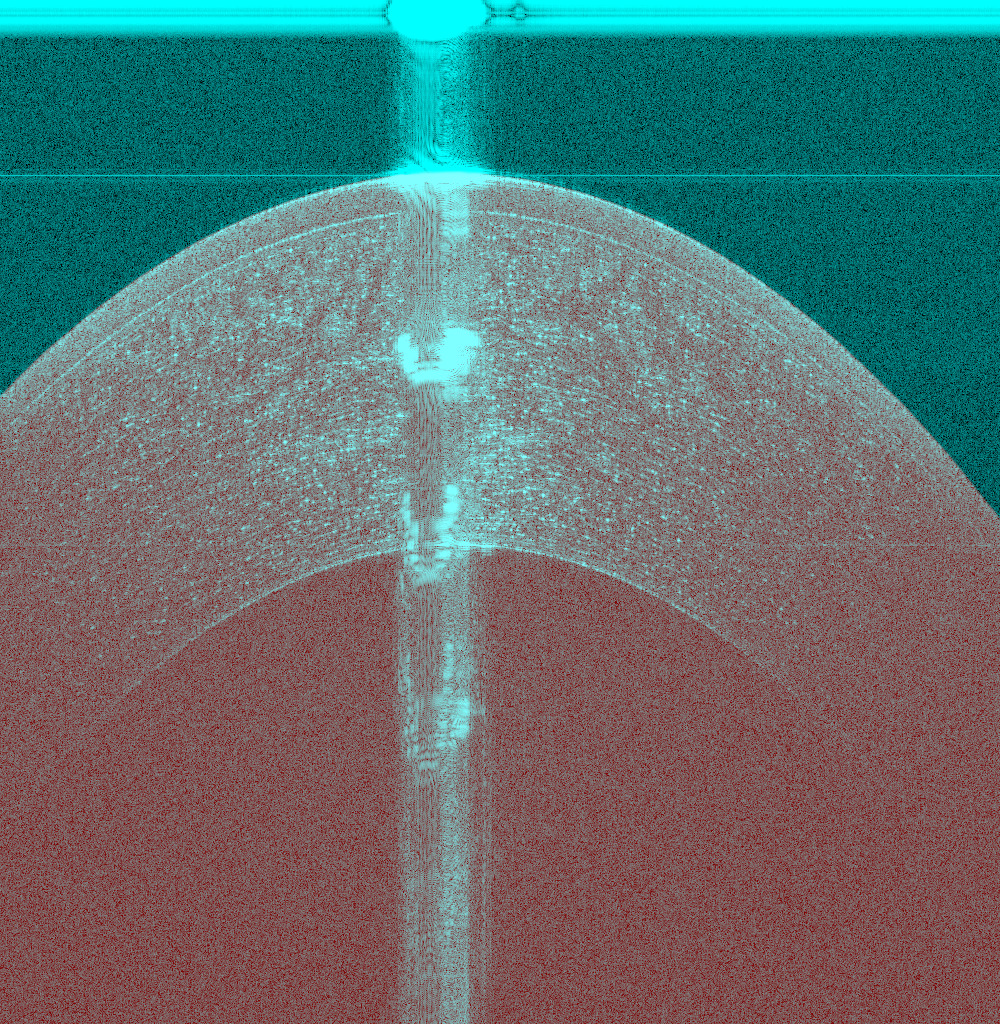}\\\medskip
\includegraphics[height=3cm,width=0.4\columnwidth]{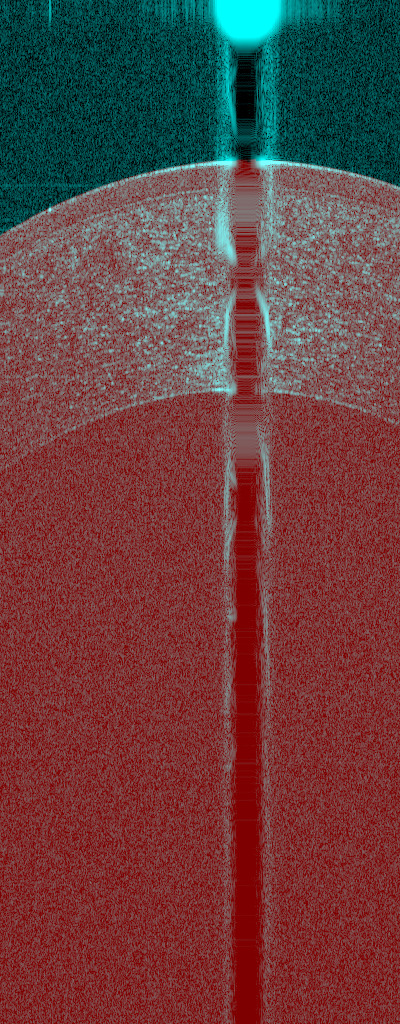}\\\medskip
\includegraphics[height=3cm,width=0.4\columnwidth]{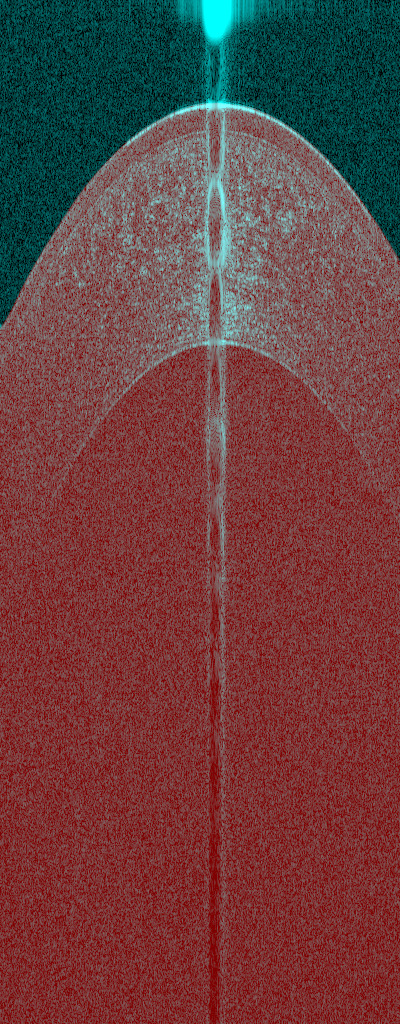}
\centerline{(c)}
\end{subfigure}
\begin{subfigure}[b]{0.245\columnwidth}
\centering
\includegraphics[height=3cm,width=0.9\columnwidth]{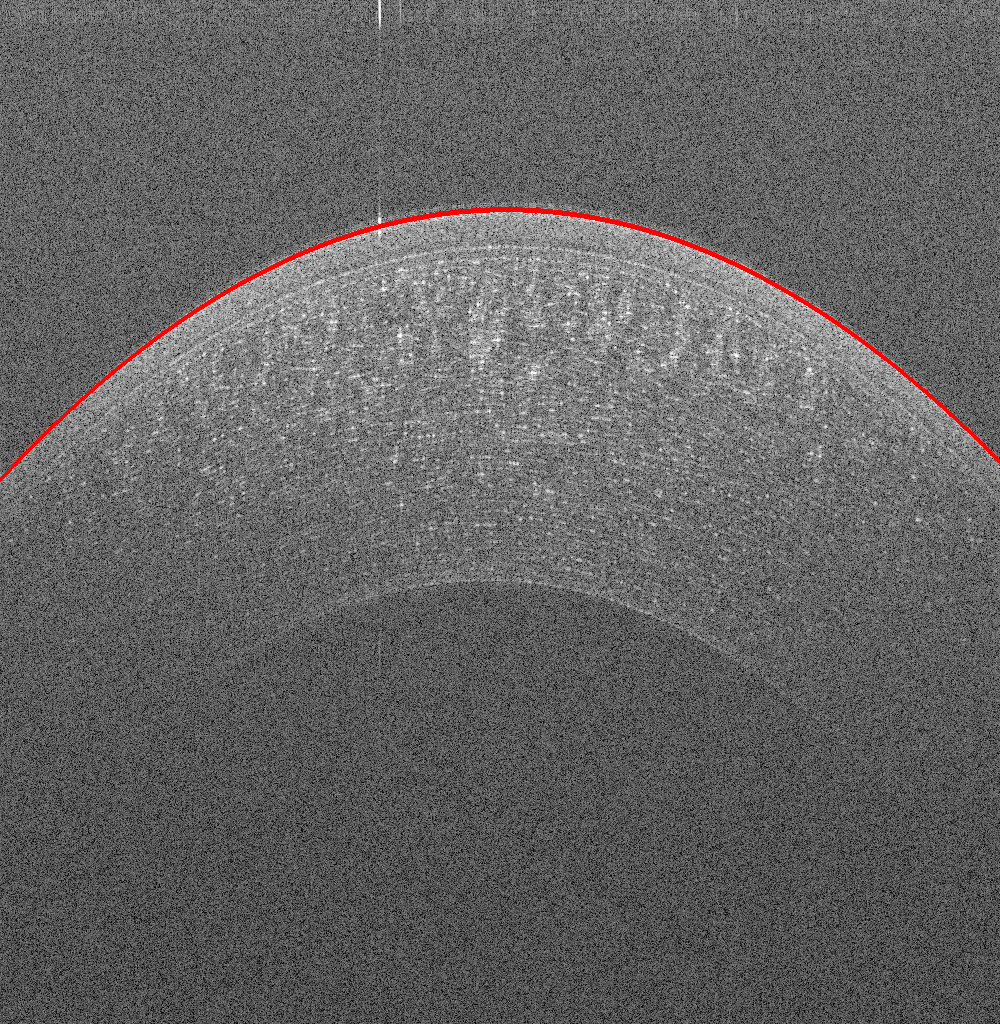}\\\medskip
\includegraphics[height=3cm,width=0.9\columnwidth]{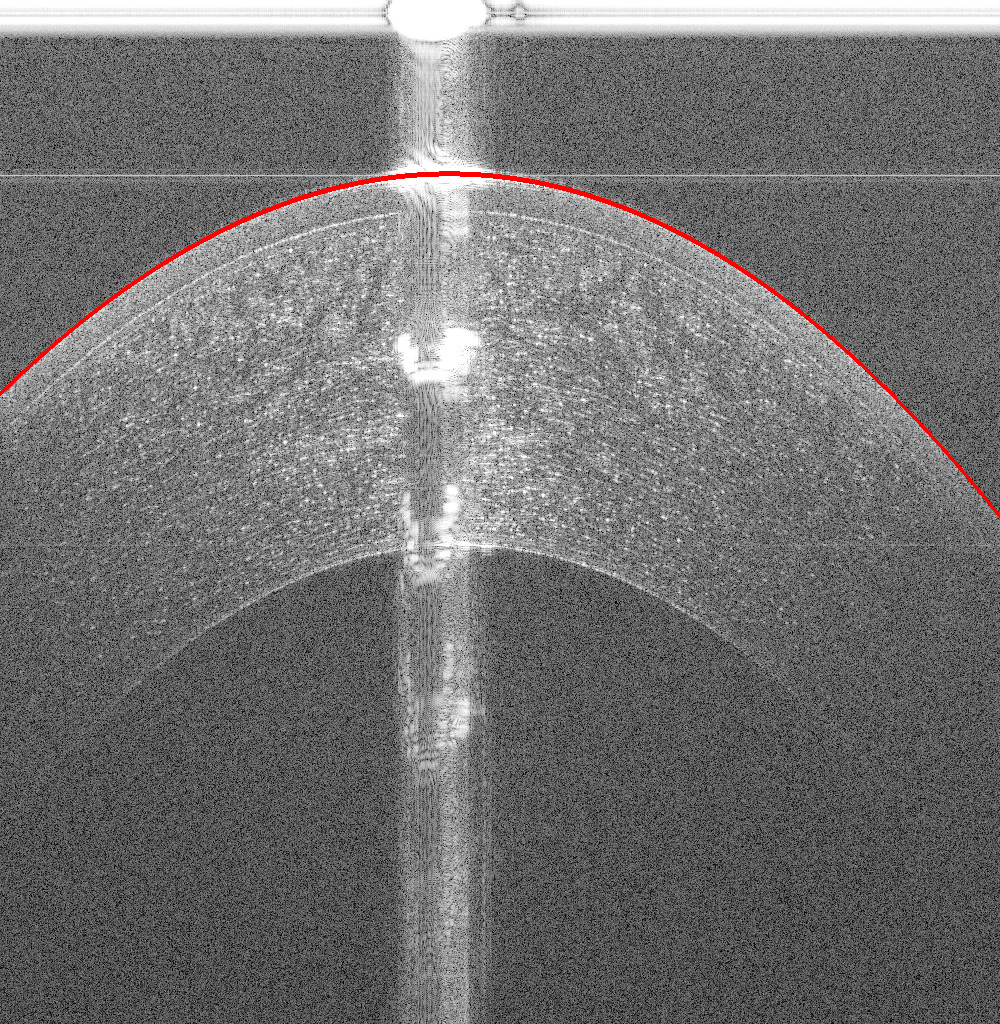}\\\medskip
\includegraphics[height=3cm,width=0.4\columnwidth]{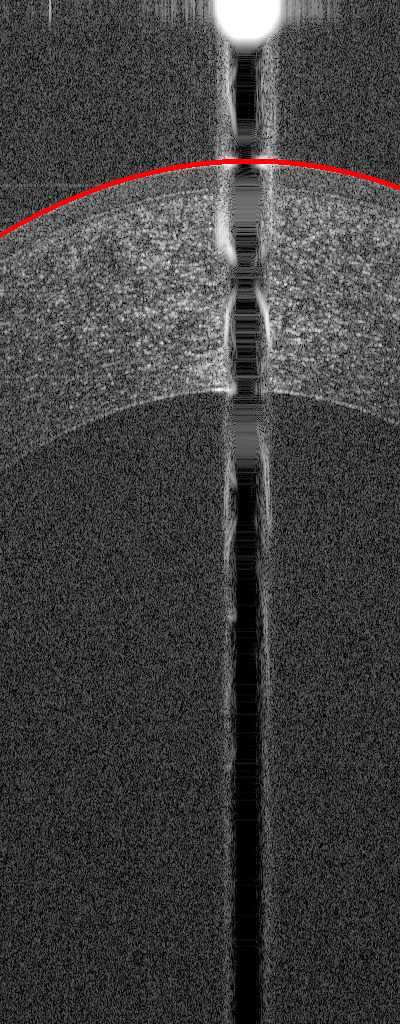}\\\medskip
\includegraphics[height=3cm,width=0.4\columnwidth]{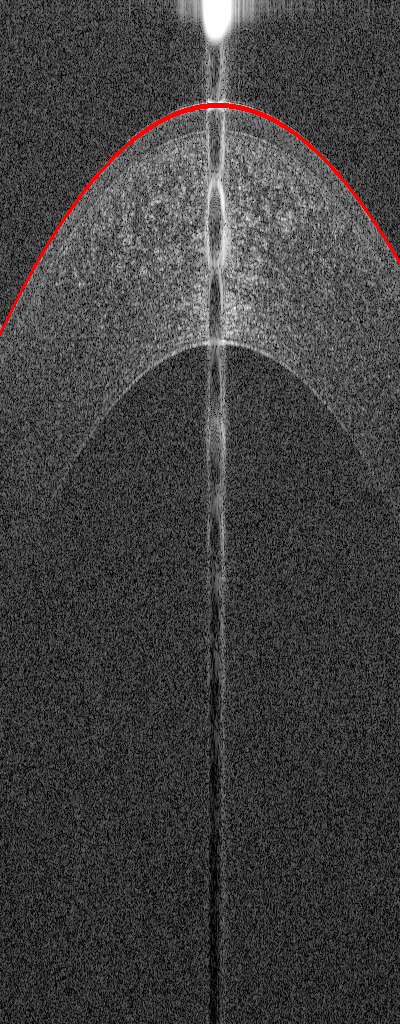}
\centerline{(d)}
\end{subfigure}
\caption{Corneal interface segmentation results for datasets acquired using Devices 1 and 2. Columns from left to right: (a) Original B-scans in corneal OCT datasets, (b) Pre-segmented OCT images from the cGAN with the specular artifact and speckle noise patterns removed just prior to the shallowest tissue interface, (c) Binary segmentation from the TISN overlaid in false color (red - foreground, turquoise - background) on the original B-scan, (d) Curve fit to the shallowest interface (red contour).}
\label{fig:res_cornea}
\end{figure}

\begin{figure}[!h]
\centering
\begin{subfigure}[b]{0.245\columnwidth}
\centering
\includegraphics[height=3cm,width=0.9\columnwidth]{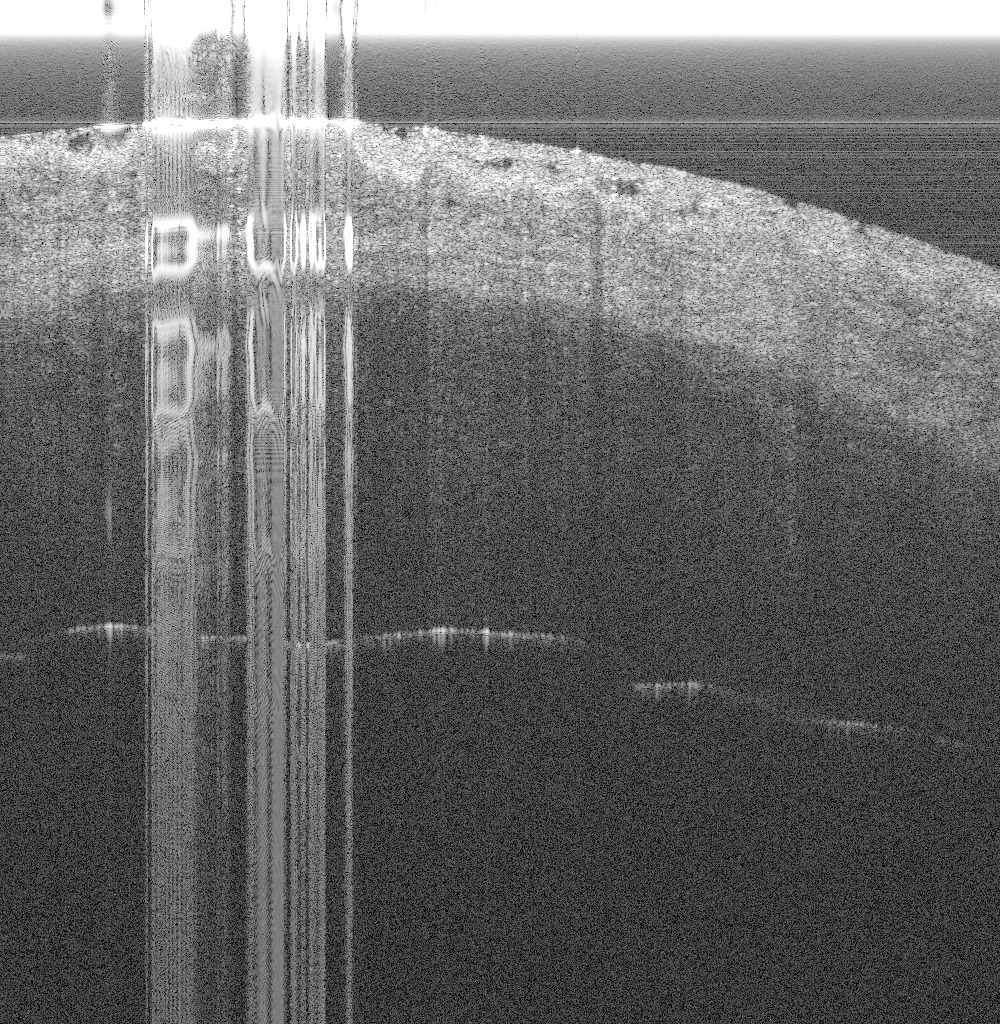}\\\medskip
\includegraphics[height=3cm,width=0.5\columnwidth]{dp2_i0026}\\\medskip
\includegraphics[height=3cm,width=0.4\columnwidth]{dp3_i00100}\\\medskip
\includegraphics[height=3cm,width=0.4\columnwidth]{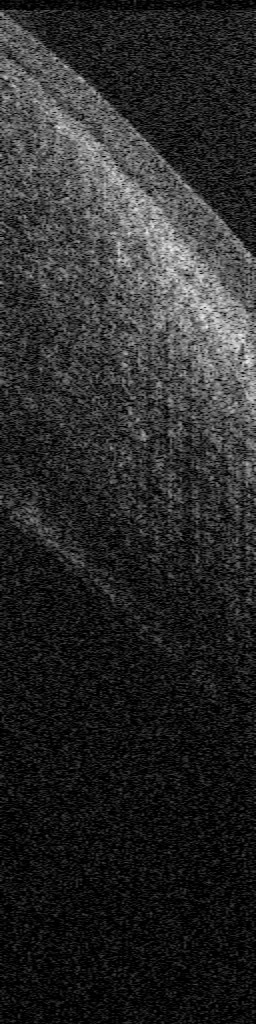}\\\medskip
\includegraphics[height=3cm,width=0.4\columnwidth]{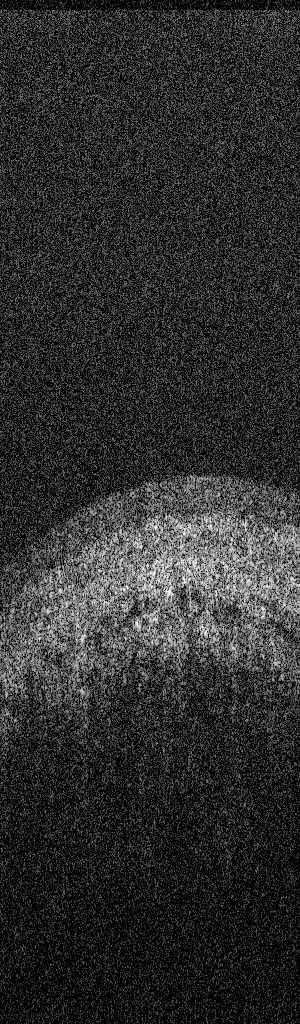}
\centerline{(a)}
\end{subfigure}
\begin{subfigure}[b]{0.245\columnwidth}
\centering
\includegraphics[height=3cm,width=0.9\columnwidth]{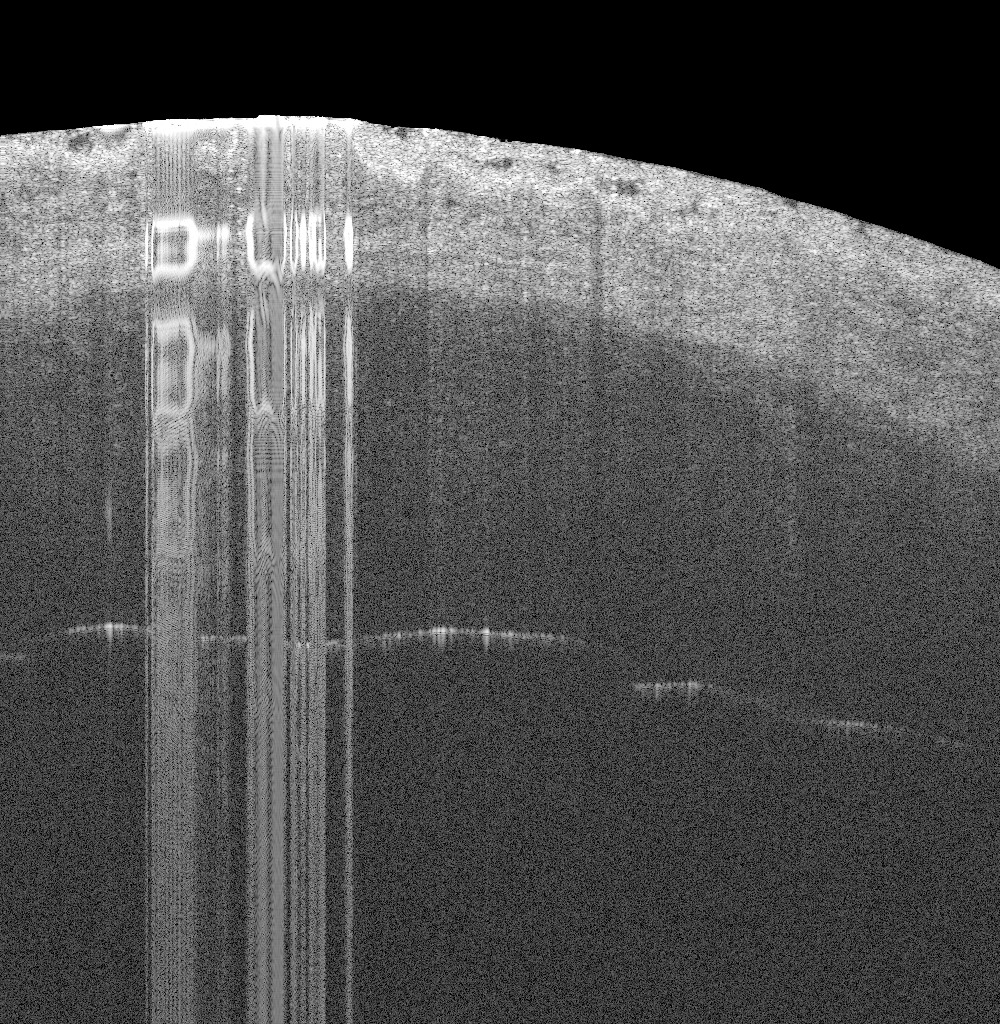}\\\medskip
\includegraphics[height=3cm,width=0.5\columnwidth]{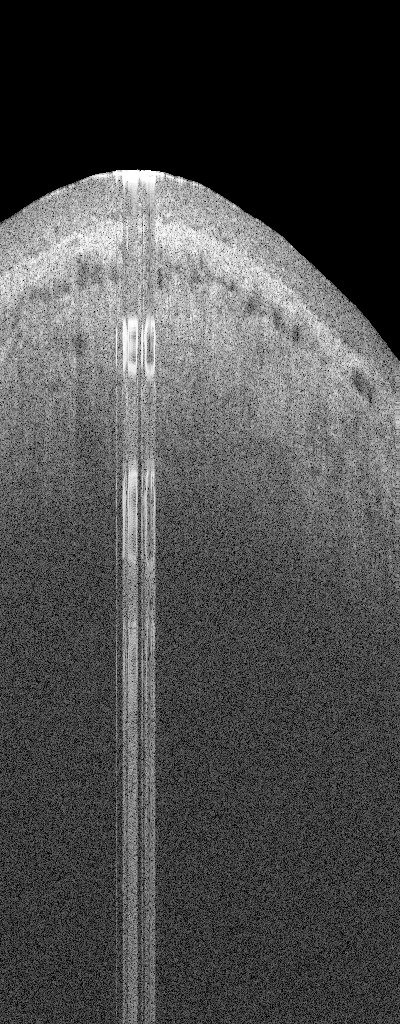}\\\medskip
\includegraphics[height=3cm,width=0.4\columnwidth]{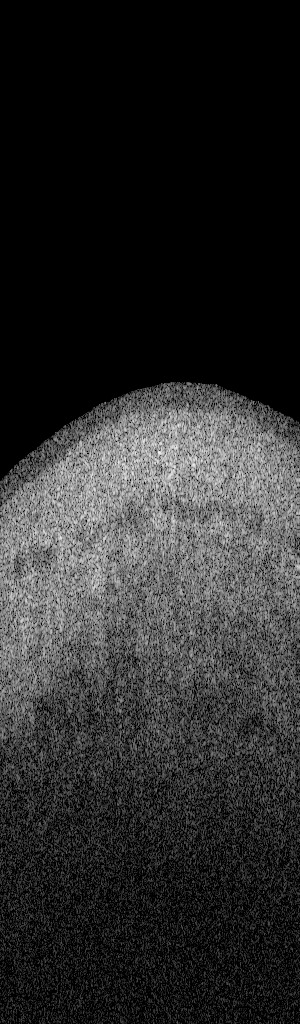}\\\medskip
\includegraphics[height=3cm,width=0.4\columnwidth]{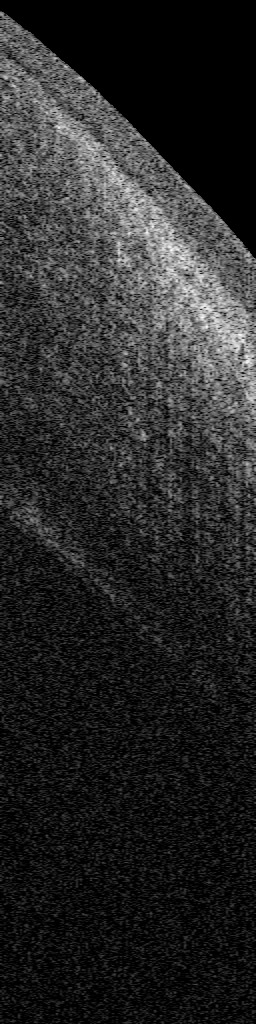}\\\medskip
\includegraphics[height=3cm,width=0.4\columnwidth]{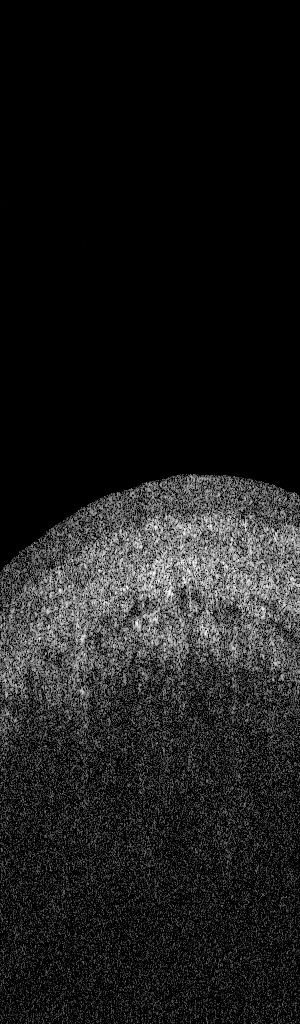}
\centerline{(b)}
\end{subfigure}
\begin{subfigure}[b]{0.245\columnwidth}
\centering
\includegraphics[height=3cm,width=0.9\columnwidth]{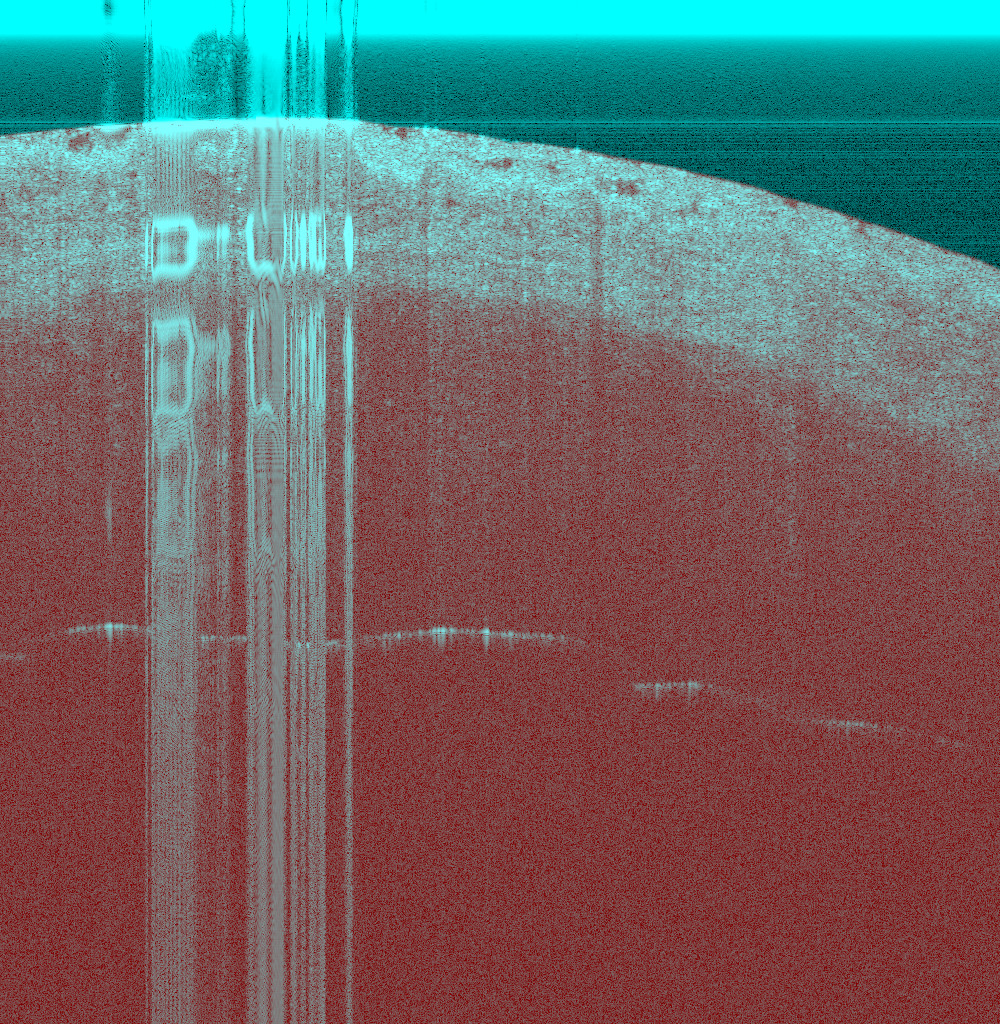}\\\medskip
\includegraphics[height=3cm,width=0.5\columnwidth]{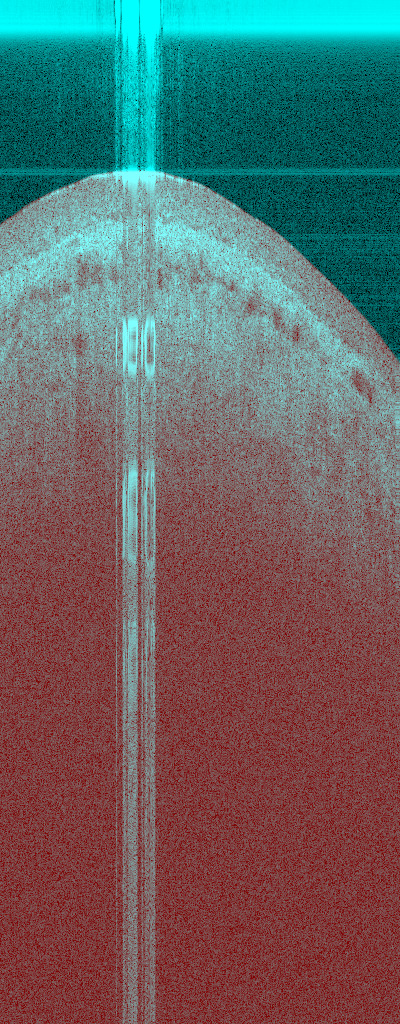}\\\medskip
\includegraphics[height=3cm,width=0.4\columnwidth]{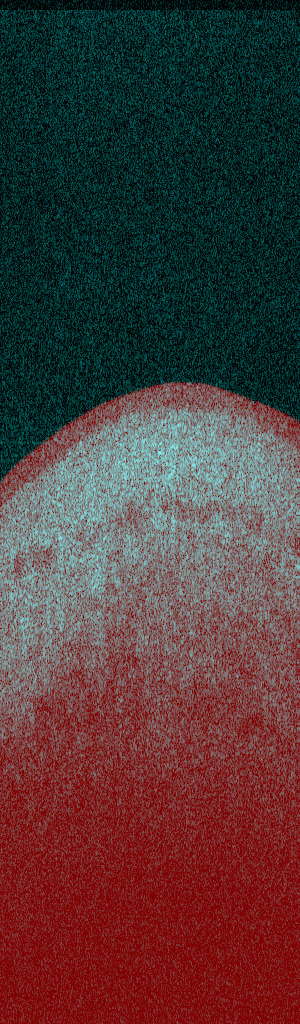}\\\medskip
\includegraphics[height=3cm,width=0.4\columnwidth]{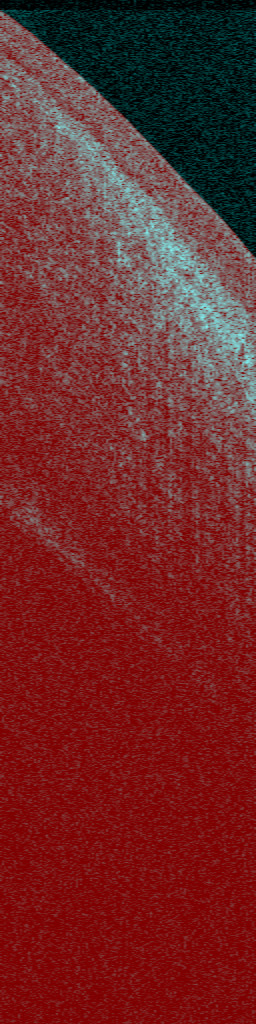}\\\medskip
\includegraphics[height=3cm,width=0.4\columnwidth]{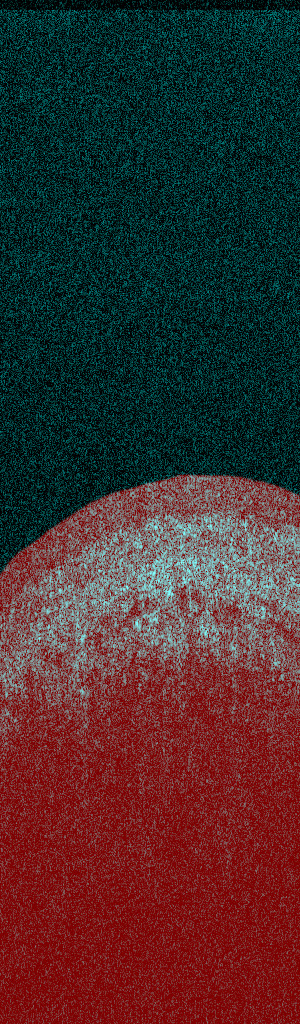}
\centerline{(c)}
\end{subfigure}
\begin{subfigure}[b]{0.245\columnwidth}
\centering
\includegraphics[height=3cm,width=0.9\columnwidth]{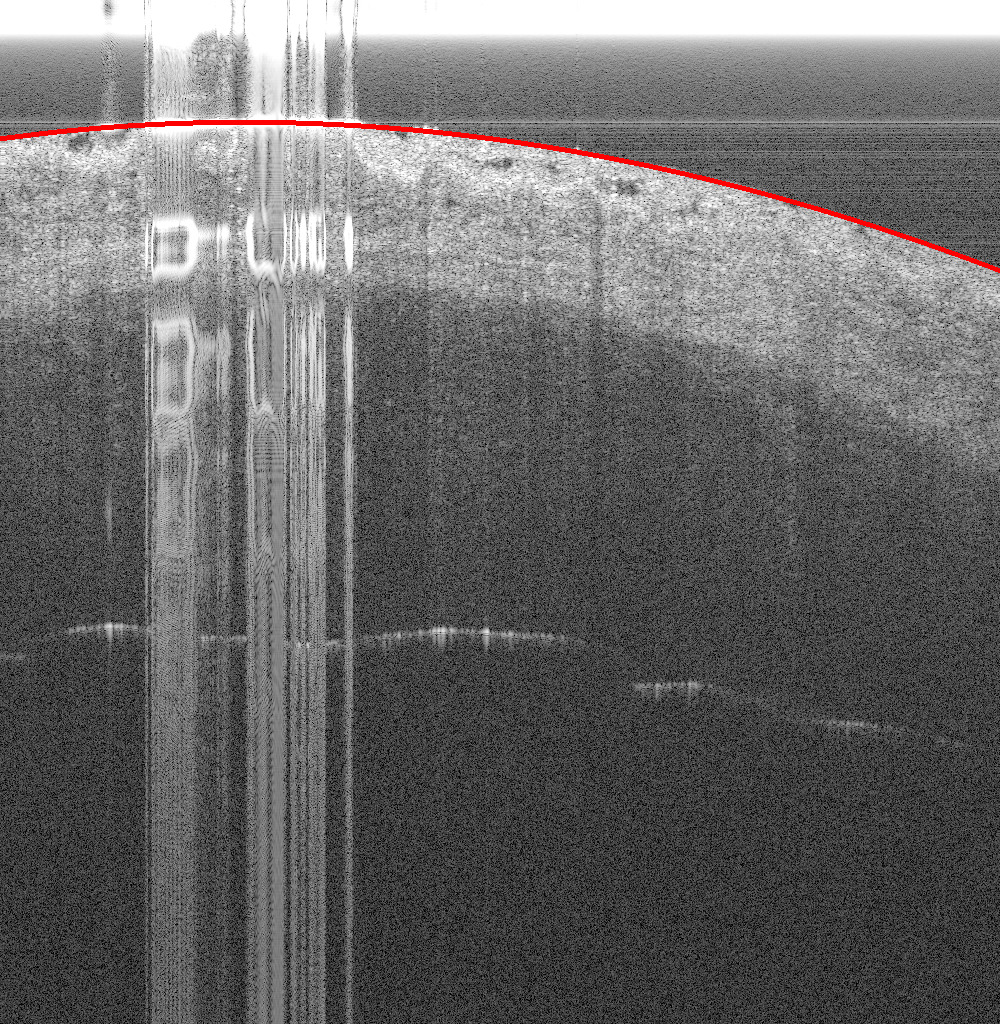}\\\medskip
\includegraphics[height=3cm,width=0.5\columnwidth]{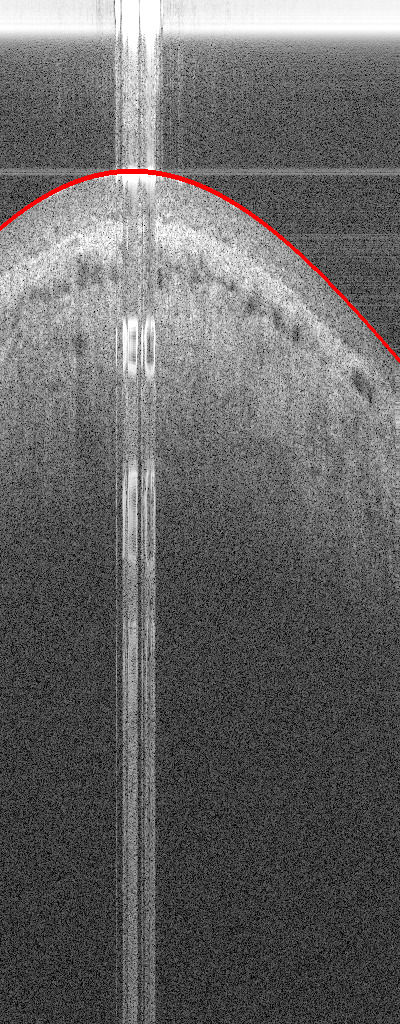}\\\medskip
\includegraphics[height=3cm,width=0.4\columnwidth]{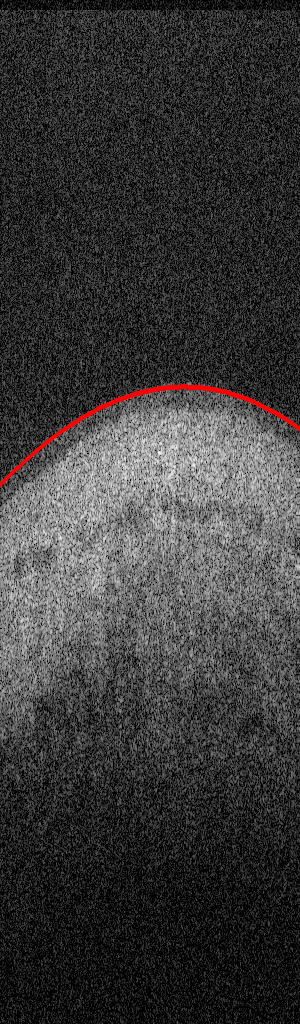}\\\medskip
\includegraphics[height=3cm,width=0.4\columnwidth]{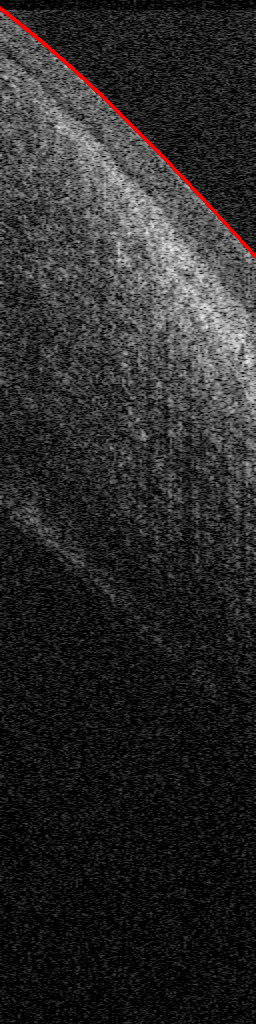}\\\medskip
\includegraphics[height=3cm,width=0.4\columnwidth]{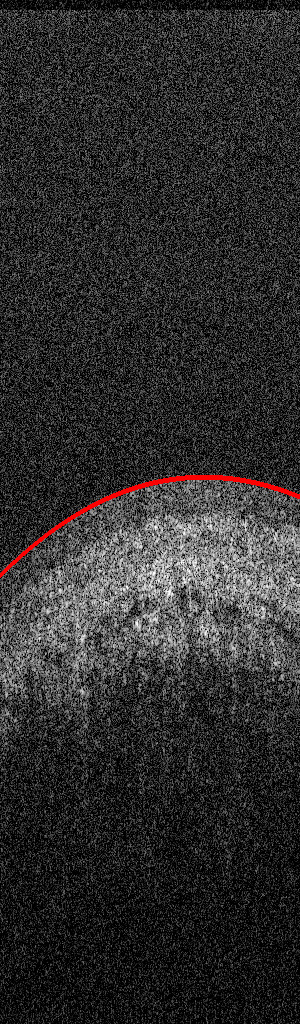}
\centerline{(d)}
\end{subfigure}
\caption{Limbal interface segmentation results for datasets acquired using Devices 2 and 3. Columns from left to right: (a) Original B-scans in the limbal OCT datasets, (b) Pre-segmented OCT images from the cGAN with the specular artifact and speckle noise patterns removed above the shallowest tissue interface, (c) Binary segmentation from the TISN overlaid in false color (red - foreground, turquoise - background) on the original B-scan, (d) Curve fit to the shallowest interface (red contour).}
\label{fig:res_limbus}
\end{figure}

The feature maps in the final layer of the network are activated using the softmax function to produce a two-channel output. Once the network was trained, it was used to segment the shallowest interface in our testing datasets. At test time, the TISN yielded a two-channel output; the first channel corresponded to the foreground tissue segmentation, and the second channel corresponded to the background pixel segmentation (above the tissue interface). The foreground pixels corresponded to the boundary of the interface and those pixels below it, while the pixels above the tissue boundary denoted the background. Finally, the predicted segmentation was fitted with a curve \cite{Lowess1981} after the tissue interface was identified using a fast GPU-based method \cite{Mathai2014}. We show our final results in Figs. \ref{fig:res_cornea}, \ref{fig:res_limbus} and \ref{fig:res_cor_to_limbus} along with the supplementary video visualizations. 

%--------------
%--------------

%--------------
%--------------

%-------------------------------------------------------------------
%-------------------------------------------------------------------
\subsection{Baseline Comparisons}
%-------------------------------------------------------------------
%-------------------------------------------------------------------

Extensive evaluation of the performance of our approach was conducted across all the testing datasets. First, we wanted to investigate the accuracy of a traditional image analysis-based algorithm \cite{Mathai2018} that directly segmented the interface in our test datasets. Briefly, this algorithm filtered the OCT image to reduce speckle noise and artifacts, extracted the monogenic signal \cite{Felsberg2001}, and segmented the tissue interface. We denote this baseline in the rest of the paper by the acronym: Traditional WithOut Pre-Segmentation (TWOPS).

Second, we designed a hybrid framework, where the pre-segmented OCT image from the cGAN is used by the traditional image analysis-based algorithm \cite{Mathai2018} to segment the shallowest interface. We wanted to determine the improvement in segmentation accuracy when the traditional algorithm used the pre-segmentation instead of the original OCT image. Going forward, we denote this baseline by the acronym: Traditional With Pre-Segmentation (TWPS).

Third, we trained a CorNet architecture \cite{Mathai2018_2} to directly segment the foreground in the input OCT image, \textit{without} including the cGAN pre-segmentation as an additional input channel. We compared the direct segmentation result against our cascaded framework. Henceforth, in the remainder of the paper, we refer to the direct deep learning-based segmentation approach by the acronym: Deep Learning WithOut Pre-Segmentation (DLWOPS). Finally, we call our cascaded framework as: Deep Learning With Pre-Segmentation (DLWPS). % wherein the latter framework contains the cGAN that generates the pre-segmentation and passes it to the TISN for final segmentation

To summarize, the following baseline methods were considered for performance evaluation:
\begin{enumerate} %itemize
\item TWOPS - A traditional image analysis-based algorithm \cite{Mathai2018} that directly segmented the tissue interface.
\item TWPS - The hybrid framework. 
\item DLWOPS - A deep learning-based approach \cite{Mathai2018_2} that directly segmented the tissue interface.
\item DLWPS - The cascaded framework.
\end{enumerate}

%-------------------------------------------------------------------
%-------------------------------------------------------------------
\subsection{Evaluation}
%-------------------------------------------------------------------
%-------------------------------------------------------------------

%-------------------------------------------------------------------
%-------------------------------------------------------------------
\subsubsection{Annotation}
%-------------------------------------------------------------------
%-------------------------------------------------------------------

Each corneal dataset was annotated by an expert grader (G1; Grader 1) and a trained grader (G2; Grader 2). However, only expert annotations were available for the limbal datasets in the research database. The graders were asked to annotate the shallowest interface in all test datasets. For each dataset, the graders annotated the interface using a 5-pixel width band with an admissible annotation error of 3 pixels. All the annotations were fitted with a curve for comparison with the different baselines. We also estimated the inter-grader annotation variability for the corneal datasets, and refer to it in the rest of the paper by the acronym: IG.

%-------------------------------------------------------------------
%-------------------------------------------------------------------
\subsubsection{Metrics}
%-------------------------------------------------------------------
%-------------------------------------------------------------------

In order to compare the segmentation accuracy across the different baselines, we calculated the following metrics: 1) Mean Absolute Difference in Layer Boundary Position (MADLBP) and 2) Hausdorff Distance (HD) between the fitted curves. These metric values were determined over all testing datasets, and only for the shallowest interface. In Eqs. \eqref{eq_MADLBP} and \eqref{eq_HD}, the sets of points that represent the gold standard annotation and the segmentation to which it is compared (each fitted with curves) are denoted by $G$ and $S$ respectively. We denote by $y_{G}(x)$ the Y-coordinate (rounded down after curve fitting) of the point in $G$ whose X-coordinate is $x$, and $y_{S}(x)$ is the Y-coordinate (rounded down) of the point in $S$. $d_{S}(p)$ is the distance of a point $p$ in $G$ to the closest point in $S$, and similarly for $d_{G}(p)$.

We chose MADLBP in Eq. \eqref{eq_MADLBP} as one of our error metrics since it was used in \cite{Mathai2018} to compare the segmentation accuracy between the automatic segmentations and grader annotations. Although MADLBP quantifies error in pixels, it did not measure the Euclidean distance error; instead, it simply measured the positional distance between the detected boundary location and the annotation along the same A-scan. On the other hand, the Hausdorff distance in Eq. \eqref{eq_HD} captured the greatest of all distances between the points in the segmentation and annotation. Therefore, it quantitatively describes the worst segmentation error in microns as it is more clinically relevant (e.g. to detect structural changes over time). In this work, we did not compute Dice similarity as it did not provide segmentation error in microns. 

% ^^^^^^^^
\begin{align}
%\small
	% ===
  	\textnormal{MADLBP} &= \frac{1}{X} \sum\limits^{X-1}_{x=0} \abs{y_{G}(x) - y_{S}(x)} \label{eq_MADLBP}\\
    % ===
	\textnormal{HD} &= \max\bigg(\underset{p \in G}{\max} \ d_{S}(p), \ \underset{p \in S}{\max} \ d_{G}(p) \bigg) \label{eq_HD}
\end{align}
% ^^^^^^^^

\begin{figure}[!hb]
\begin{subfigure}[b]{0.04\columnwidth}
\includegraphics[height=0.5cm,width=\columnwidth]{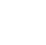}\\
\vfill\vfill\vfill
\includegraphics[height=4cm,width=\columnwidth]{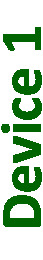}\\
\vfill\vfill\vfill
\includegraphics[height=4cm,width=\columnwidth]{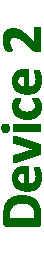}\\
\vfill\vfill\vfill
\includegraphics[height=4cm,width=\columnwidth]{Device2ce}\\
\vfill\vfill\vfill
\end{subfigure}\hfill
\begin{subfigure}[b]{0.45\columnwidth}
\includegraphics[height=0.5cm,width=\columnwidth]{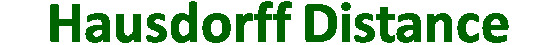}\\
\vfill\vfill\vfill
\includegraphics[height=4cm,width=\columnwidth]{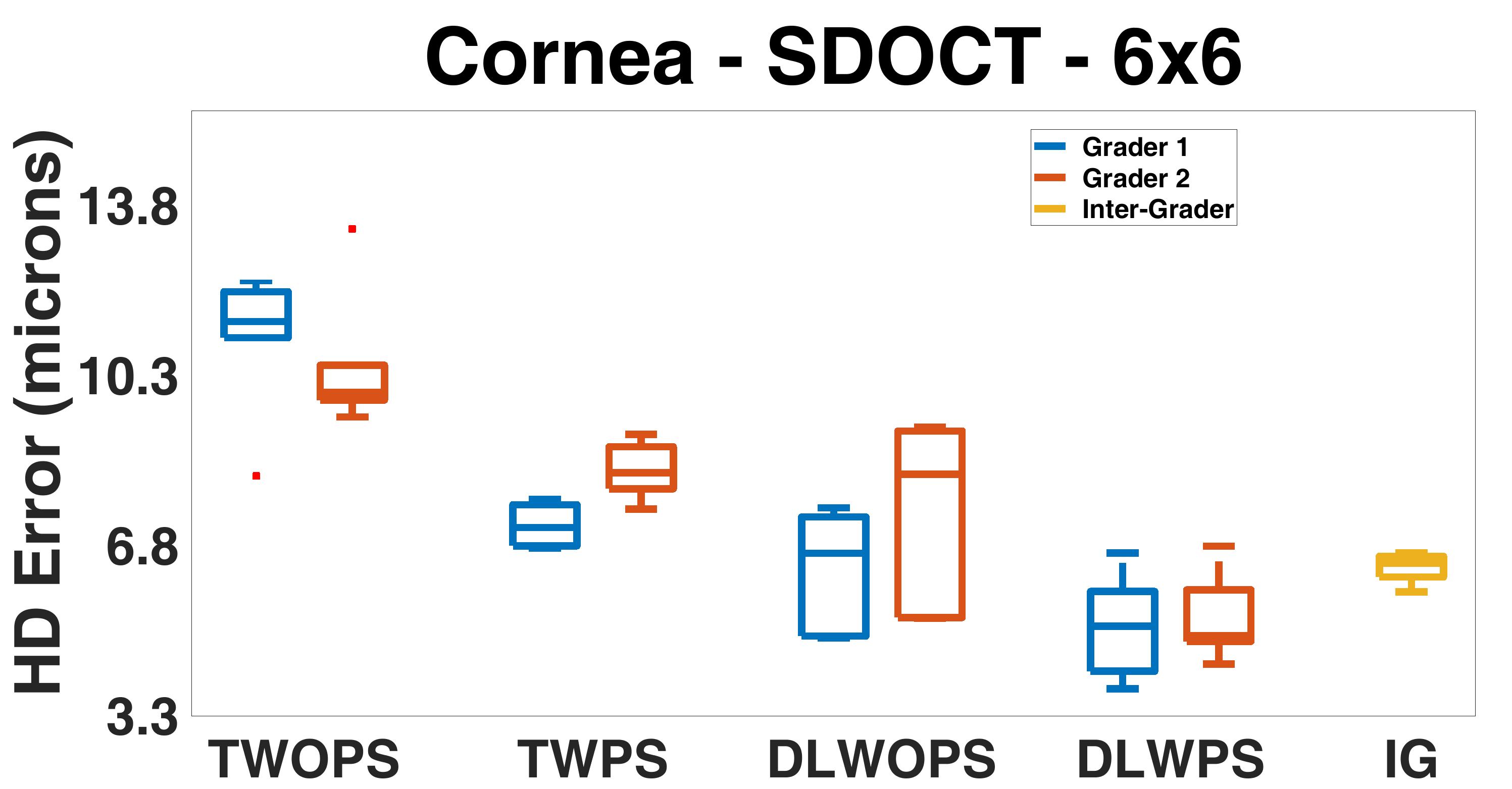}\\
\centering{(a)}\vfill
\includegraphics[height=4cm,width=\columnwidth]{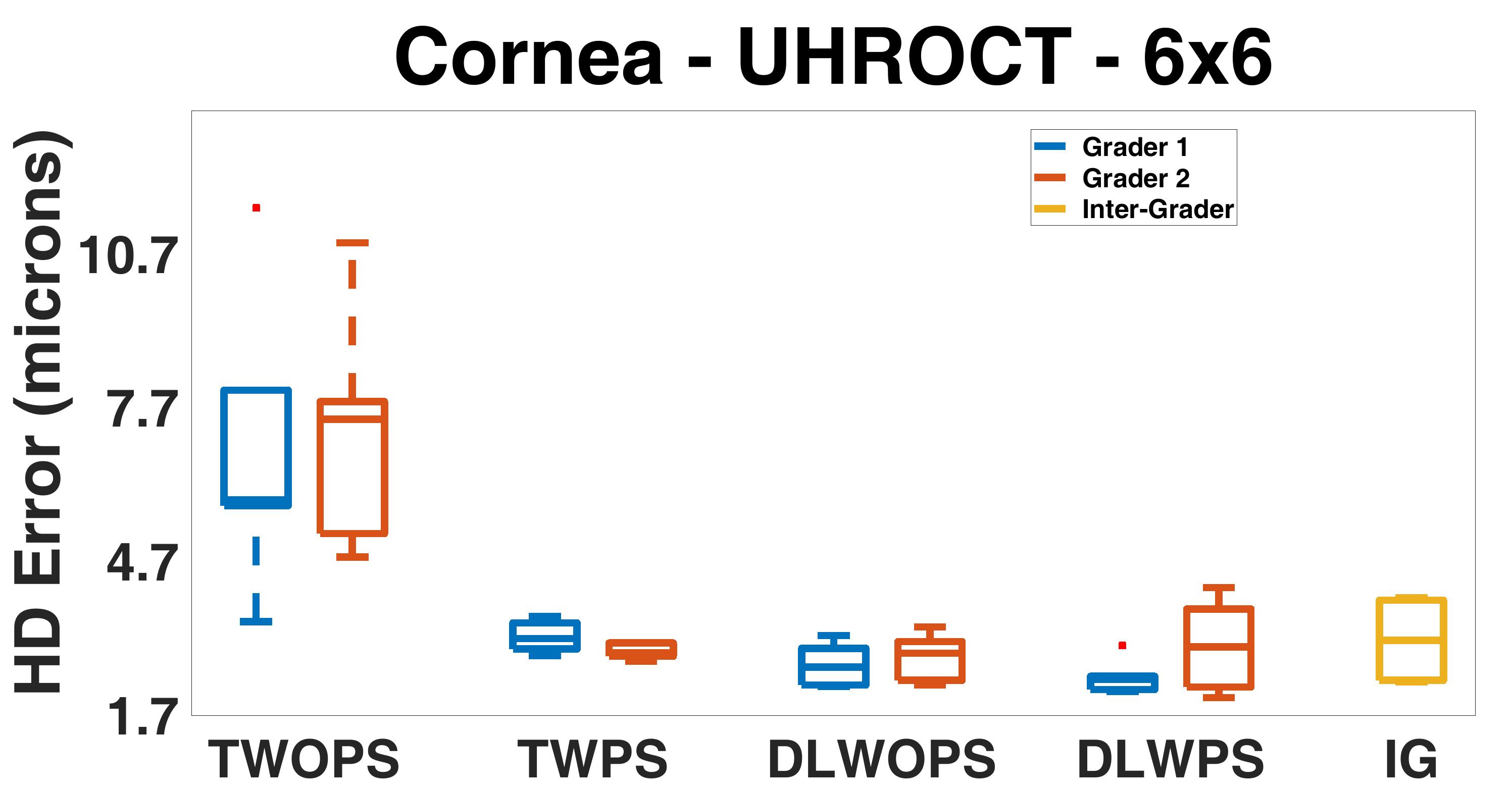}\\
\centering{(b)}\vfill
\includegraphics[height=4cm,width=\columnwidth]{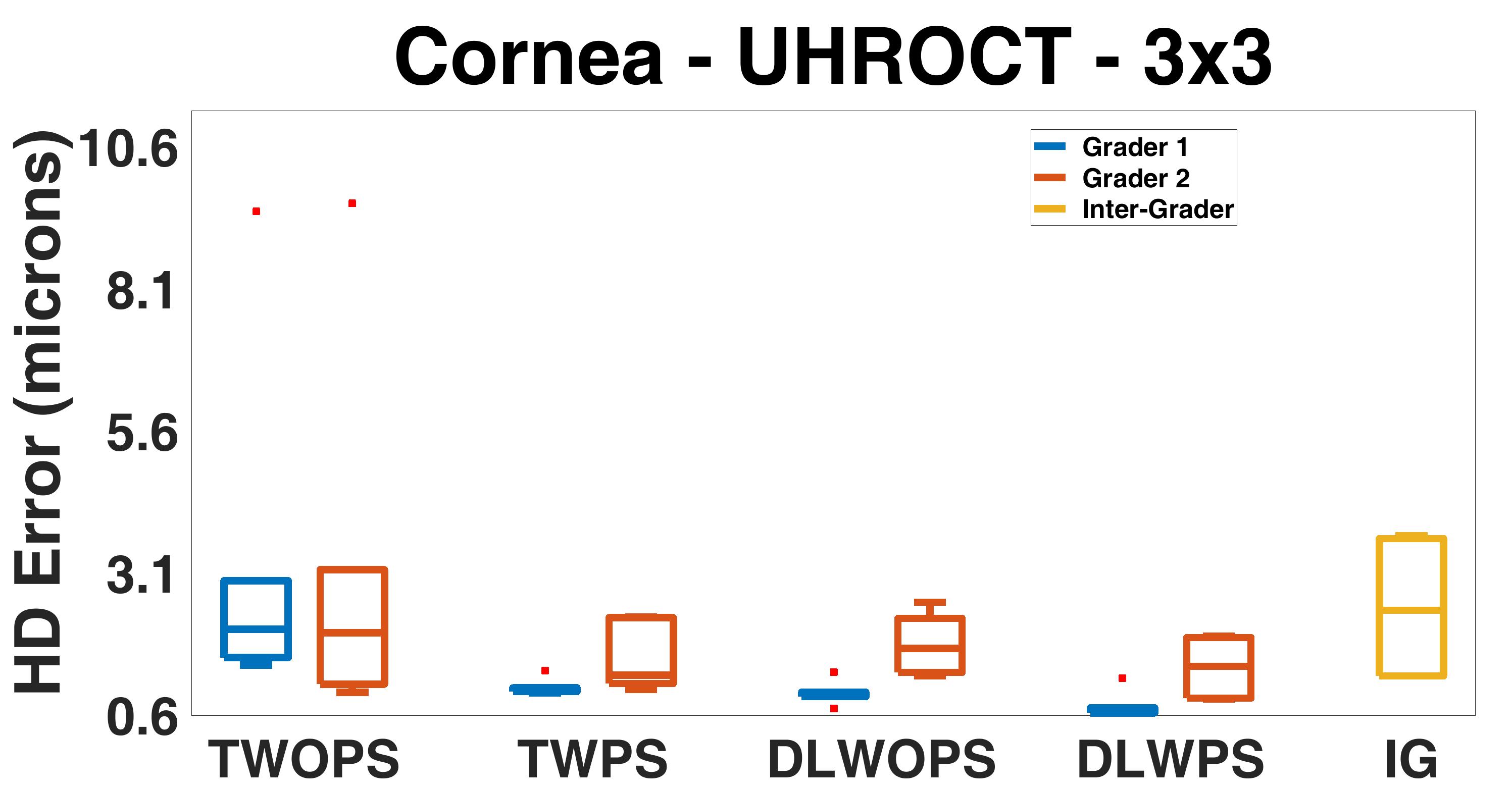}\\
\centering{(c)}
\end{subfigure}\hfill
\begin{subfigure}[b]{0.45\columnwidth}
\includegraphics[height=0.5cm,width=\columnwidth]{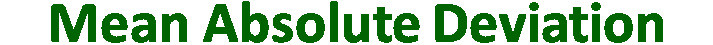}\\
\vfill\vfill\vfill
\includegraphics[height=4cm,width=\columnwidth]{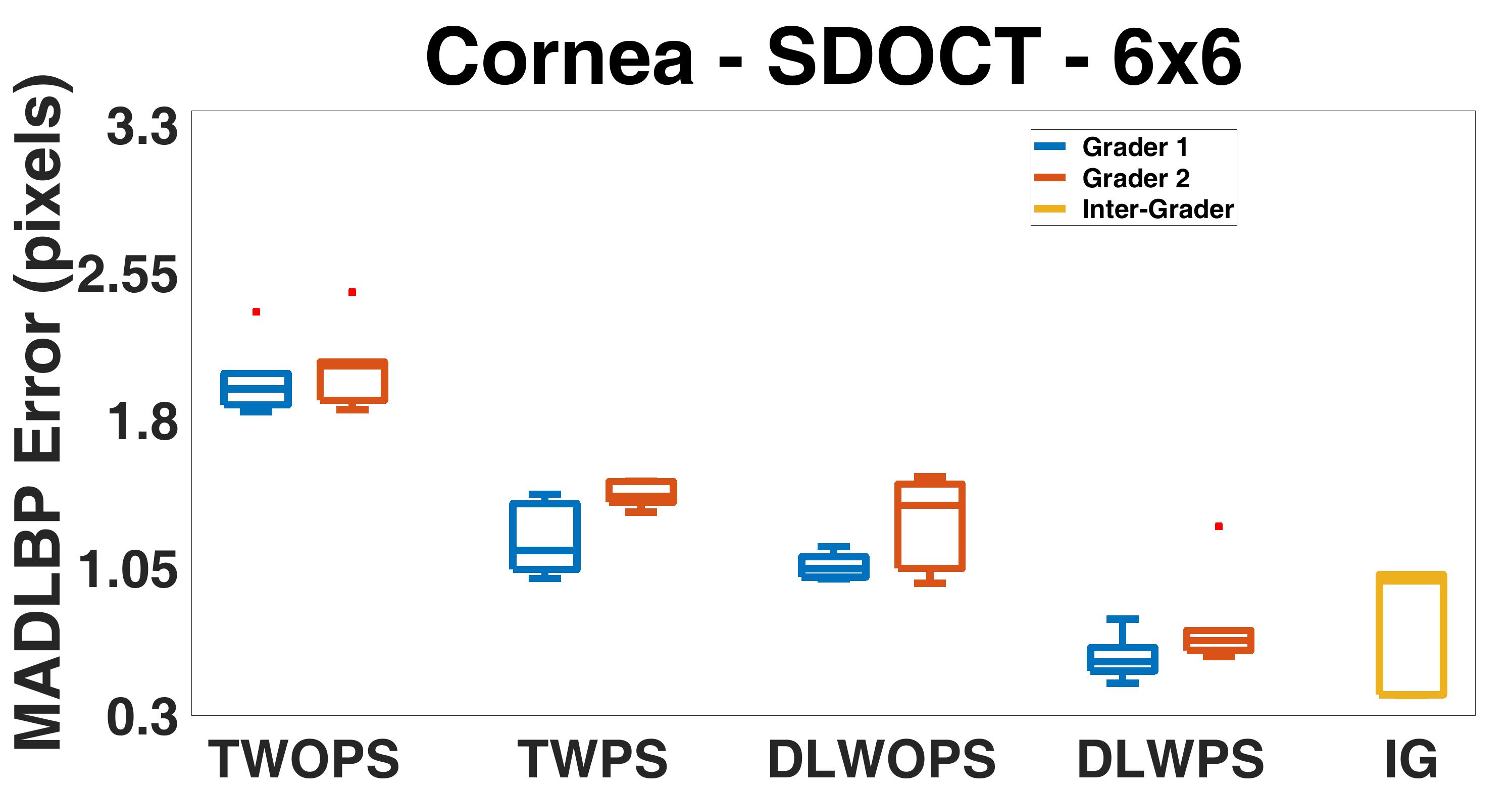}\\
\centering{(d)}\vfill
\includegraphics[height=4cm,width=\columnwidth]{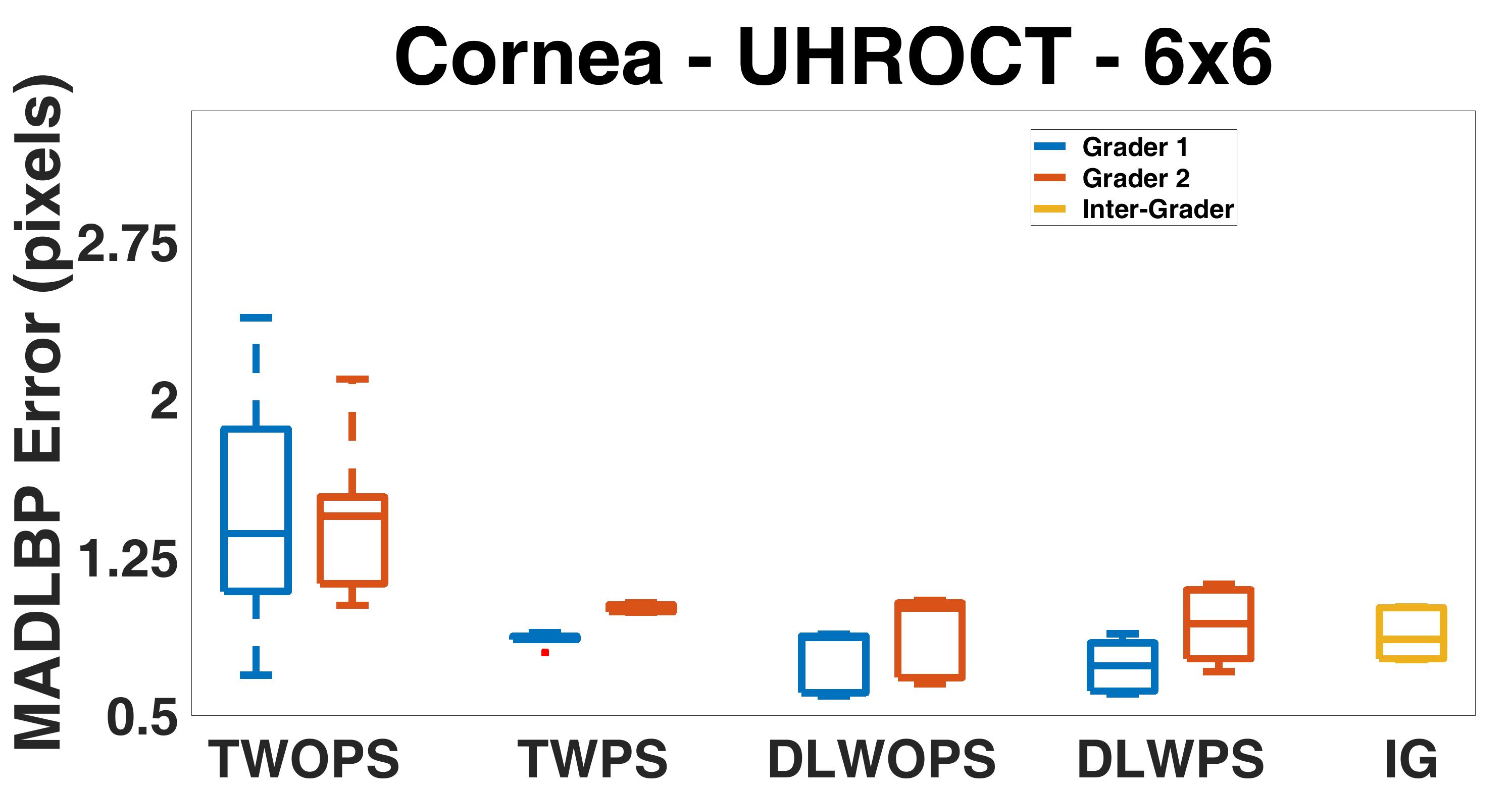}\\
\centering{(e)}\vfill
\includegraphics[height=4cm,width=\columnwidth]{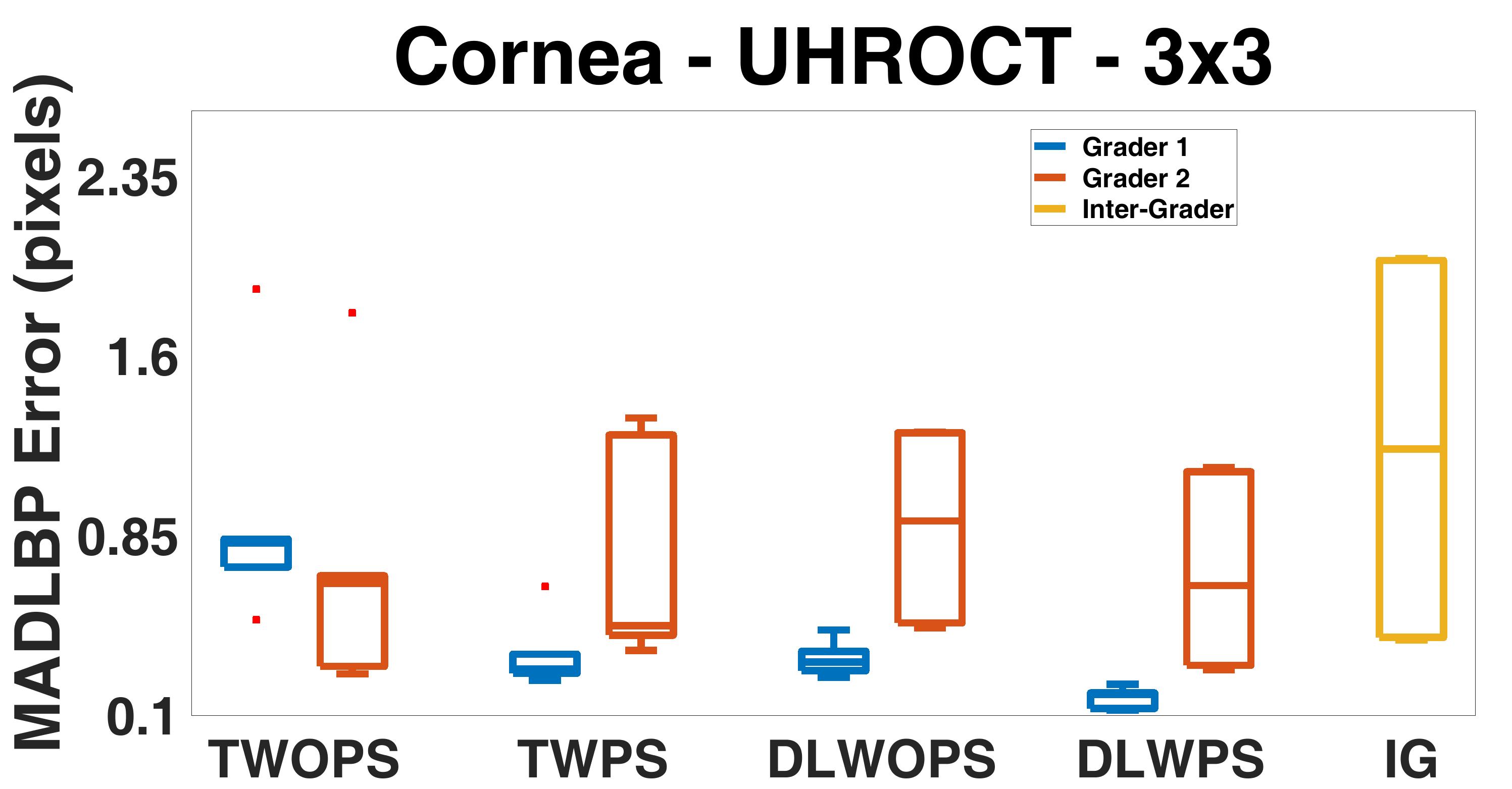}\\
\centering{(f)}
\end{subfigure}
\caption{(a)-(c) HD error and (d)-(f) MADLBP error comparison for the corneal datasets acquired with Devices 1 and 2 respectively. In the boxplots, the segmentation results obtained for each baseline method are contrasted against expert grader (blue) and trained grader (red) annotations, while the Inter-Grader (IG) variability is shown in yellow.}
\label{cor_HD_MAD}
\end{figure}

\begin{figure}[!hb]
\begin{subfigure}[b]{0.04\columnwidth}
\includegraphics[height=4cm,width=\columnwidth]{Device1ce}\\
\vfill\vfill\vfill
\includegraphics[height=4cm,width=\columnwidth]{Device2ce}\\
\vfill\vfill\vfill
\includegraphics[height=4cm,width=\columnwidth]{Device2ce}\\
\vfill\vfill\vfill
\end{subfigure}\hfill
\begin{subfigure}[b]{0.45\columnwidth}
\includegraphics[height=0.5cm,width=\columnwidth]{HDce}\\
\vfill\vfill\vfill
\includegraphics[height=4cm,width=\columnwidth]{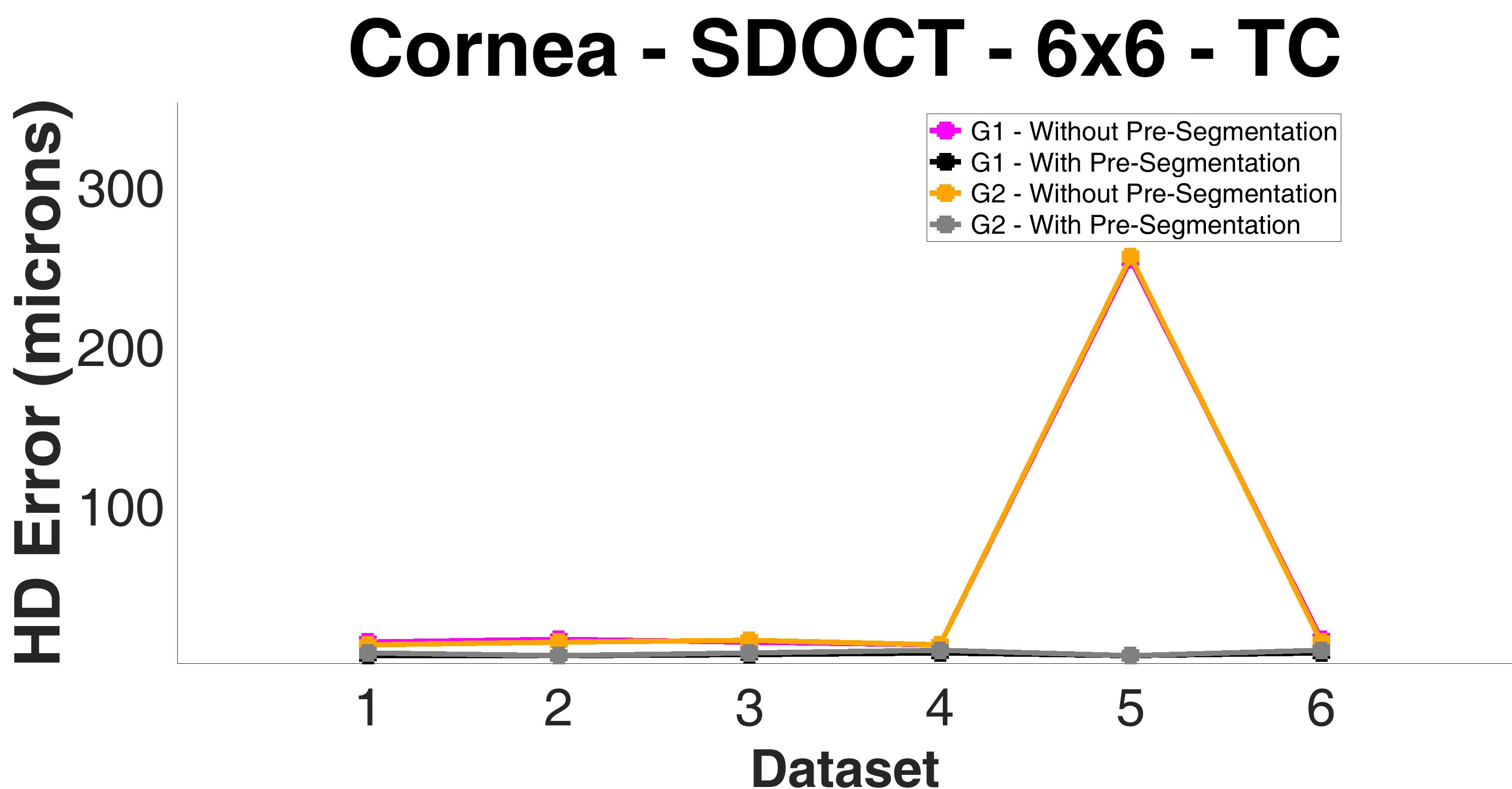}\\
\centering{(a)}\vfill
\includegraphics[height=4cm,width=\columnwidth]{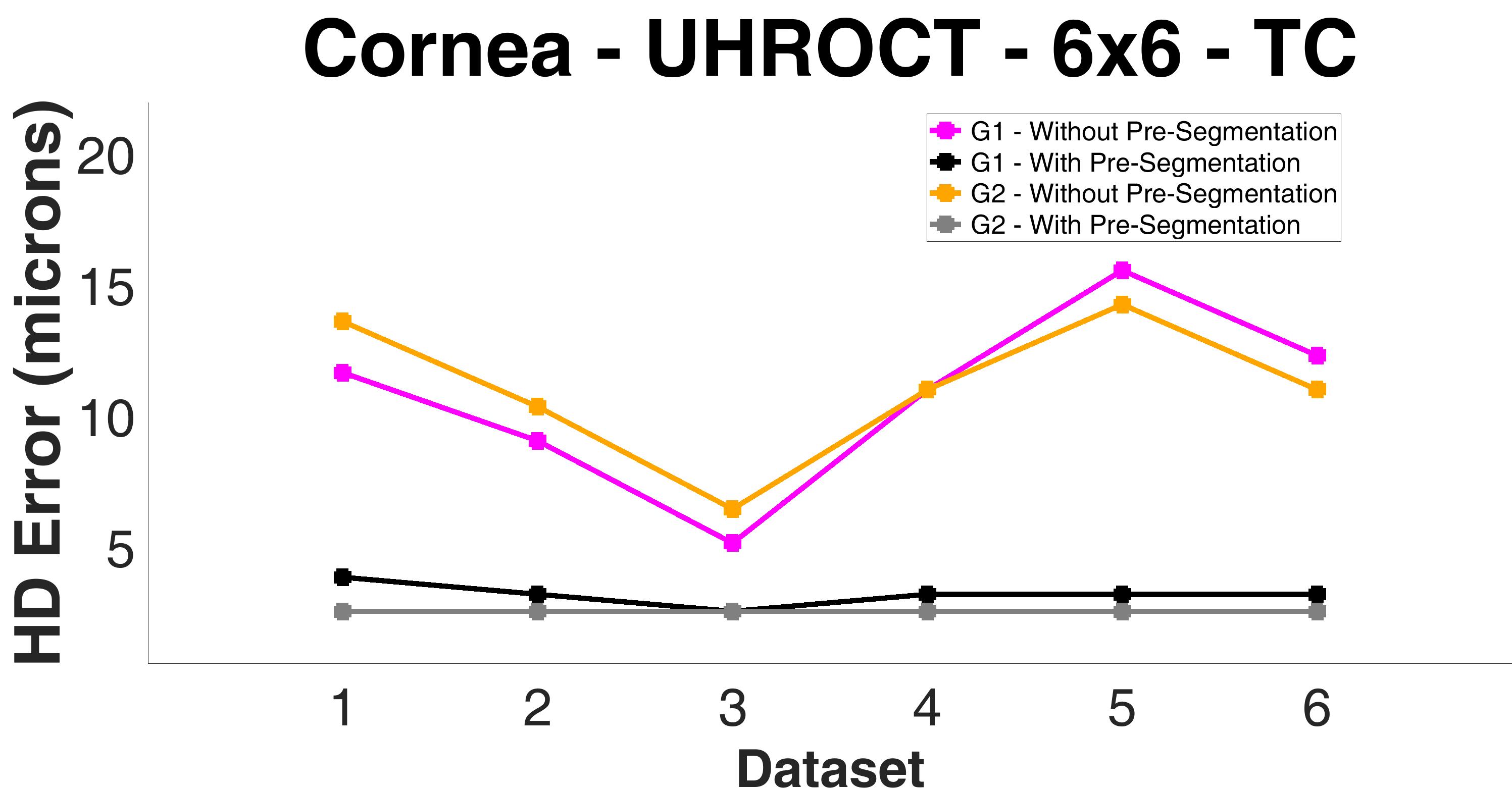}\\
\centering{(c)}\vfill
\includegraphics[height=4cm,width=\columnwidth]{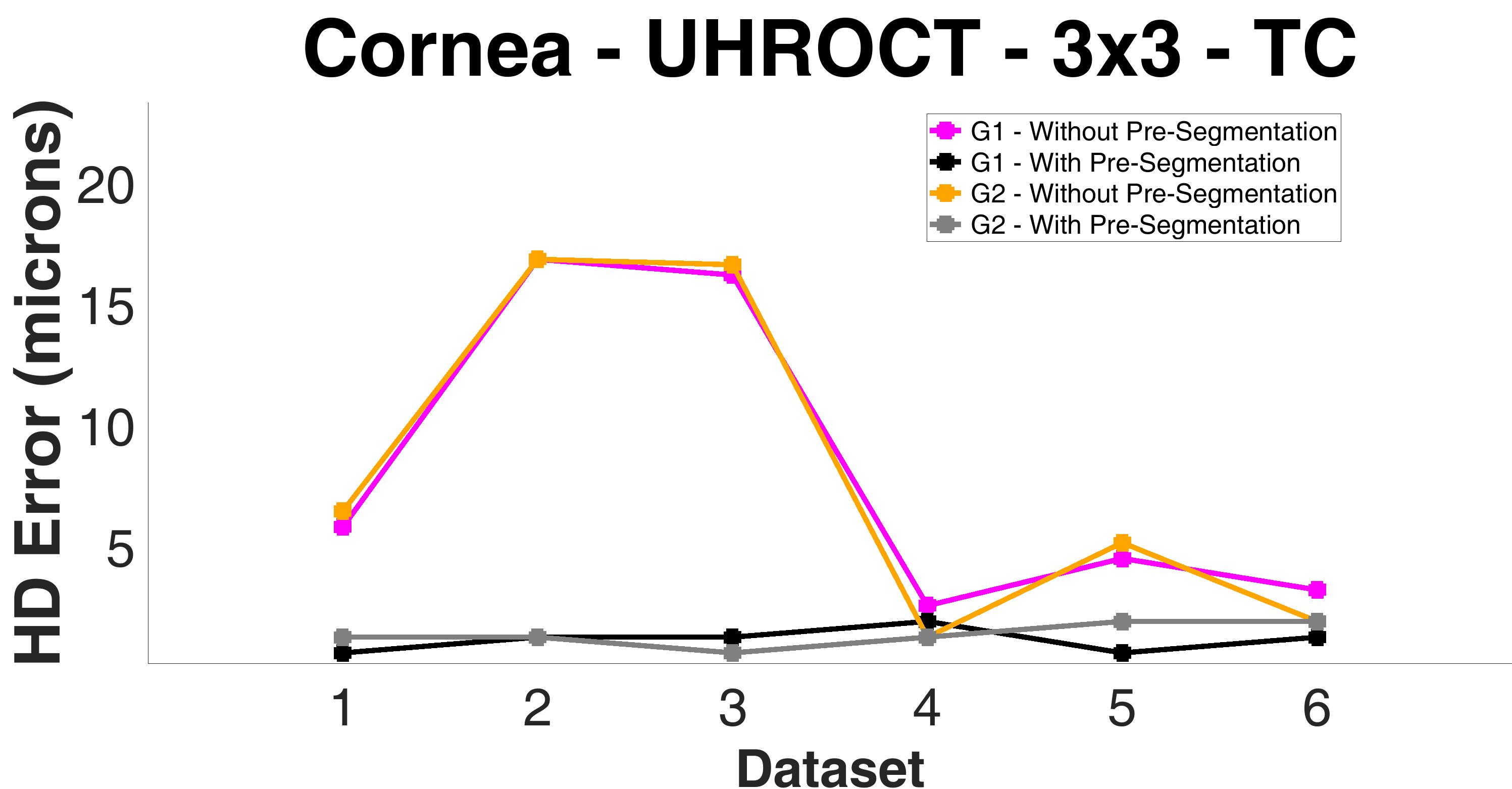}\\
\centering{(e)}
\end{subfigure}\hfill
\begin{subfigure}[b]{0.45\columnwidth}
\includegraphics[height=0.5cm,width=\columnwidth]{HDce}\\
\vfill\vfill\vfill
\includegraphics[height=4cm,width=\columnwidth]{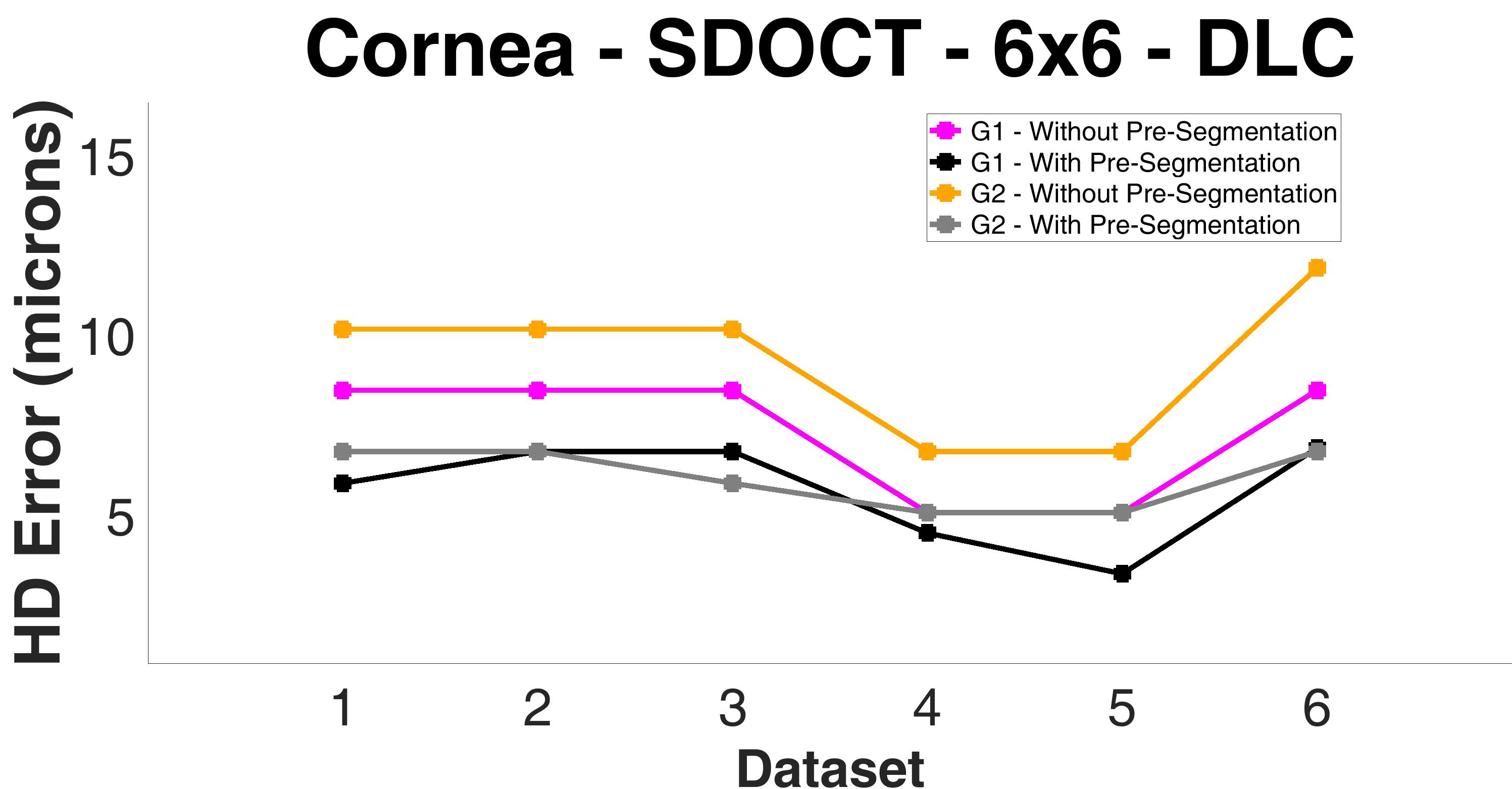}\\
\centering{(b)}\vfill
\includegraphics[height=4cm,width=\columnwidth]{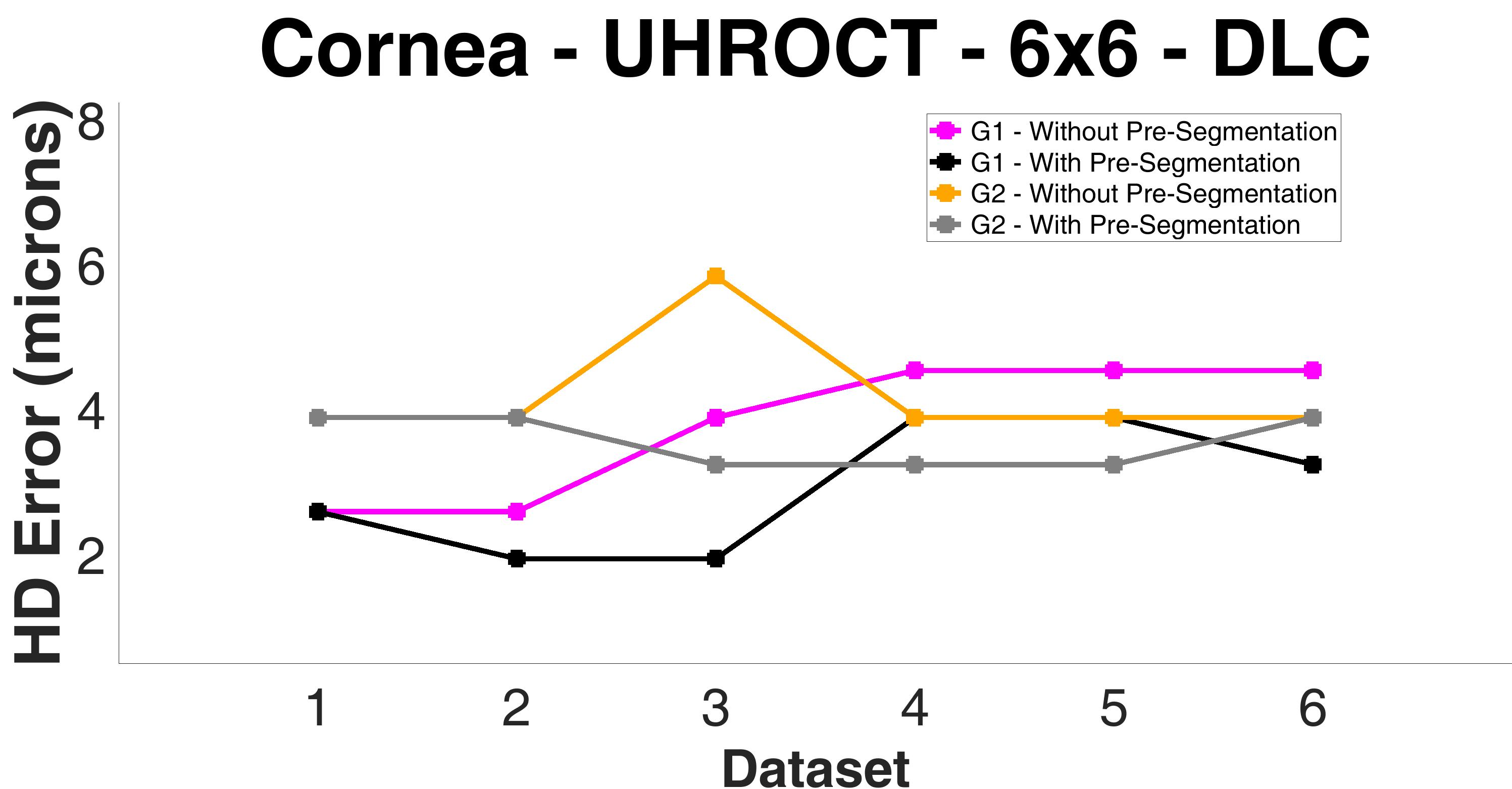}\\
\centering{(d)}\vfill
\includegraphics[height=4cm,width=\columnwidth]{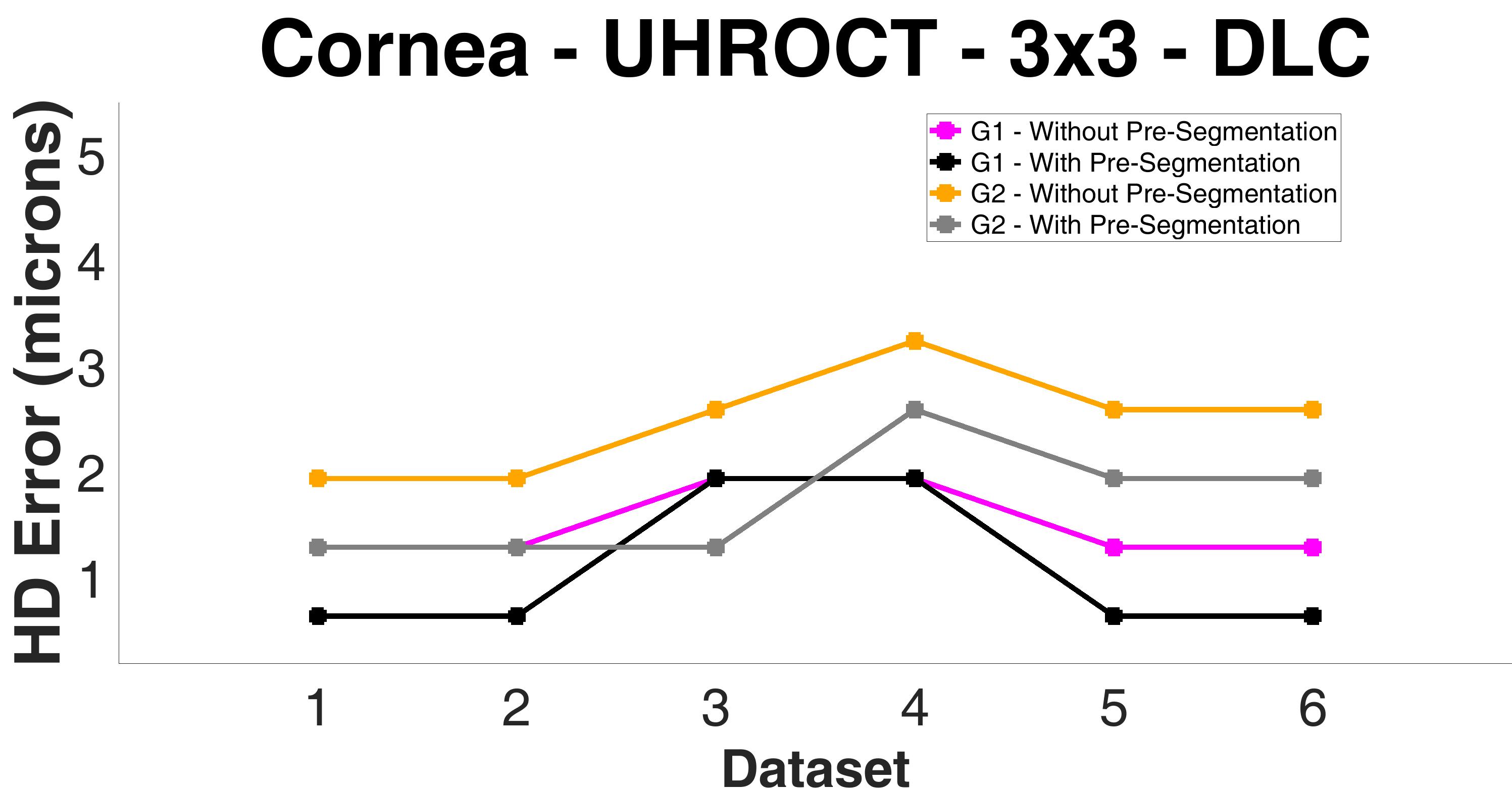}\\
\centering{(f)}
\end{subfigure}
\caption{Quantitative estimation of the benefit of pre-segmenting the corneal OCT image. All the baselines were grouped into two categories: Traditional Comparison (TC; TWOPS vs TWPS), and Deep Learning Comparison (DLC; DLWOPS vs DLWPS). The first column corresponds to the former, and the second column corresponds to the latter. For each corneal test dataset, the image with the maximum HD error was found over all images in the sequence, and the image location in the sequence was stored. This was done only for the TWOPS and DLWOPS baselines respectively. The stored location indicies were then used to retrieve the corresponding HD errors from the TWPS and DLWPS baselines respectively. This procedure was repeated for each grader and plotted. G1 : without pre-segmentation (purple curve), with pre-segmentation (black curve). G2 : without pre-segmentation (yellow curve), with pre-segmentation (gray curve).}
\label{me_cor_HD}
\end{figure}

\begin{figure}[!h]
\begin{subfigure}[b]{0.04\columnwidth}
\includegraphics[height=0.5cm,width=\columnwidth]{empty}\\
\vfill\vfill\vfill
\includegraphics[height=4cm,width=\columnwidth]{Device2ce}\\
\vfill\vfill\vfill
\includegraphics[height=4cm,width=\columnwidth]{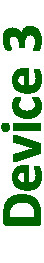}\\
\vfill\vfill\vfill
\end{subfigure}\hfill
\begin{subfigure}[b]{0.45\columnwidth}
\includegraphics[height=0.5cm,width=\columnwidth]{HDce}\\
\vfill\vfill\vfill
\includegraphics[height=4cm,width=\columnwidth]{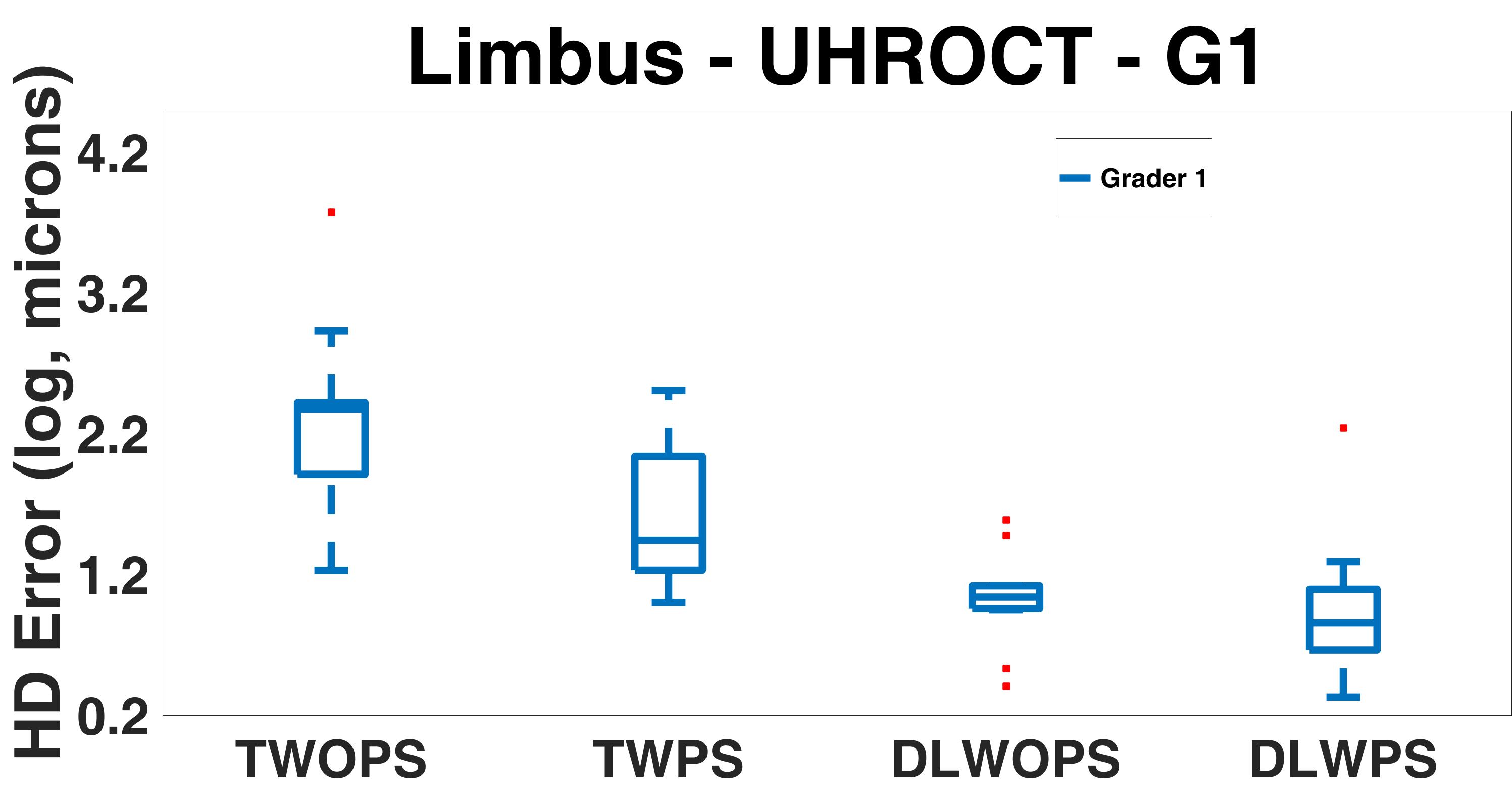}\\
\centering{(a)}\vfill
\includegraphics[height=4cm,width=\columnwidth]{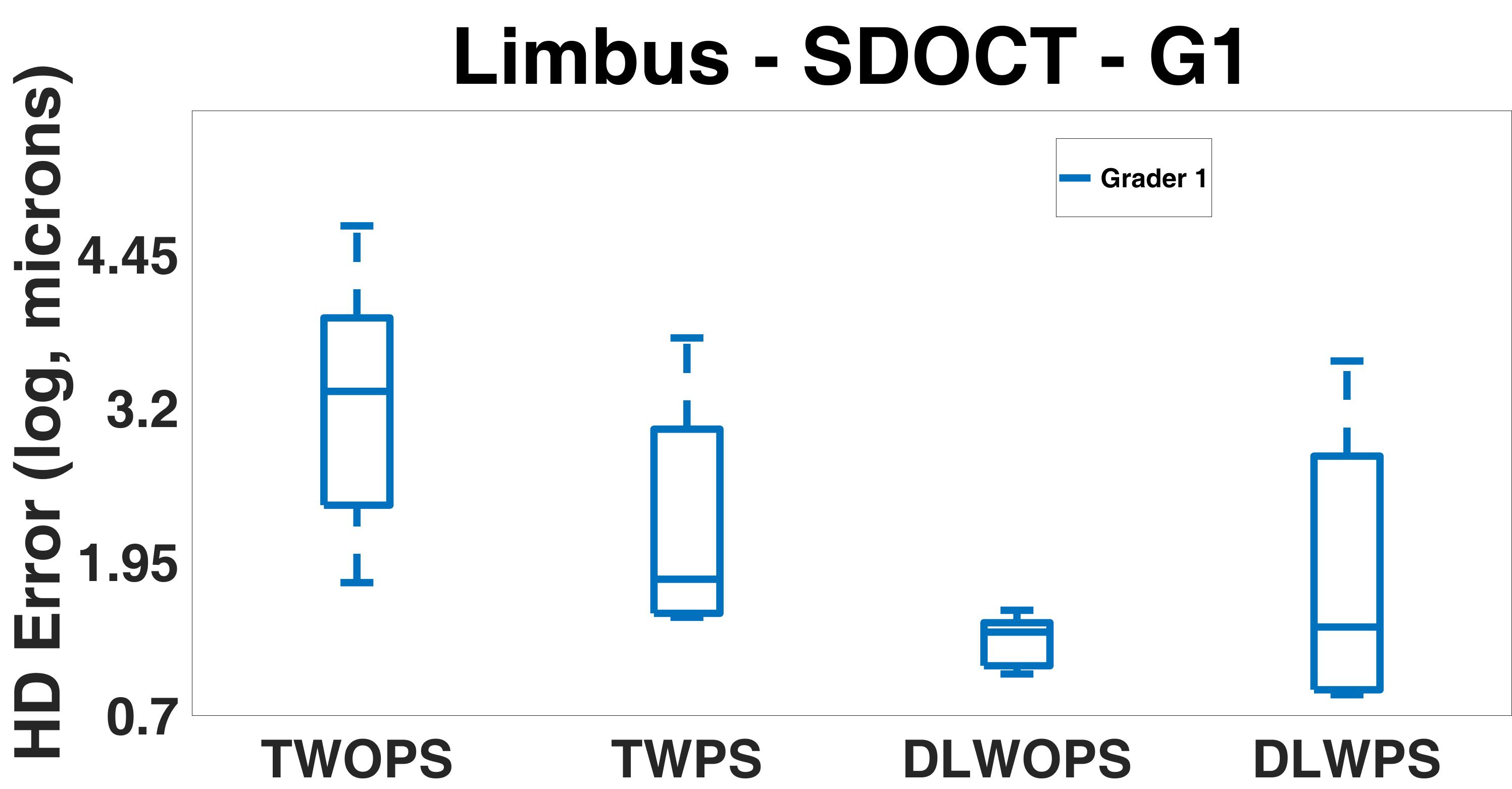}\\
\centering{(b)}
\end{subfigure}\hfill
\begin{subfigure}[b]{0.45\columnwidth}
\includegraphics[height=0.5cm,width=\columnwidth]{MADce}\\
\vfill\vfill\vfill
\includegraphics[height=4cm,width=\columnwidth]{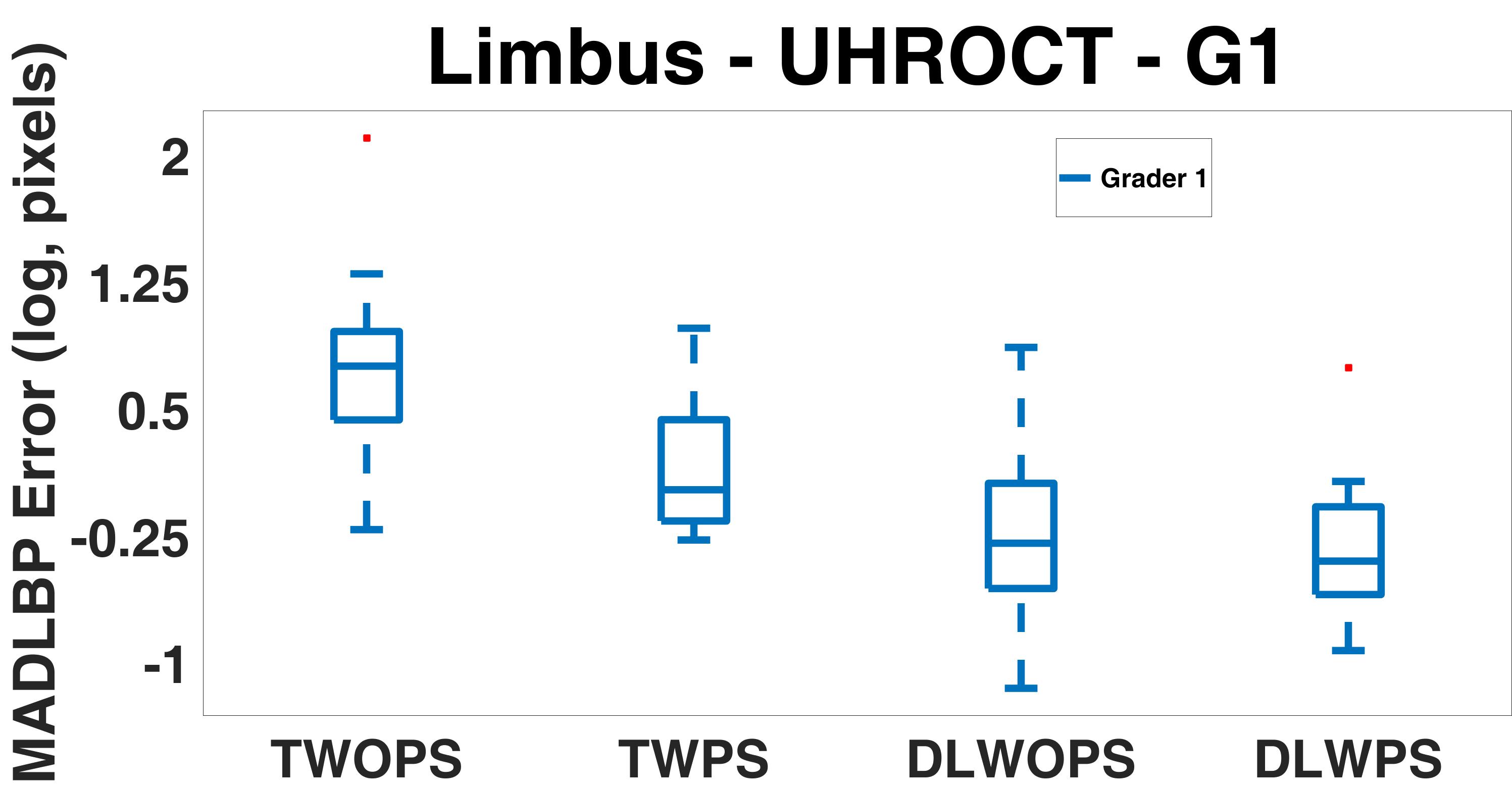}\\
\centering{(c)}\vfill
\includegraphics[height=4cm,width=\columnwidth]{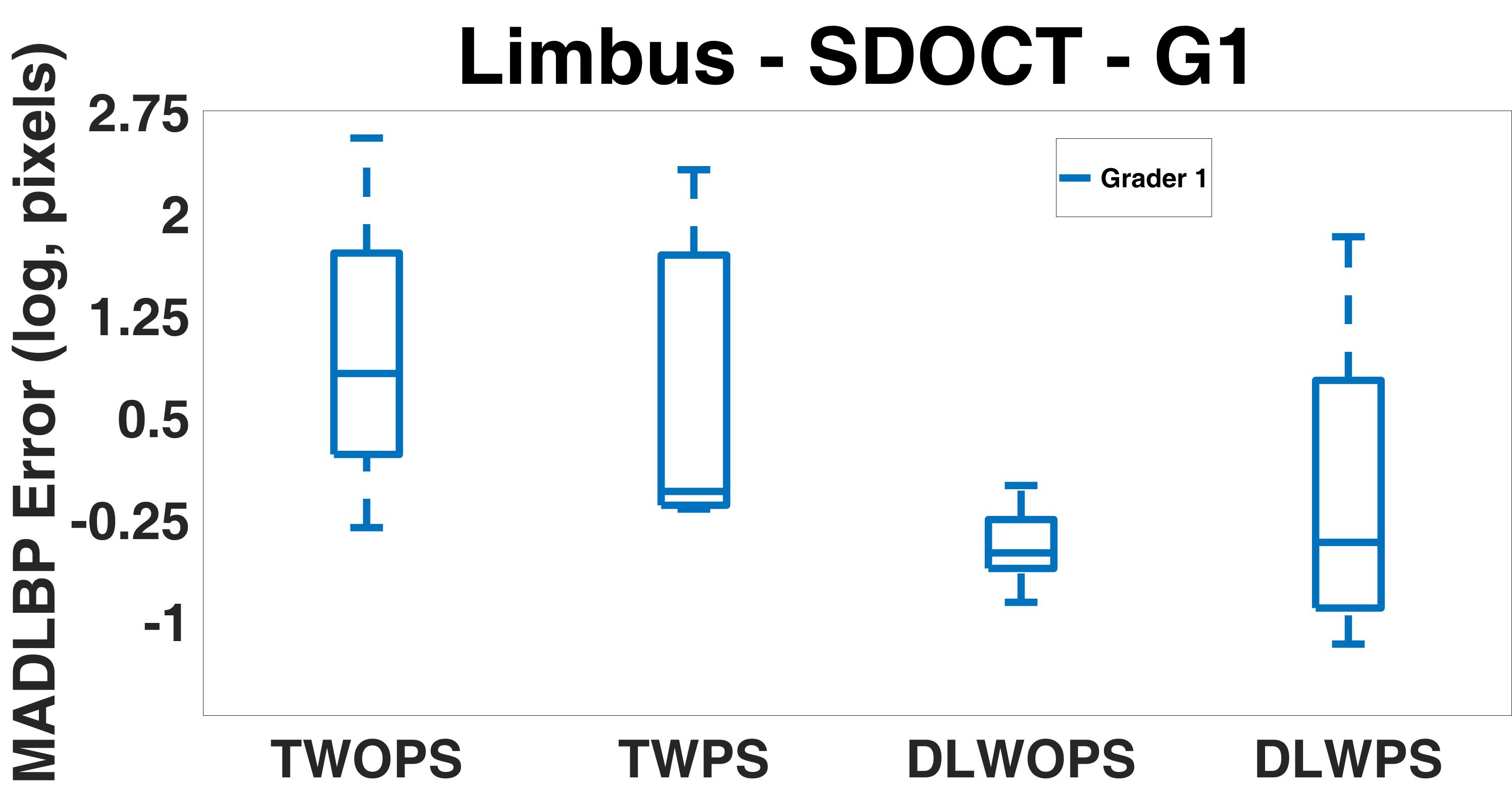}\\
\centering{(d)}
\end{subfigure}
\caption{(a)-(b) HD error and (c)-(d) MADLBP error comparison for the limbal datasets acquired with Devices 2 and 3 respectively. For the limbal datasets, the segmentation results obtained for each baseline method were contrasted exclusively against the expert annotations (G1). This graph plots the errors across all limbal datasets, including the failure cases. In contrast to Fig. \ref{limbus_HD_MAD}, note the increased segmentation error in the DLWPS baseline due to imprecise pre-segmentations.}
\label{bad_limbus_HD_MAD}
\end{figure}

\begin{figure}[!hb]
\begin{subfigure}[b]{0.04\columnwidth}
\includegraphics[height=4cm,width=\columnwidth]{Device2ce}\\
\vfill\vfill\vfill
\includegraphics[height=4cm,width=\columnwidth]{Device3ce}\\
\vfill\vfill\vfill
\end{subfigure}\hfill
\begin{subfigure}[b]{0.45\columnwidth}
\includegraphics[height=0.5cm,width=\columnwidth]{HDce}\\
\vfill\vfill\vfill
\includegraphics[height=4cm,width=\columnwidth]{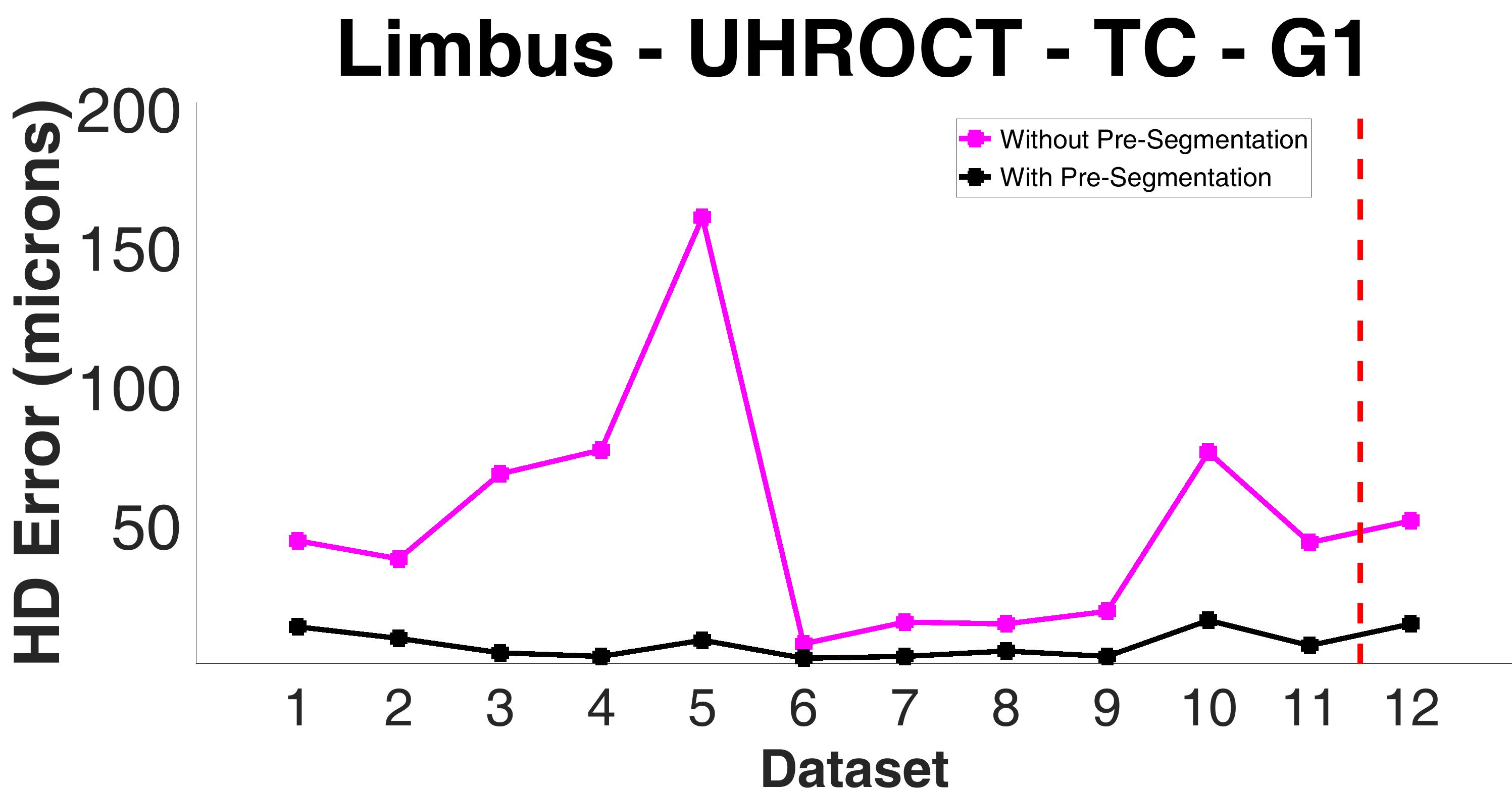}\\
\centering{(a)}\vfill
\includegraphics[height=4cm,width=\columnwidth]{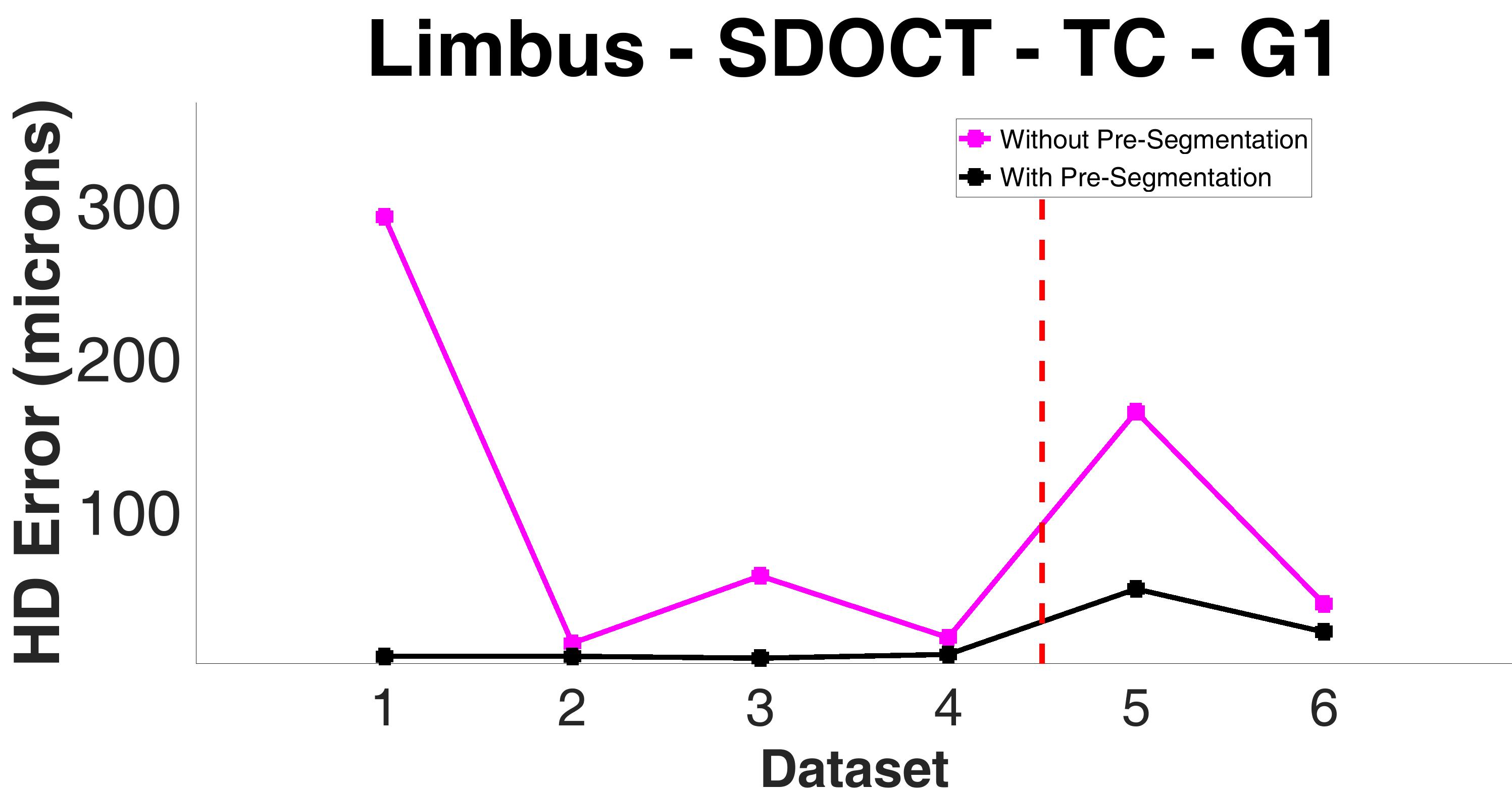}\\
\centering{(c)}
\end{subfigure}\hfill
\begin{subfigure}[b]{0.45\columnwidth}
\includegraphics[height=0.5cm,width=\columnwidth]{HDce}\\
\vfill\vfill\vfill
\includegraphics[height=4cm,width=\columnwidth]{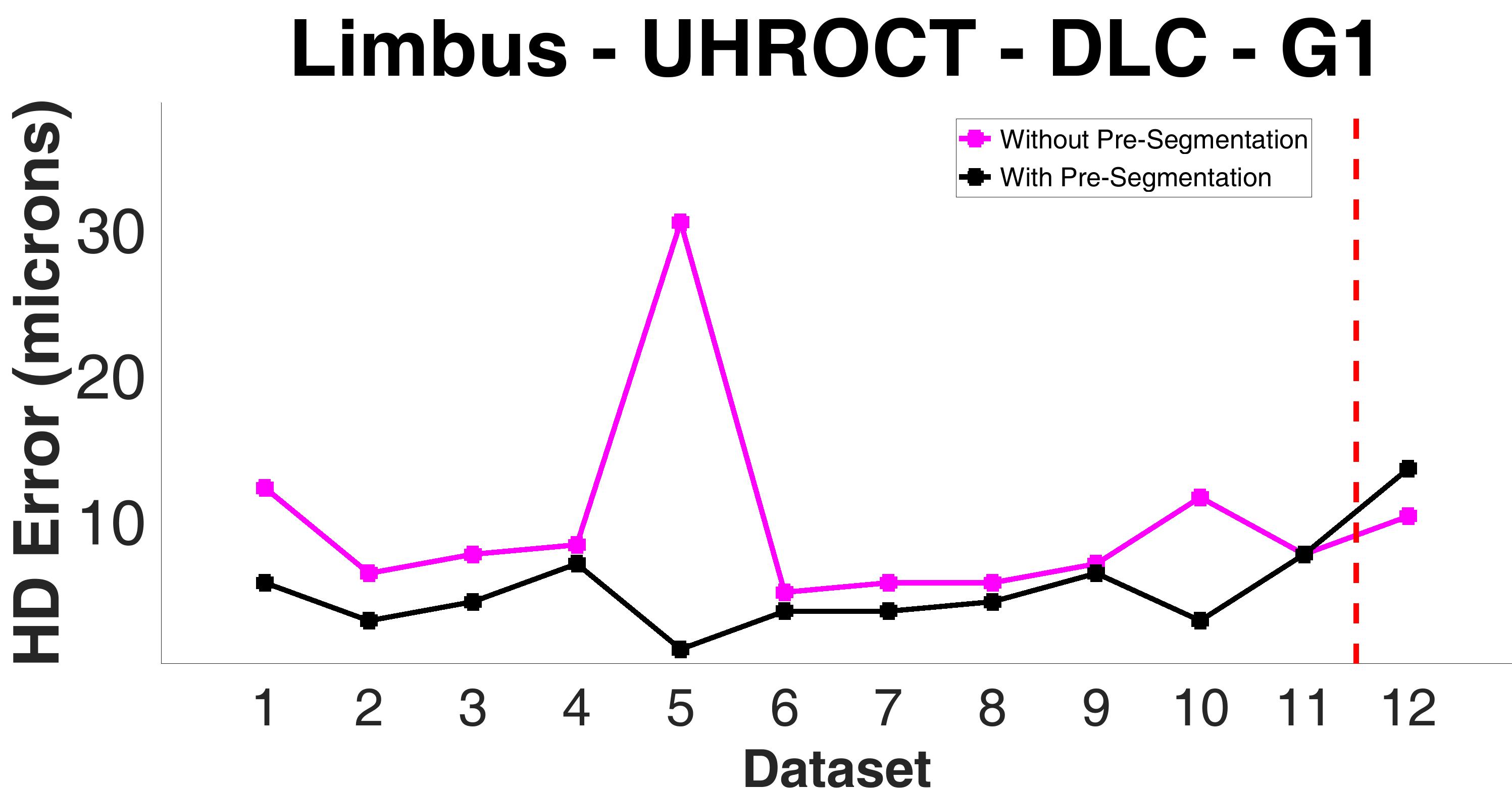}\\
\centering{(b)}\vfill
\includegraphics[height=4cm,width=\columnwidth]{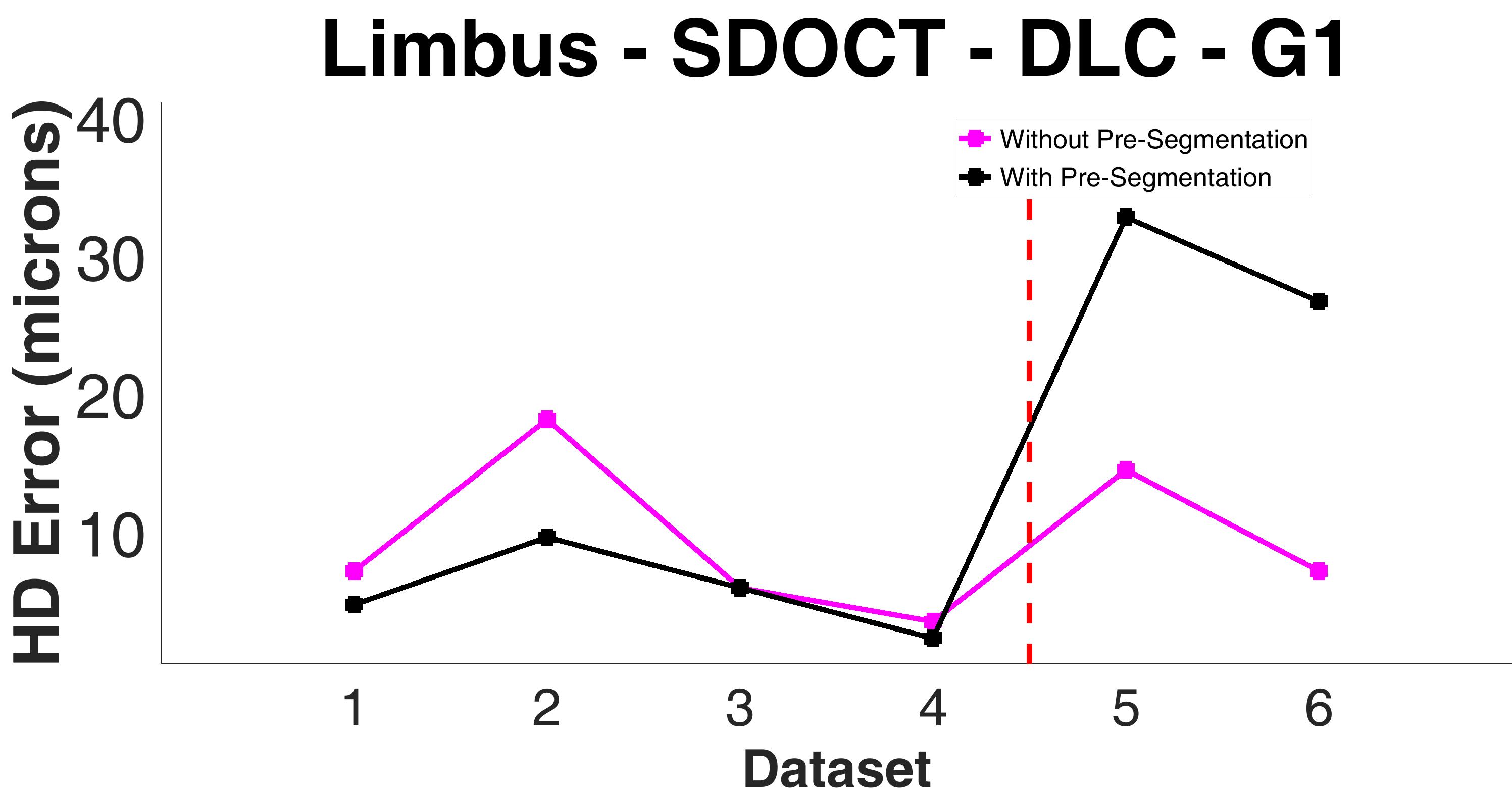}\\
\centering{(d)}
\end{subfigure}
\caption{Quantitative estimation of the benefit of pre-segmenting the corneal OCT image. All the baselines were grouped into two categories: TC (TWOPS vs TWPS), and DLC (DLWOPS vs DLWPS). The first column corresponds to the former, and the second column corresponds to the latter. For each test dataset, the image with the maximum HD error was found over all images in the sequence, and the image location in the sequence was stored. This was done only for the TWOPS and DLWOPS baselines respectively. The stored location indicies were then used to retrieve the corresponding HD errors from the TWPS and DLWPS baselines respectively. This procedure was done for only the expert grader and plotted. G1 : without pre-segmentation (purple curve), with pre-segmentation (black curve). Errors shown after red vertical line correspond to the failure cases of our approach.}
\label{me_limbus_HD}
\end{figure}

In Fig. \ref{cor_HD_MAD}, the HD error and the MADLBP error across all baselines for the corneal datasets acquired from devices 1 and 2 were compared. In Fig. \ref{me_cor_HD}, the benefit of pre-segmenting the OCT image was verified by first grouping the baselines into two categories - Traditional Comparison (TC; TWOPS vs TWPS) and Deep Learning Comparison (DLC; DLWOPS vs DLWPS) - and then contrasting the maximum HD error per dataset for each category and for each grader. We also determined the HD and MADLBP error across the limbal datasets in Figs. \ref{bad_limbus_HD_MAD} and \ref{limbus_HD_MAD}. Again in Fig. \ref{me_limbus_HD}, we estimated the benefit of pre-segmenting limbal datasets by grouping baselines into two categories, TC and DLC, and comparing maximum HD error per dataset for each category. Moreover, we found a few instances where our cascaded framework failed to correctly segment the tissue interface as seen in Fig. \ref{me_limbus_HD} (results after the red vertical line). 

%--------------
%--------------
%--------------
%--------------
%--------------
%--------------
%--------------
%--------------

%-------------------------------------------------------------------
%-------------------------------------------------------------------
\section{Discussion}
\label{sec_discussion}
%-------------------------------------------------------------------
%-------------------------------------------------------------------

%-------------------------------------------------------------------
%-------------------------------------------------------------------
\subsection{Segmentation Accuracy of Corneal Interface}
\label{discuss_seg_acc_cor}
%-------------------------------------------------------------------
%-------------------------------------------------------------------

From the HD and MADLBP errors in Figs. \ref{cor_HD_MAD}, the error is worse for the TWOPS baseline method, where the traditional algorithm \cite{Mathai2018} used the original OCT image (without the pre-segmentation) to directly segment the interface. The hand-crafted features in this baseline algorithm failed to handle severe specular artifacts and noise patterns as seen in Fig. \ref{fig:fig_prior_algo_comparison}. In contrast, the TWPS baseline (hybrid framework), which uses the pre-segmented image instead of the original OCT image, produced a lower segmentation error. To quantify these observations, a paired t-test between the TWOPS and TWPS baselines was computed for each error metric, and we estimated that the results were statistically significant ($p_{\scaleto{\textnormal{HD}}{4pt}}$ = 4.2747e-05, $p_{\scaleto{\textnormal{MADLBP}}{4pt}}$ = 1.2859e-05). From these results, we concluded that the traditional algorithm fared better in the hybrid framework when the pre-segmented OCT image was used to segment the corneal tissue interface. 
%In Fig. \ref{cor_HD_MAD}(f), the MADLBP error for the TWOPS baseline is lower for the trained grader as opposed to the expert grader, and we attribute this to the large inter-grader variability between the expert and trained graders for the 3$\times$3mm datasets from Device 2. 

The DLWOPS baseline in Fig. \ref{cor_HD_MAD} had lower HD and MADLBP errors than the TWPS baseline for the expert grader annotations. However, the errors were higher for the trained grader especially on the 3$\times$3mm datasets from Device 2, as seen in Figs. \ref{cor_HD_MAD}(c) and \ref{cor_HD_MAD}(f), due to the large inter-grader variability. On the other hand, our DLWPS baseline approach, which used the pre-segmented image, fared better in contrast to the other three baselines. Again, we computed paired t-tests between the DLWPS approach and all other baselines to determine the improvement in segmentation accuracy for each error metric. From the $p$-values in Table. \ref{table_cor_pValue} and Fig. \ref{cor_HD_MAD}, the cascaded framework generated results that were an improvement upon the other baselines, and indicated statistically significant results across all corneal datasets (p $<$ 0.05). 

%--------------
\begin{table}[!ht]
\centering\fontsize{9}{11}\selectfont
\caption{Statistical significance between our cascaded framework (DLWPS) against each baseline method for all the corneal datasets from Devices 1 and 2.}
\begin{tabular}{cccc}\hline
\toprule
					                        & TWOPS                 & TWPS                  & DLWOPS \\ 	\midrule

$p_{\scaleto{\textnormal{HD}}{4pt}}$		& 5.1929e-06            & 2.2079e-04            & 5.1454e-04   \\
$p_{\scaleto{\textnormal{MADLBP}}{4pt}}$    & 2.6848e-06            & 1.9264e-04            & 2.0734e-04   \\ \bottomrule
\end{tabular}
\label{table_cor_pValue}
\end{table}
%--------------

We also wanted to determine the improvement in segmentation accuracy on an per-image basis in each of the corneal test datasets. To do so, we first grouped the baselines into two categories: only traditional image analysis-based approaches (TWOPS vs. TWPS), and only deep learning-based approaches (DLWOPS vs. DLWPS). Next, we searched for the image in each corneal dataset that had the maximum HD error over all images in that dataset, and noted its index in the sequence. This was done only for the TWOPS and DLWOPS baselines respectively, and we plotted these maximum HD errors for each grader in Fig. \ref{me_cor_HD} (purple and yellow colored curves). Then, we queried the errors for the same images (using the image indicies) in the TWPS and DLWPS baseline approaches respectively, and plotted the corresponding HD errors for each grader in Fig. \ref{me_cor_HD} (black and gray curves). From Fig. \ref{me_cor_HD}, we noted that the baselines incorporating the pre-segemented OCT image performed better than one that did not include the pre-segmentation. The pre-segmentation always improved the segmentation performance of the traditional image-analysis based approach when incorporated into a hybrid framework, and also improved the accuracy of a deep learning-based approach in a majority of corneal datasets when used in the cascaded framework. This quantitatively attests to the benefit of utilizing the pre-segmented OCT image as part of a segmentation framework.

%-------------------------------------------------------------------
%-------------------------------------------------------------------
\subsection{Segmentation Accuracy of Limbal Interface}
%-------------------------------------------------------------------
%-------------------------------------------------------------------

\begin{figure}[!h]
\begin{subfigure}[b]{0.04\columnwidth}
\includegraphics[height=0.5cm,width=\columnwidth]{empty}\\
\vfill\vfill\vfill
\includegraphics[height=4cm,width=\columnwidth]{Device2ce}\\
\vfill\vfill\vfill
\includegraphics[height=4cm,width=\columnwidth]{Device3ce}\\
\vfill\vfill\vfill
\end{subfigure}\hfill
\begin{subfigure}[b]{0.45\columnwidth}
\includegraphics[height=0.5cm,width=\columnwidth]{HDce}\\
\vfill\vfill\vfill
\includegraphics[height=4cm,width=\columnwidth]{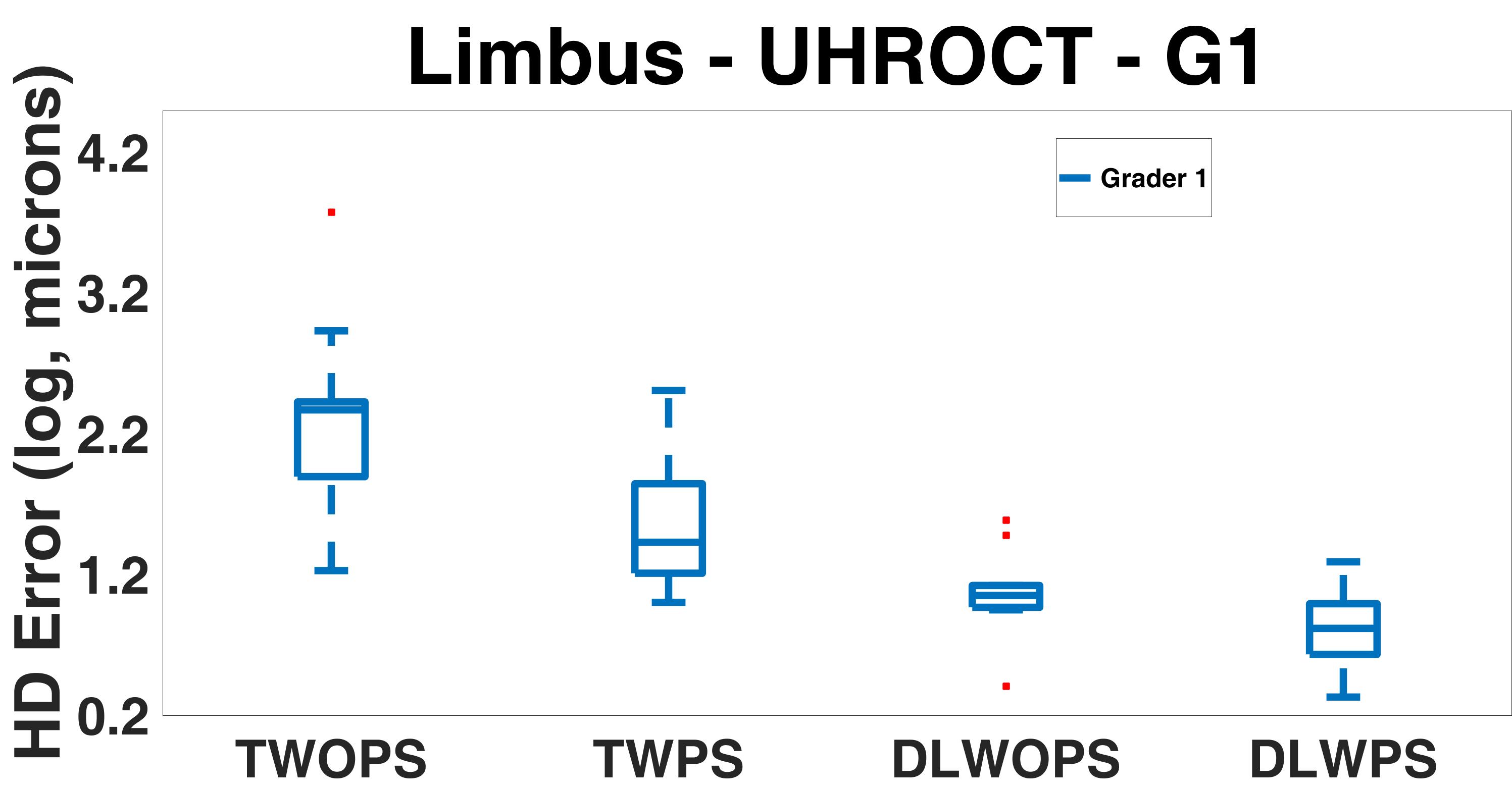}\\
\centering{(a)}\vfill
\includegraphics[height=4cm,width=\columnwidth]{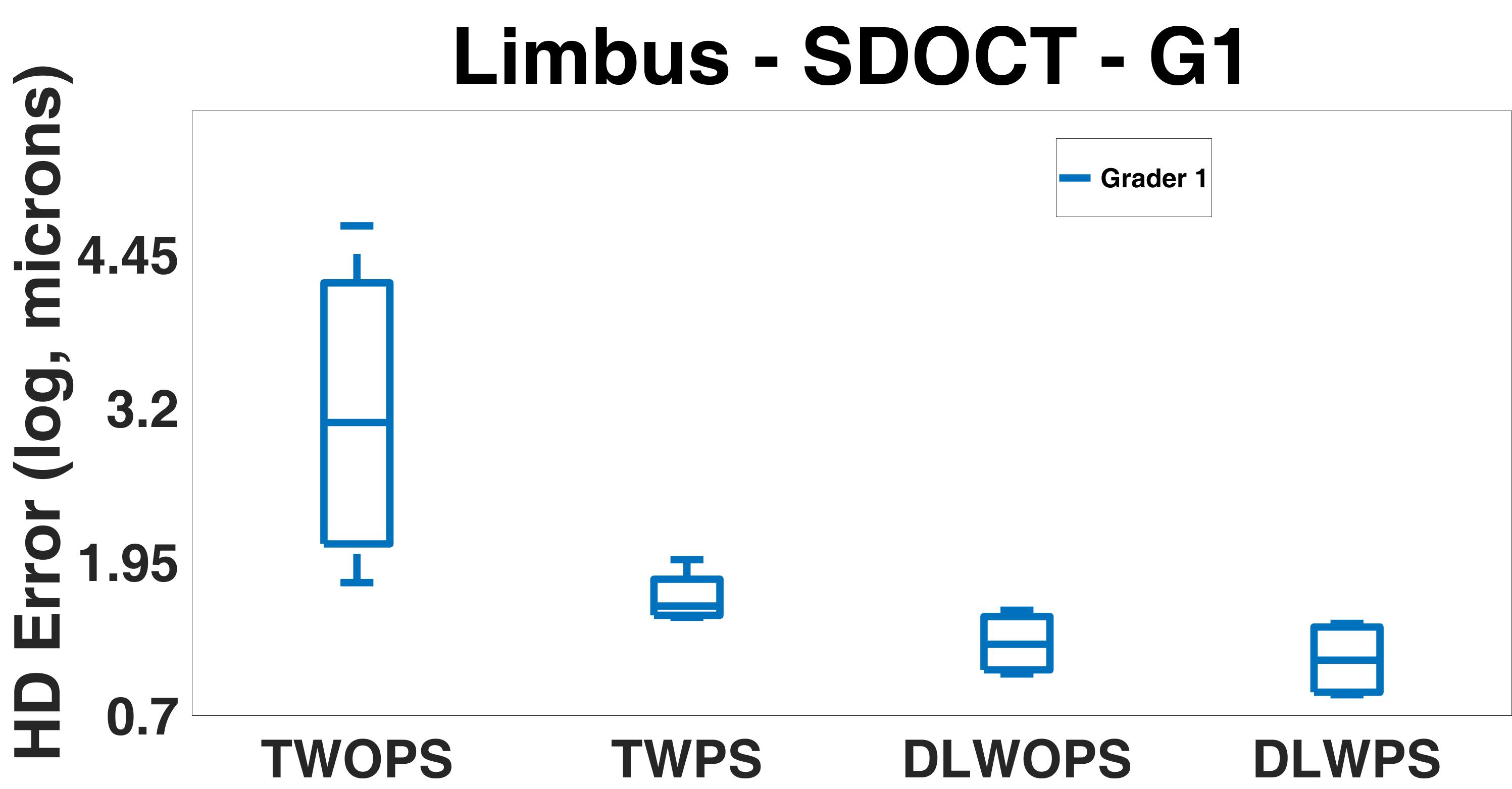}\\
\centering{(b)}
\end{subfigure}\hfill
\begin{subfigure}[b]{0.45\columnwidth}
\includegraphics[height=0.5cm,width=\columnwidth]{MADce}\\
\vfill\vfill\vfill
\includegraphics[height=4cm,width=\columnwidth]{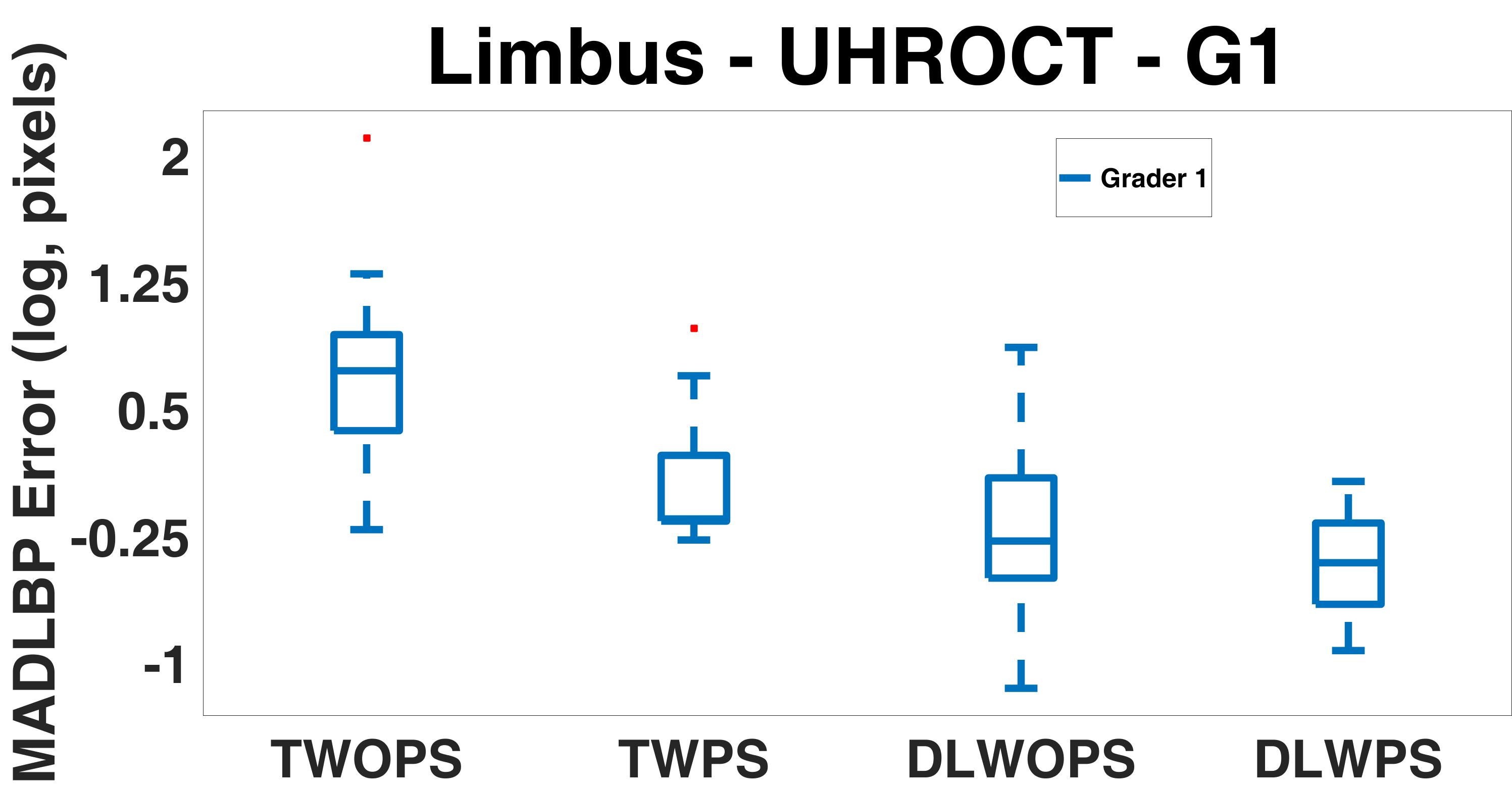}\\
\centering{(c)}\vfill
\includegraphics[height=4cm,width=\columnwidth]{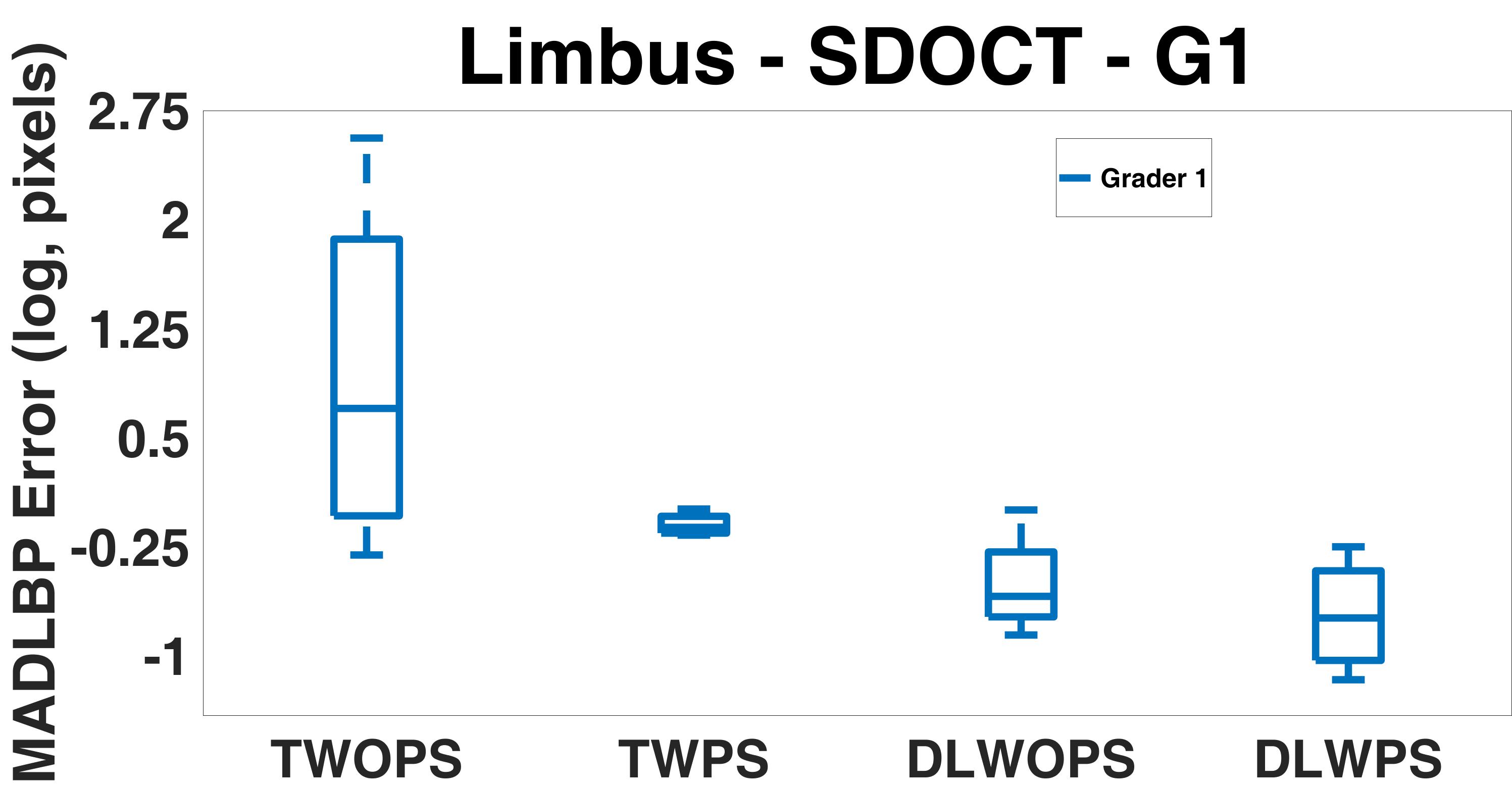}\\
\centering{(d)}
\end{subfigure}
\caption{(a)-(b) HD error and (c)-(d) MADLBP error comparison for the limbal datasets acquired with Devices 2 and 3 respectively. For the limbal datasets, the segmentation results obtained for each baseline method were contrasted exclusively against the expert annotations (G1). These graphs plot errors for the successful segmentation results on 15 limbal test datasets.}
\label{limbus_HD_MAD}
\end{figure}

We plotted the segmentation error for the baseline methods executed on limbal datasets in Figs. \ref{bad_limbus_HD_MAD}, \ref{me_limbus_HD} and \ref{limbus_HD_MAD}. In Fig. \ref{bad_limbus_HD_MAD}, we plotted the errors across all limbal test datasets, including the instances when the cascaded and hybrid frameworks failed to accurately segment the shallowest interface. In Fig. \ref{limbus_HD_MAD}, we plot the errors only for the successful instances of interface segmentation. From Figs. \ref{bad_limbus_HD_MAD} and \ref{limbus_HD_MAD}, the error for the TWOPS baseline is the worst amongst all baselines as it failed to handle strong specular artifacts and severe speckle noise. On the other hand, the TWPS baseline fared better with lower errors than the TWOPS baseline. Similar to Sec. \ref{discuss_seg_acc_cor}, we also assessed the improvement in segmentation accuracy on a per-image basis for each of the 18 limbal datasets. We plotted these errors in Fig. \ref{me_limbus_HD}. From the errors (after the red vertical dashed line) in Figs. \ref{me_limbus_HD}(a) and \ref{me_limbus_HD}(c), the hybrid framework (TWPS baseline) was able to reduce the segmentation error even with an incorrect OCT image pre-segmentation. Therefore, the incorporation of the pre-segmented OCT image in the hybrid framework lead to lower errors for the traditional image analysis-based approach.
%discuss the errors for datasets that our cascaded framework performed well on in Figs. \ref{limbus_HD_MAD} and \ref{me_limbus_HD} (before the red vertical dashed line), and examine the errors on datasets where our framework failed in Figs. \ref{me_limbus_HD} (after the red vertical dashed line) and \ref{bad_limbus_HD_MAD}. 
%the three datasets containing faintly visible tissue boundaries, with pixel intensities of the same amplitude as speckle noise and being obscured by it, and washed out tissue boundaries due to the image being saturated by specular artifacts. 

\begin{figure}[!h]
\centering
\begin{subfigure}[b]{0.245\columnwidth}
\centering
\includegraphics[height=3cm,width=0.5\columnwidth]{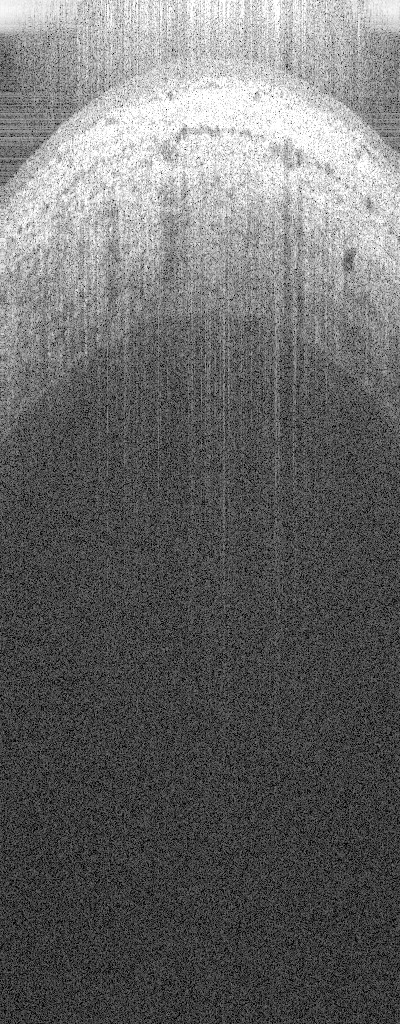}\\\medskip
\includegraphics[height=3cm,width=0.4\columnwidth]{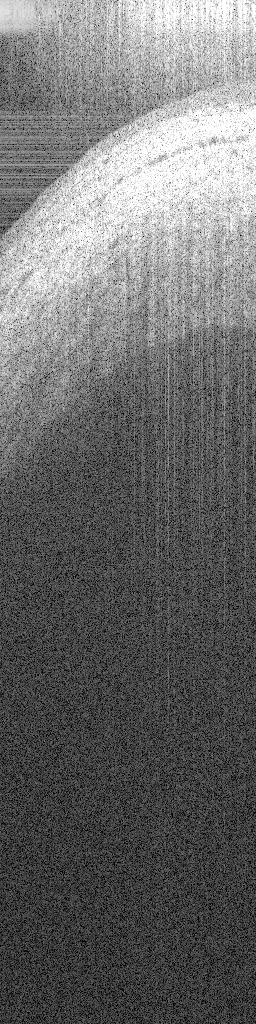}\\\medskip
\includegraphics[height=3cm,width=0.4\columnwidth]{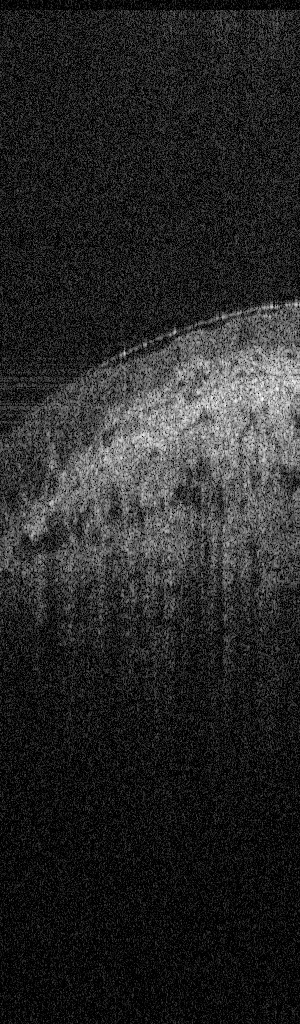}
\centerline{(a)}
\end{subfigure}
\begin{subfigure}[b]{0.245\columnwidth}
\centering
\includegraphics[height=3cm,width=0.5\columnwidth]{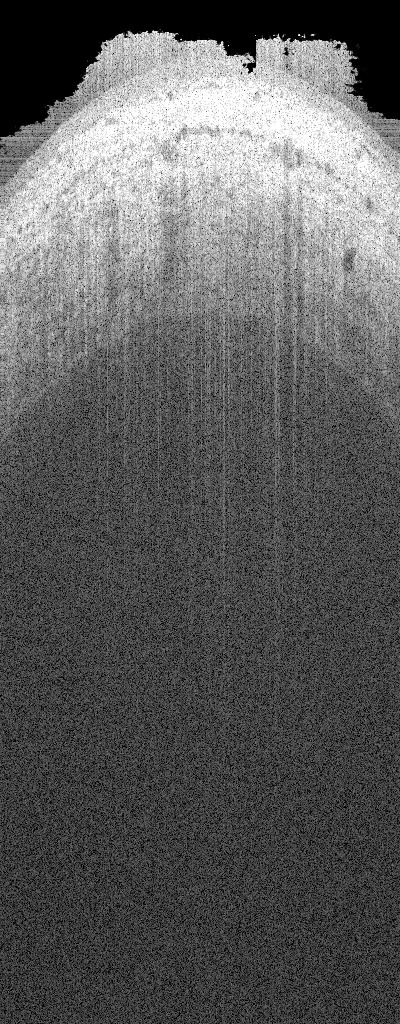}\\\medskip
\includegraphics[height=3cm,width=0.4\columnwidth]{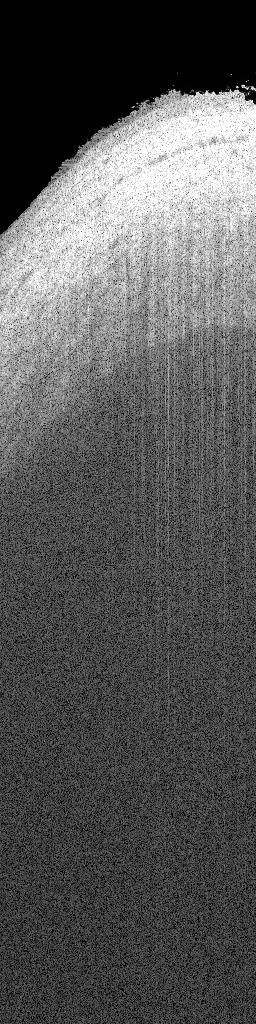}\\\medskip
\includegraphics[height=3cm,width=0.4\columnwidth]{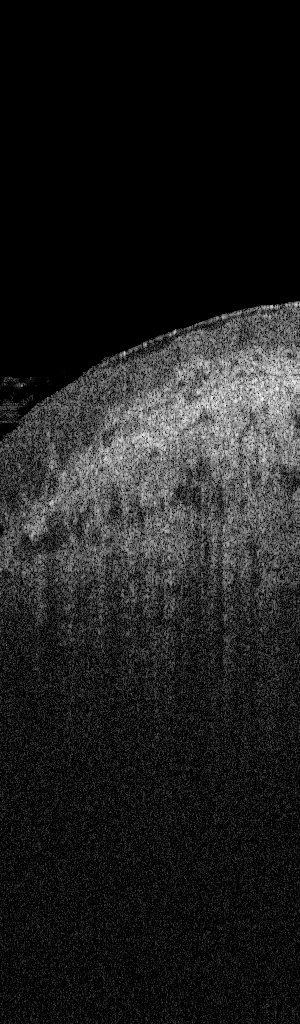}
\centerline{(b)}
\end{subfigure}
\begin{subfigure}[b]{0.245\columnwidth}
\centering
\includegraphics[height=3cm,width=0.5\columnwidth]{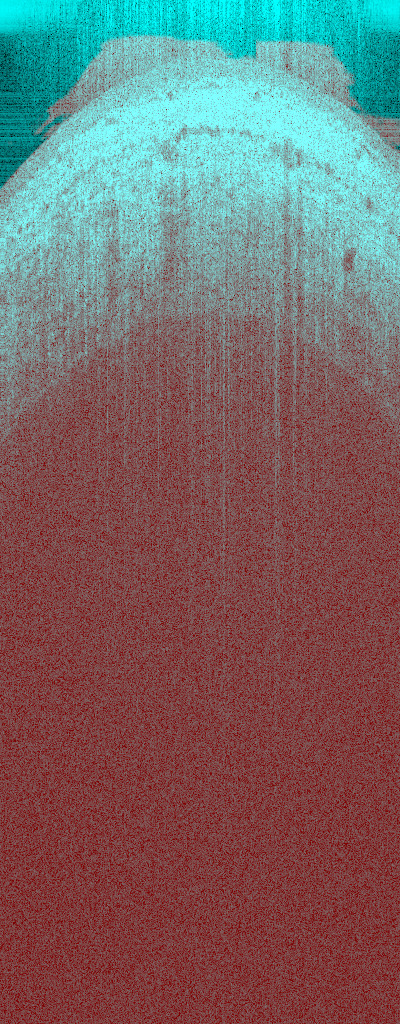}\\\medskip
\includegraphics[height=3cm,width=0.4\columnwidth]{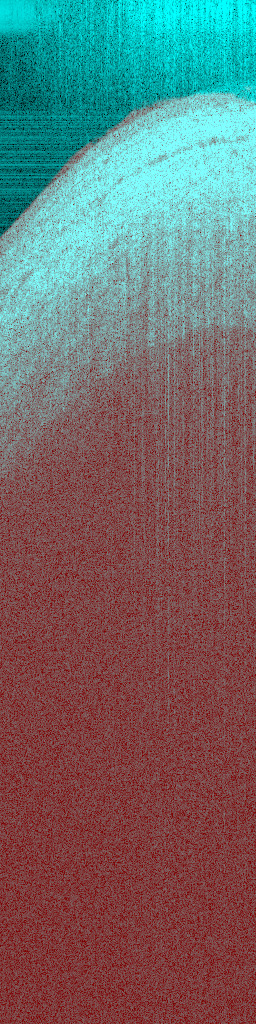}\\\medskip
\includegraphics[height=3cm,width=0.4\columnwidth]{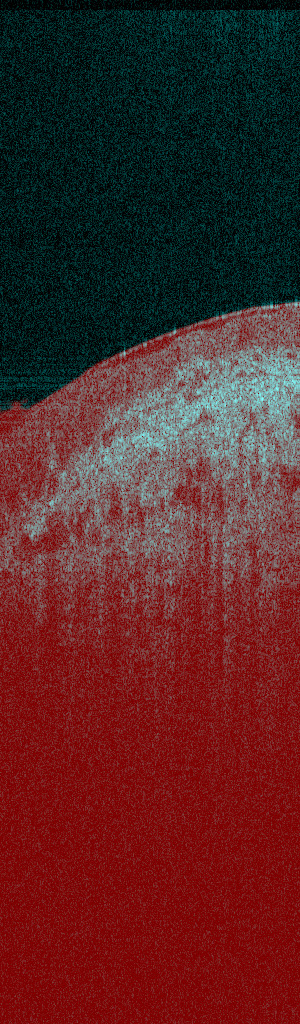}
\centerline{(c)}
\end{subfigure}
\begin{subfigure}[b]{0.245\columnwidth}
\centering
\includegraphics[height=3cm,width=0.5\columnwidth]{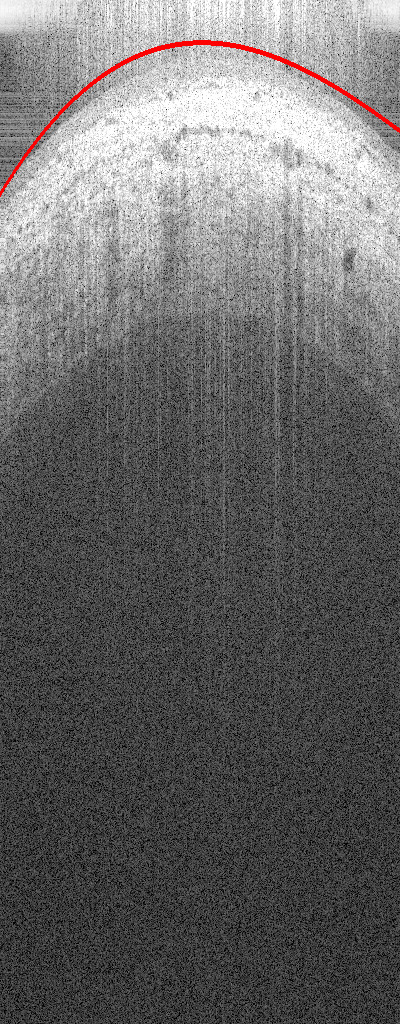}\\\medskip
\includegraphics[height=3cm,width=0.4\columnwidth]{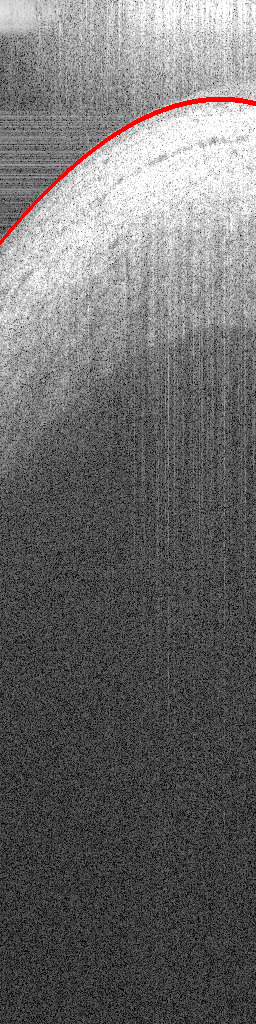}\\\medskip
\includegraphics[height=3cm,width=0.4\columnwidth]{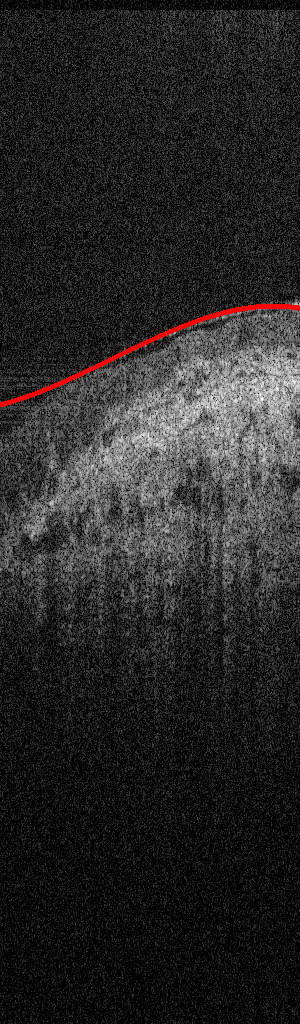}
\centerline{(d)}
\end{subfigure}
\caption{Failure cases of our cascaded framework on three challenging limbal OCT datasets. Columns from left to right: (a) Original B-scans in the limbal OCT volumes, (b) cGAN pre-segmentation results that imprecisely removed speckle noise patterns and specular artifacts above the shallowest tissue interface, (c) The binary segmentation masks from the TISN overlaid in false color (red - foreground, turquoise - background) on the original B-scans, (d) Curve fit to the shallowest interface (red contour).}
\label{fig:res_bad_limbus}
\end{figure}

The DLWOPS baseline had lower errors as shown in Figs. \ref{bad_limbus_HD_MAD} and \ref{limbus_HD_MAD} as compared to the TWOPS and TWPS baselines. But, at an image level, it sometimes yielded higher segmentation errors as seen in Figs. \ref{me_limbus_HD}(b) and \ref{me_limbus_HD}(d). On the other hand, the DLWPS baseline (cascaded framework) improved the segmentation error in a majority of the datasets, with the exception of three datasets, which are our failure cases. As shown in Fig. \ref{fig:res_bad_limbus}, two datasets presented with saturated tissue regions, which were washed out by specular artifacts. Another dataset contained regions where the interface was barely visible due to being obfuscated by speckle noise of the same amplitude. Due to these reasons, the incorrect pre-segmented OCT image degraded the segmentation performance of the TISN. Consequently, the segmentation error of the TWPS (hybrid framework) and DLWPS (cascaded framework) baselines was increased. As seen in Fig. \ref{me_limbus_HD} (after the red vertical dashed line), the DLWOPS baseline performed the best among all other baselines for these datasets. 
%obscured by speckle noise patterns or by specular artifacts. The images in these datasets were useful test cases as we could assess the behavior of the GAN and the TISN when tissue boundaries were 

We expound on the aforementioned reasons for segmentation failure. First, the contextual information available to the cGAN to remove the speckle noise patterns and specular artifacts is hindered when the pixel intensities on the tissue interface are either washed out due to saturation of the line scan camera \cite{LaRocca2011,Mathai2018,Mathai2018_2} as shown in Fig. \ref{fig:res_bad_limbus}(a) (top two rows), or blend in with the background and specular artifacts of the same amplitude \cite{LaRocca2011} as seen in Fig. \ref{fig:res_bad_limbus}(a) (bottom). In such outlier cases, the boundary becomes difficult to delineate across multiple scales through downsampling and upsampling operations in the encoder and decoder blocks, such that even the dilated convolutions and dense connections employed in the network are insufficient to recover context from surrounding boundary regions when localizing the interface.

Second, the TISN over-relied on the pre-segmentation in order to generate the final segmenation. During training of the TISN, the original image was coupled with the gold standard pre-segmentation output (see Fig. \ref{fig:fig_anno_shiftUp}) into a two-channel input. The TISN learned that the tissue boundary in the gold standard pre-segmentation was the location of the start of the true boundary. However, the TISN was not trained with gold standard pre-segmented images that were artificially induced to be corrupted and noisy, such as the images shown in Fig. \ref{fig:res_bad_limbus}(b). Hence, the performance of the TISN on such incorrectly pre-segmented OCT images is poor.

One way to address this issue is to re-train the framework with gold standard pre-segmentations that have corrupted boundaries. In this pilot work, we did not introduce any corruption to the gold standard pre-segmentation used during training as we wanted to directly measure the performance of the TISN when provided with a pre-segmentation from the cGAN (without regard to any imprecise pre-segmentation). Another option is to exploit the temporal correlation between B-scans in the dataset through recurrent neural networks, which retain long-term information in memory in order to deal with such challenging datasets. We intend to pursue these ideas in our future work. 

In this work, we set aside these three challenging failure cases, and estimated the improvement in segmentation accuracy across the remaining 15 limbus datasets. We conducted a paired t-test between the TWOPS and TWPS baselines for each error metric, and determined that our errors were statistically significant ($p_{\scaleto{\textnormal{HD}}{4pt}}$ = 0.0471, $p_{\scaleto{\textnormal{MADLBP}}{4pt}}$ = 0.0313). We also calculated paired t-tests between the DLWPS baseline and all other baselines to determine the statistical significance of our results for each error metric. As seen in Table. \ref{table_limbus_pValue}, our DLWPS cascaded framework generated statistically significant results (p $<$ 0.05). 

%--------------
\begin{table}[!ht]
\centering\fontsize{9}{11}\selectfont
\caption{Statistical significance between our cascaded framework (DLWPS) against each baseline method for 15 (out of 18) limbal datasets acquired from Devices 2 and 3.}
\begin{tabular}{cccc}\hline
\toprule
				                            & TWOPS         & TWPS          & DLWOPS \\ 	\midrule

$p_{\scaleto{\textnormal{HD}}{4pt}}$		& 0.0240        & 0.0014        & 1.0335e-04   \\
$p_{\scaleto{\textnormal{MADLBP}}{4pt}}$    & 0.0126        & 0.0012        & 0.0344   \\ \bottomrule
\end{tabular}
\label{table_limbus_pValue}
\end{table}
%--------------
%--------------
%--------------
%--------------
%--------------

%-------------------------------------------------------------------
%-------------------------------------------------------------------
\subsection{Interface Segmentation at Limbal Junction}
%-------------------------------------------------------------------
%-------------------------------------------------------------------

\begin{figure}[!h]
\centering
\begin{subfigure}[b]{0.245\columnwidth}
\centering
\includegraphics[height=3.5cm,width=0.4\columnwidth]{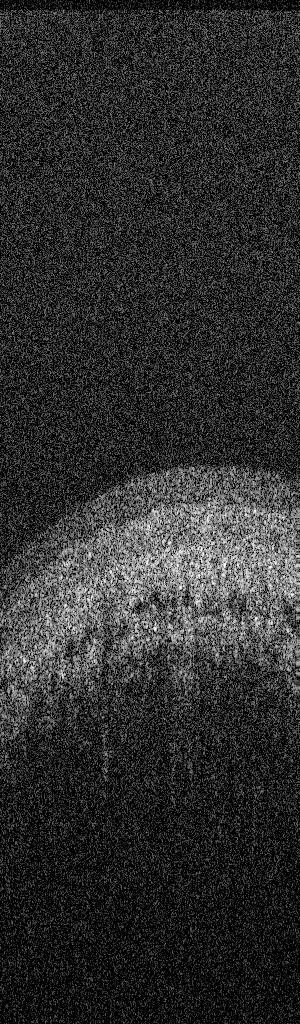}
\centerline{(a)}\\\medskip
\includegraphics[height=3.5cm,width=0.4\columnwidth]{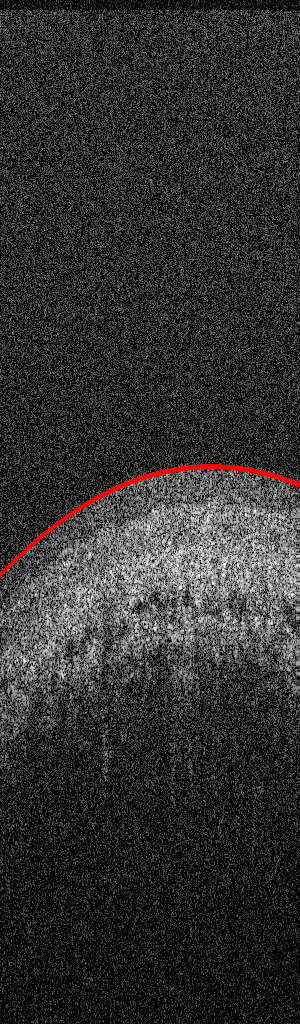}
\centerline{(e)}
\end{subfigure}
\begin{subfigure}[b]{0.245\columnwidth}
\centering
\includegraphics[height=3.5cm,width=0.4\columnwidth]{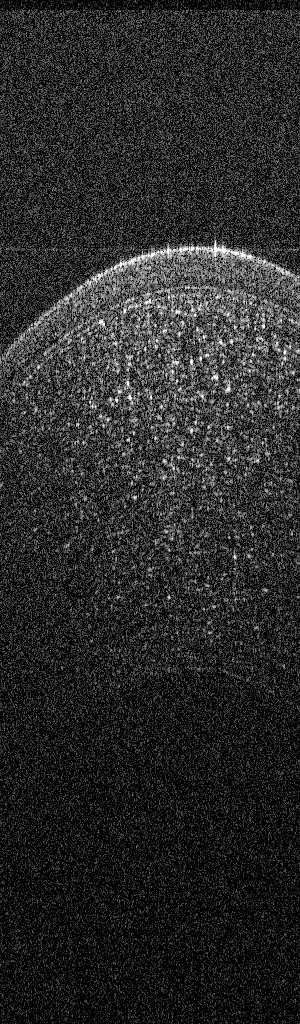}
\centerline{(b)}\\\medskip
\includegraphics[height=3.5cm,width=0.4\columnwidth]{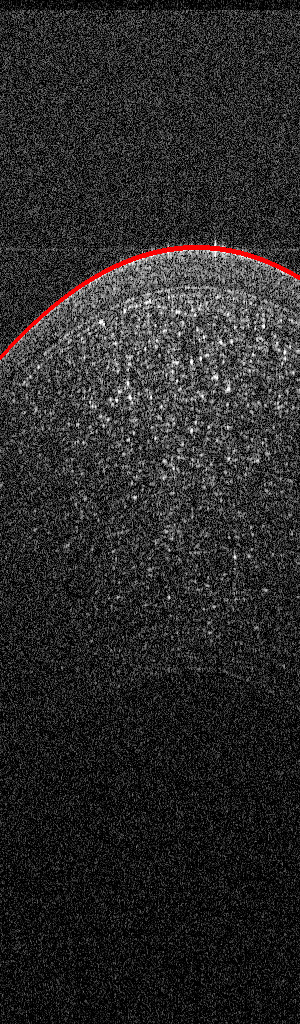}
\centerline{(f)}
\end{subfigure}
\begin{subfigure}[b]{0.245\columnwidth}
\centering
\includegraphics[height=3.5cm,width=0.4\columnwidth]{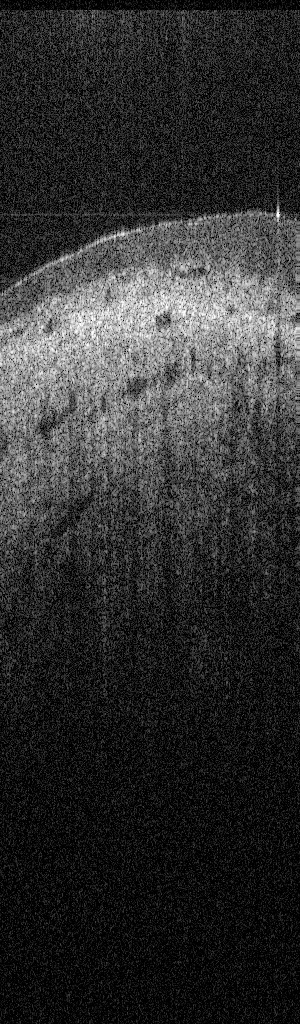}
\centerline{(c)}\\\medskip
\includegraphics[height=3.5cm,width=0.4\columnwidth]{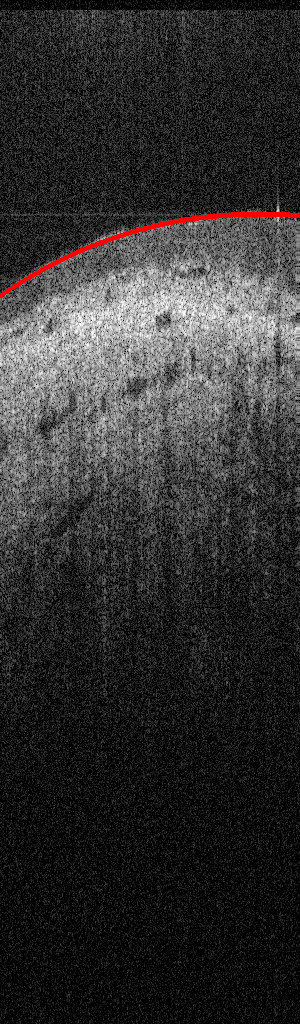}
\centerline{(g)}
\end{subfigure}
\begin{subfigure}[b]{0.245\columnwidth}
\centering
\includegraphics[height=3.5cm,width=0.4\columnwidth]{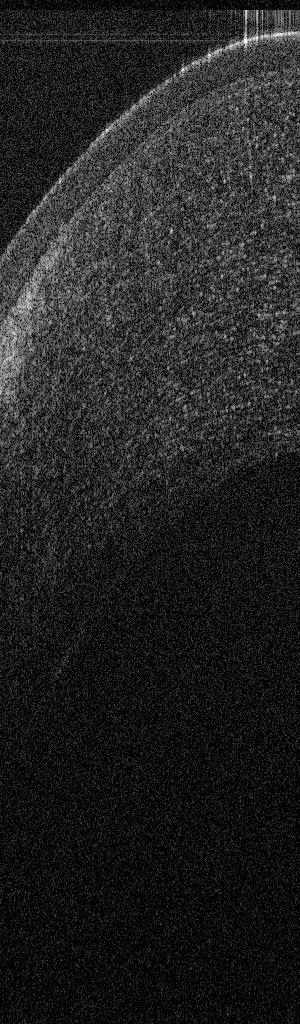}
\centerline{(d)}\\\medskip
\includegraphics[height=3.5cm,width=0.4\columnwidth]{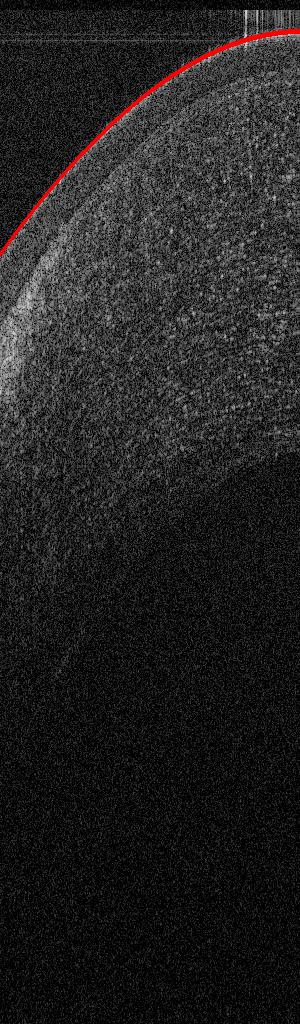}
\centerline{(h)}
\end{subfigure}
\caption{Segmenting the shallowest tissue interface in OCT datasets, wherein the OCT scanner commenced imaging from the limbus and crossed over into the cornea, thereby encompassing the limbal junction. (a),(b) B-scans \#1 and \#300 in an OCT dataset corresponding to the limbus and the cornea respectively. (c),(d) B-scans \#1 and \#220 in a different OCT dataset corresponding to the limbus and the cornea respectively. (e),(f),(g),(h) Segmentation (red curve) of the shallowest tissue interface in images shown in (a),(b),(c) and (d) respectively. Note the partial overlap of the limbal (left) and corneal (right) region in the B-scan in (d), and the correct identification of the shallowest interface in (h).}
\label{fig:res_cor_to_limbus}
\end{figure}

During imaging of the limbal region, it is very common to acquire B-scans of the cornea and the limbus in the same dataset. This is because the scan pattern of the OCT scanner that is used to acquire the dataset will sometimes encompass sections of the limbus and the cornea. Bulk tissue motion between B-scans in a dataset is also customary during image acquisition. Therefore, it is crucial to capture the shallowest tissue interface of the limbus and the cornea as it enables distinguishing between these two distinct regions. By correctly locating these interfaces, a registration algorithm can be used to potentially align regions at and below these interfaces, while compensating for bulk tissue motion. To the best of our knowledge, we believe our approach is the first to accurately detect the shallowest corneal and limbal interface in OCT images acquired at the limbal junction even in the presence of severe speckle noise patterns and specular artifacts. Results of our approach are shown in Fig. \ref{fig:res_cor_to_limbus}, wherein the shallowest interface is identified in B-scans that partially overlap both the cornea and the limbus.

%--------------
%--------------

%-------------------------------------------------------------------
%-------------------------------------------------------------------
\subsection{Choice of Framework Design}
%-------------------------------------------------------------------
%-------------------------------------------------------------------

In this work, we proposed to generate an intermediate representation of the OCT image, i.e. the cGAN pre-segmentation, that can influence the performance of a segmentation algorithm. To this end, we proposed a cascaded and hybrid segmentation framework. However, we theorized that there are other framework designs that can be implemented instead of the proposed approaches. Amongst other approaches, for example, we could have utilized a GAN directly for segmenting the tissue interface from the OCT image, or trained a multi-task neural network framework (CNN, GAN, etc.) to provide both the pre-segmentation and the final interface segmentation. We reiterate our motivations next that laid the groundwork for the proposed frameworks over the aforementioned design choices. As mentioned before, our motivations were: 1) To generate a pre-segmentation that could be utilized in a hybrid framework, 2) Integrate the pre-segmentation into the image acquisition pipeline of custom-built OCT scanners, and 3) Incorporate the pre-segmentation in a cascaded framework and compare its segmentation performance against that of a state-of-the-art CNN-based segmentation method \cite{Mathai2018_2}.

Utilizing the GAN to directly yield the final interface segmentation does not provide an intermediate output, which can be integrated in a hybrid framework. Similarly, the multi-task framework would provide both the pre-segmented OCT image and the final interface segmentation. The pre-segmentation can be directly used in the hybrid framework and imaging pipeline respectively. However, the final segmentation may only be influenced by the shared weights of the multi-task network, and not by the pre-segmentation, which will be different from the final segmentation. Thus, if the pre-segmentation must influence the final interface segmentation (as it should), it may be necessary to train a new framework again (in a cascaded fashion with the multi-task framework) that would include the pre-segmentation. For these reasons and in line with our motivations, we believe that our choice of framework design was warranted. 

%-------------------------------------------------------------------
%-------------------------------------------------------------------
%\subsection{Hausdorff Error is a better metric than MADLBP}
%-------------------------------------------------------------------
%-------------------------------------------------------------------

%An observation drawn from this work is that the MADLBP error is not a reliable metric to quantify distance errors in segmentation; it only takes into account the mean distance error across the set of all segmented points being compared against the set of grader annotation points, while providing very little information on the distance to the farthest point in the annotated set. On the other hand, the Hausdorff Distance captures the maximum distance of \textit{any} segmented point to the annotation points, thereby allowing it to robustly determine the segmentation error. Bearing in mind that the MADLBP error is a popular error metric of choice in state-of-the-art segmentation algorithms, we have incorporated it in this work to compare our results against other baselines. 

%-------------------------------------------------------------------
%-------------------------------------------------------------------
\section{Conclusion}
%-------------------------------------------------------------------
%-------------------------------------------------------------------

In this paper, we generated an intermediate OCT image representation which can influence the performance of a segmentation algorithm. The intermediate representation is a pre-segmentation, generated by a cGAN, wherein speckle noise patterns and specular artifacts are eliminated just prior to the shallowest tissue interface in the OCT image. We proposed two frameworks that incorporate the intermediate representation: a cascaded framework and a hybrid framework. The cascaded framework comprised of a cGAN and a TISN; the cGAN pre-segmented the OCT image by removing the undesired specular artifacts and speckle noise patterns that confounded boundary segmentation, while the TISN segmented the final tissue interface by combining the original image and the pre-segmentation as inputs. The hybrid framework contained an image analysis-based segmentation method, among other state-of-the-art methods, that exploited the cGAN pre-segmentation to generate the final tissue interface segmentation. The frameworks were trained on corneal and limbal datasets acquired from three different OCT scanners with different scan protocols. They were able to handle varying degrees of specular artifacts, speckle noise patterns, and bulk tissue motion, and deliver consistent segmentation results. We compared the results of our frameworks against those from the state-of-the-art image analysis-based and deep learning-based algorithms. To the best of our knowledge, this is the first approach for OCT-based tissue interface segmentation that integrated the cGAN component of our framework in a hybrid fashion. We have shown the benefit of pre-segmenting the OCT image through the lower segmentation errors that were yielded. Finally, we have shown the utility of our algorithm in being able to segment the tissue interface at the limbal junction. We believe that the cGAN pre-segmentation output can be easily integrated into the image acquisition pipelines of custom-built OCT scanners.   

%-------------------------------------------------------------------
%-------------------------------------------------------------------
\section*{Acknowledgments}
%-------------------------------------------------------------------
%-------------------------------------------------------------------

We thank our funding sources: NIH 1R01EY021641, Core Grant for Vision Research EY008098-28, DOD awards W81XWH-14-1-0371 and W81XWH-14-1-0370, CMU GSA. We thank NVIDIA Corporation for their GPU donations.

%-------------------------------------------------------------------
%-------------------------------------------------------------------
\section*{Disclosures}
%-------------------------------------------------------------------
%-------------------------------------------------------------------

The authors declare that there are no conflicts of interest related to this article.

\end{document}